%% file: SUS-24-001_temp.tex
\pdfoutput=1
\documentclass[11pt,twoside,a4paper,cmspaper,final,collab]{cms-tdr}
\def\svnVersion{1c662e0}\def\svnDate{2025/06/18}\def\cmsCernNoTag{CERN-EP-2024-335}\def\cmsCernDate{\today}\def\cmsMessage{Published in the Journal of High Energy Physics as \href{http://dx.doi.org/10.1007/JHEP06(2025)144}{\doi{10.1007/JHEP06(2025)144}.}}
\begin{document}\cmsNoteHeader{SUS-24-001}

\cmsNoteHeader{SUS-24-001}

\newcommand{\cosba}{\ensuremath{\cos(\beta-\alpha)}\xspace}
\newcommand{\typetwo}{\textit{Type-II}\xspace}
\newcommand{\flipped}{\textit{Flipped}\xspace}
\newcommand{\mA}{\ensuremath{m_{\PSA}}\xspace}
\newcommand{\mh}{\ensuremath{m_{\Ph}}\xspace}
\newcommand{\mH}{\ensuremath{m_{\PH}}\xspace}
\newcommand{\mAH}{\ensuremath{m_{\PSA,\PH}}\xspace}
\newcommand{\mPhi}{\ensuremath{m_{\phi}}\xspace}
\newcommand{\mZ}{\ensuremath{m_{\PZ}}\xspace}
\newcommand{\mjj}{\ensuremath{M_{12}}\xspace}
\newcommand{\mhmax}{\ensuremath{m_{\Ph}^{\text{max}}}\xspace}
\newcommand{\mhmod}{\ensuremath{m_{\Ph}^{\text{mod}}}\xspace}
\newcommand{\mhmodp}{\ensuremath{m_{\Ph}^{\text{mod+}}}\xspace}
\newcommand{\mhmodm}{\ensuremath{m_{\Ph}^{\text{mod--}}}\xspace}
\newcommand{\mhmass}{\ensuremath{M_{\Ph}^{125}}\xspace}
\newcommand{\CP}{\ensuremath{CP}\xspace}

\newlength\cmsTabSkip\setlength{\cmsTabSkip}{1ex}

\title{Search for bosons of an extended Higgs sector in \PQb quark final states in proton-proton collisions at \texorpdfstring{$\sqrt{s} = 13\TeV$}{sqrt(s) = 13 TeV}}

\date{\today}

\abstract{
A search for beyond-the-standard-model neutral Higgs bosons decaying to a pair of bottom quarks, and produced in association with at least one additional bottom quark, is performed with the CMS detector. The data were recorded in proton-proton collisions at a centre-of-mass energy of 13\TeV at the CERN LHC and correspond to an integrated luminosity of 36.7--126.9\fbinv, depending on the probed mass range. No signal above the standard model background expectation is observed. Upper limits on the production cross section times branching fraction are set for Higgs bosons in the mass range of 125--1800\GeV. The results are interpreted in benchmark scenarios of the minimal supersymmetric standard model, as well as suitable classes of two-Higgs-doublet models.
}

\hypersetup{
pdfauthor={CMS Collaboration},
pdftitle={Search for bosons of an extended Higgs sector in b quark final states in proton-proton collisions at sqrt(s) = 13 TeV},
pdfsubject={CMS},
pdfkeywords={CMS, Higgs boson, bottom quark, MSSM, 2HDMs}
}

\maketitle 

\section{Introduction}
\label{sec:Intro}
The confirmation of the Higgs boson's existence by the ATLAS and CMS Collaborations~\cite{Aad:2012tfa,Chatrchyan:2012xdj,CMS:2013btf} at the CERN LHC, with a mass of 125\GeV, reinforces the standard model (SM) as the most promising theory to date for understanding the fundamental constituents of matter and their mutual forces. The Higgs boson has been experimentally observed in its main production modes, namely gluon fusion and vector boson fusion, as well as production in association with vector bosons and top quark pairs~\cite{ATLAS:2022vkf, CMS:2022dwd}. A recent CMS study has placed stringent constraints on its production in association with b quark pairs~\cite{CMS:2024goa}. Also, the most prominent decay modes have been observed, including those to massive fermions of the third family~\cite{ATLAS:2022vkf, CMS:2022dwd}. Measurements are being extended to include decays to the second family of fermions. Evidence for Higgs boson decays to muon pairs has been established~\cite{CMS:2020xwi}, and upper limits on its branching fraction to charm quark pairs have been determined~\cite{ATLAS:2022ers,CMS:2022psv}. The measurements of Higgs boson properties have so far been consistent with the SM expectations~\cite{ATLAS:2022vkf,CMS:2022dwd,ATLAS:2019nkf,CMS:2018uag}.

Although the SM has been remarkably successful, it is nevertheless regarded as an incomplete theory. Various theories beyond the SM have been proposed that extend its Higgs sector~\cite{Steggemann:2020egv}. The addition of a second complex Higgs doublet leads to two-Higgs-doublet models (2HDMs)~\cite{Branco:2011iw}. The minimal supersymme\-tric extension MSSM~\cite{Nilles19841} also features a Higgs sector with two complex Higgs doublets. Both cases result in five physical states of the Higgs bosons: two charged Higgs bosons \PHpm and three neutral ones, jointly denoted as $\phi$. Under the assumption of charge and parity (\CP) conservation, the neutral Higgs bosons are one \CP-odd (\PSA) and two \CP-even (\Ph, \PH) states, where \Ph denotes the lighter \CP-even state, usually associated with the Higgs boson discovered at a mass of 125\GeV.

In the SM, the Higgs boson has large couplings to \PW and \PZ bosons. The Higgs boson mass of 125\GeV being below the threshold for \PW and \PZ pairs prevents those channels from dominating the decay width, resulting in \bbbar being the most abundant decay mode. However, the situation changes for additional Higgs bosons. In a wide class of extended Higgs sector models, including those under consideration in this paper, a sum rule at lowest order implies that the squared couplings to gauge bosons of all neutral Higgs bosons sum up to the squared coupling of the SM Higgs boson to gauge bosons~\cite{Drechsel:2018mgd}. As the Higgs boson at 125\GeV is known to have gauge boson couplings within less than 10\% from their SM values, the sum rule is close to saturated, and the gauge boson couplings of any additional scalar Higgs bosons must be significantly suppressed compared with those of an SM Higgs boson of the same mass. On the other hand, there is no such sum rule limitation for fermions. Even very heavy additional Higgs bosons may thus have large and dominant branching fractions to fermion pairs. For this reason, third-generation fermion pairs can be considered most promising channels in the search for heavy Higgs bosons.

In the most general case, 2HDMs allow for flavour-changing neutral currents and \CP-violating terms at tree level. Since these effects have not been observed experimentally, it is common to suppress both by imposing a discrete $\mathbb{Z}_2$ symmetry. This leads to four types of models with natural flavour conservation: \textit{Type-I}, \typetwo, \textit{Lepton-specific} (or \textit{Type-X}), and \flipped (or \textit{Type-Y}) models~\cite{Branco:2011iw}. In the \textit{Type-I} model, every charged fermion interacts with the same Higgs doublet. The \typetwo model distinguishes between up-type quarks (\PQu, \PQc, \PQt), which couple to one doublet, and down-type fermions (\PQd, \PQs, \PQb, \Pe, \PGm, \PGt), which couple to the other. This structure is also present in the MSSM. The \textit{Lepton-specific} model assigns all charged leptons to interact with one doublet and all quarks with the other. In contrast, the \flipped model inverts this arrangement of the down-type fermions, with charged leptons and up-type quarks coupled to one doublet and down-type quarks to the other. While the \textit{Type-I} and \typetwo models have undergone extensive study, the \flipped model has not been explored as thoroughly by experiments. The decay process $\phi \to \bbbar$ is particularly well suited for investigating this model due to the potentially large branching fraction of the Higgs boson to \PQb quark pairs, but experimentally very challenging because of overwhelming backgrounds. 

The 2HDMs with \CP conservation are characterised by seven free parameters. These parameters include the masses of the various Higgs bosons (\mh, \mH, \mA, $m_{\PHpm}$), the mixing angle between the \CP-even Higgs bosons ($\alpha$), the ratio of the vacuum expectation values of the two Higgs doublets defined as $\tanb = v_2/v_1$, and the parameter $m_{12}$, which describes the potential mixing between the two Higgs doublets. In the limit $\cosba \to 0$, referred to as the alignment limit, the lighter \CP-even Higgs particle, \Ph, manifests properties that are identical to those of the SM Higgs boson at the same mass across all four types of models.

In the case of the MSSM, the Higgs sector has the structure of a \typetwo 2HDM. At tree level, the Higgs boson masses and $\alpha$ are constrained by the fermion-boson symmetry. Such constraints simplify the model to have only two free parameters at this level. These parameters are commonly chosen to be \mA and \tanb. Since the discovery of the Higgs boson at the LHC, the MSSM benchmark scenarios have been refined for better agreement with experimental observations~\cite{Carena:2013qia, Carena:2005ek}. In this analysis, interpretations in the context of the \mhmass~\cite{Bagnaschi:2018ofa,Bahl:2020kwe}, the \mhmod~\cite{Carena:2013qia}, and the hMSSM~\cite{Maiani:2013hud,Djouadi:2013uqa,Djouadi:2015jea} scenarios are considered. 

For \tanb values larger than one, the couplings of the heavy neutral Higgs bosons to \PQb quarks are enhanced both in the \typetwo and \flipped models, and thus also in the MSSM. This increases the cross section for Higgs boson production in association with \PQb quarks, which is thus a promising signature to study. In addition, there is an approximate mass degeneracy between the \PSA and \PH bosons in the MSSM for the studied range of \mA. For the 2HDMs, we are considering scenarios in which the masses of these bosons are assumed to be equal~\cite{Haber:2015pua}. These effects enhance the combined cross section for producing these Higgs bosons in association with \PQb quarks by a factor of up to $2\tanbsq$ with respect to the SM. The decay $(\PSA, \PH)\to\bbbar$ is expected to have a relatively high branching fraction, even at large values of the Higgs boson mass~\cite{Heinemeyer:2013tqa}. 

Searches for additional Higgs bosons in the \bbbar decay mode have been performed previously at the CERN LEP~\cite{Schael:2006cr} and by the CDF and \DZERO Collaborations at the Fermilab Tevatron~\cite{PhysRevD.86.091101}. At the LHC, the analyses in this decay mode with associated \PQb jets have initially been performed by the CMS Collaboration using the 7 and 8\TeV data sets~\cite{Chatrchyan:2013qga,Khachatryan:2015tra}. Thereafter, the ATLAS and CMS Collaborations have published first results from data at 13\TeV~\cite{ATLAS:2019tpq, CMS:2018hir}, with corresponding integrated luminosities of 27.8\fbinv (recorded in 2015 and 2016) and 35.7\fbinv (recorded in 2016), respectively. In the absence of a significant excess of events, upper limits on the $\Pp\Pp \to \PQb\phi(\to \bbbar) + \mathrm{X}$ cross section have been provided by the ATLAS and CMS Collaborations in the 450--1400 and 300--1300\GeV mass ranges, respectively. The results were also interpreted within the 2HDMs and the MSSM benchmark scenarios.
 
\begin{figure}[htbp]\centering
	\includegraphics[width=0.28\textwidth]{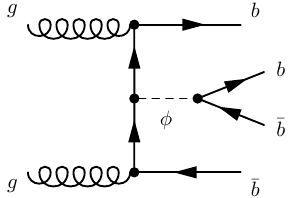}
	\includegraphics[width=0.28\textwidth]{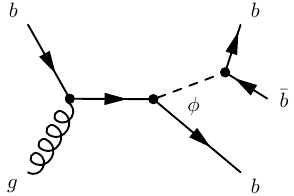}
	\\
	\includegraphics[width=0.28\textwidth]{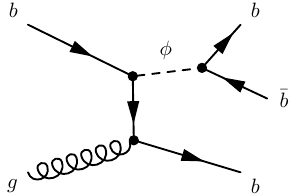}
	\caption{Example Feynman diagrams for the signal processes.}
	\label{fig:signalprocesses}
\end{figure}

In this paper we present a search for additional Higgs bosons with masses in the range 125--1800\GeV that decay into a \bbbar pair and are produced in association with at least one additional \PQb quark. The analysis focusses on a search for neutral Higgs bosons $\phi$ (\Ph, \PH, and \PSA) with mass \mPhi that are produced in association with at least one \PQb quark and decay to \bbbar, as shown by the diagrams in Fig.~\ref{fig:signalprocesses}. The data were collected in proton-proton ($\Pp\Pp$) collisions by the CMS detector at the LHC with a centre-of-mass energy of $\sqrt{s} = 13\TeV$. The analysis is based on data taken in the years 2017--2018, and eventually combined with the previously published results based on the 2016 data~\cite{CMS:2018hir}, in total comprising an integrated luminosity of 36.7--126.9\fbinv depending on the probed mass range. Compared to the previous CMS results in this final state~\cite{CMS:2018hir}, this analysis benefits from a larger data set, improved \PQb jet identification using the \textsc{DeepJet} algorithm~\cite{Bols:2020bkb}, a more robust background estimation strategy based on transfer factors, and improved track reconstruction from the upgrade of the CMS pixel detector~\cite{CMSTrackerGroup:2020edz}.

The paper is organised as follows. Section~\ref{sec:Detector} gives an overview of the CMS detector. Section~\ref{sec:Recon} describes the event reconstruction. Data and simulations used in this analysis are detailed in Section~\ref{sec:Data}, while Section~\ref{sec:Select} summarises the analysis selection. Sections~\ref{sec:Signal} and \ref{sec:Background} describe signal and background models, respectively, with parametric approaches, and Section~\ref{sec:Syst} the systematic uncertainties. Section~\ref{sec:Result} contains the results of the analysis and interpretations in the MSSM and general 2HDMs. Finally, a brief summary of the paper is provided in Section~\ref{sec:Summary}. Tabulated results are provided in the HEPData record for this analysis~\cite{hepdata}.

\section{The CMS detector}
\label{sec:Detector}
The central feature of the CMS apparatus is a superconducting solenoid of 6\unit{m} internal diameter, providing a magnetic field of 3.8\unit{T}. Within the solenoid volume are a silicon pixel and strip tracker, a lead tungstate crystal electromagnetic calorimeter (ECAL), and a brass and scintillator hadron calorimeter (HCAL), each composed of a barrel and two endcap sections. Forward calorimeters extend the pseudorapidity coverage provided by the barrel and endcap detectors. Muons are measured in gas-ionisation detectors embedded in the steel flux-return yoke outside the solenoid. More detailed descriptions of the CMS detector, together with a definition of the coordinate system used and the relevant kinematic variables, can be found in Refs.~\cite{CMS:2008xjf, CMS:2023gfb}.

Events of interest are selected using a two-tiered trigger system. The first level (L1), composed of custom hardware processors, uses information from the calorimeters and muon detectors to select events at a rate of around 100\unit{kHz} within a fixed latency of about 4\unit{$\mu$s}~\cite{CMS:2020cmk}. The second level, known as the high-level trigger, consists of a farm of processors running a version of the full event reconstruction software optimised for fast processing, and reduces the event rate to around 1\unit{kHz} before data storage~\cite{CMS:2016ngn, Hayrapetyan_2024}.

\section{Event reconstruction}
\label{sec:Recon}
The particle-flow event reconstruction~\cite{Sirunyan:2017ulk} aims to reconstruct and identify each individual particle in an event, with an optimized combination of all subdetector information. In this process, the identification of the particle type (photon, electron, muon, charged hadron, neutral hadron) plays an important role in the determination of the particle direction and energy. Photons (including those coming from \PGpz decays or from electron bremsstrahlung) are identified as ECAL energy clusters not linked to the extrapolation of any charged-particle trajectory to the ECAL. Electrons (including those coming from photon conversions in the tracker material or from \PQb hadron semileptonic decays) are identified as a primary charged-particle track and potentially many ECAL energy clusters corresponding to the primary track extrapolation to the ECAL and to possible bremsstrahlung photons emitted while passing through the tracker material. Muons (including those from \PQb hadron semileptonic decays) are identified as tracks in the central tracker consistent with either a track or several hits in the muon system, and associated with calorimeter deposits compatible with the muon hypothesis. Charged hadrons are identified as charged-particle tracks identified neither as electrons nor as muons. Finally, neutral hadrons are identified as HCAL energy clusters not linked to any charged-hadron trajectory, or as a combined ECAL and HCAL energy excess with respect to the expected charged-hadron energy deposit.
The primary vertex is taken to be the vertex corresponding to the hardest scattering in the event, evaluated using tracking information alone, as described in Section 9.4.1 of Ref.~\cite{CMS-TDR-15-02}.

For each event, jets are clustered from these reconstructed particles using the infrared- and collinear-safe anti-\kt algorithm~\cite{Cacciari:2008gp, Cacciari:2011ma} with a distance parameter of 0.4. Jet momentum is determined as the vectorial sum of all particle momenta in the jet, and is found from simulation to be, on average, within 5 to 10\% of the true momentum over the entire transverse momentum (\pt) spectrum and detector acceptance. Additional pp interactions within the same or nearby bunch crossings (pileup) can contribute additional tracks and calorimetric energy depositions to the jet momentum. To mitigate this effect, charged particles identified to originate from pileup vertices are discarded, and an offset correction is applied to correct for remaining contributions. Jet energy corrections are derived from simulation to bring the measured response of jets to that of particle level jets on average. In situ measurements of the momentum balance in dijet, $\text{photon} + \text{jet}$, $\PZ + \text{jet}$, and multijet events are used to account for any residual differences in the jet energy scale between data and simulation~\cite{CMS:2016lmd}. The jet energy resolution amounts typically to 15--20\% at 30\GeV, 10\% at 100\GeV, and 5\% at 1\TeV~\cite{CMS:2016lmd}. Additional selection criteria are applied to each jet to remove jets potentially dominated by anomalous contributions from various subdetector components or reconstruction failures~\cite{CMS:2020ebo}.

To identify jets resulting from the hadronisation of \PQb quarks~\cite{BTV-16-002}, the \textsc{DeepJet} algorithm is used, as described in Ref.~\cite{Bols:2020bkb}, and provides a unity-normalised score for different jet flavours. A specific value requirement for a jet on the score of the algorithm is referred to as \PQb tagging, and positively identified jets are referred to as \PQb-tagged jets or \PQb jets. In this analysis, the medium working point of the \textsc{DeepJet} algorithm is chosen. This working point corresponds to an expected \PQb jet identification efficiency of ${\approx}80\%$ for an expected misidentification rate for jets originating from light-flavour quarks and gluons (\cPqc quarks) of 1 (15)\%~\cite{CMS-DP-2018-058, CMS-DP-2021-004}.

Muons are measured in the pseudorapidity range $\abs{\eta} < 2.4$, with detection planes made using three technologies: drift tubes, cathode strip chambers, and resistive-plate chambers. Matching muons offline to tracks measured in the silicon tracker results in a relative \pt resolution, for muons with \pt up to 100\GeV, of 1\% in the barrel and 3\% in the endcaps. The \pt resolution in the barrel is better than 7\% for muons with \pt up to 1\TeV~\cite{CMS:2018rym}.

The missing transverse momentum vector \ptvecmiss is computed as the negative vector sum of the transverse momenta of all the PF candidates in an event, and its magnitude is denoted as \ptmiss~\cite{CMS:2019ctu}. The \ptvecmiss is modified to account for corrections to the energy scale of the reconstructed jets in the event.

Events with anomalous high-\ptmiss can be due to a variety of reconstruction failures, detector malfunctions or noncollision backgrounds. Such events are rejected by event filters that are designed to identify more than 85--90\% of the spurious high-\ptmiss events with a mistagging rate less than 0.1\%~\cite{CMS:2019ctu}. Although \ptvecmiss is not directly used in the event selection, such anomalous events can still introduce structures in the mass spectrum, potentially leading to a spurious excess.

\section{Data and simulated samples}
\label{sec:Data}
The data used for the analysis were collected with the CMS detector using LHC $\Pp\Pp$ collisions at $\sqrt{s} = 13\TeV$. The copious rate of background from SM events composed uniquely of jets produced through the strong interaction, referred to as quantum chromodynamics (QCD) multijet events, at the LHC is one of the challenges for this analysis. A dedicated trigger scheme has been employed to address this challenge. These triggers require at least two jets with large \pt, in combination with \PQb tagging requirements, and optionally a muon candidate compatible with a semileptonic \PQb hadron decay. More details can be found in Section~\ref{sec:Select}.

Simulated samples of signal and background processes are produced using different event generators. The signal samples are produced at next-to-leading order (NLO) in the strong coupling constant \alpS using the \POWHEG 2.0 implementation~\cite{Nason:2004rx, Frixione:2007vw, Alioli:2010xd, Jager:2015hka} in the four-flavour scheme (4FS)~\cite{Maltoni:2012}. Multijet background events from QCD processes are generated at leading order with the \MGvATNLO event generator v2.6.5 ~\cite{Alwall:2011uj, Alwall:2014hca, Frixione:2002ik} using the five-flavour scheme (5FS) and MLM jet merging~\cite{Alwall:2007fs}. These multijet samples are used for studying qualitative features of the background, but not for a quantitative background prediction. More details are described in Section~\ref{sec:Background}.

The PDF4LHC15~\cite{Butterworth:2015oua} parton distribution functions (PDFs) are used for the generation of the signal process, while the NNPDF3.1~\cite{Ball:2017nwa} PDFs are used for multijet background process. The description of the underlying event is parameterised according to the CP5~\cite{Sirunyan:2019dfx} tunes in both cases. For all generated samples, parton showering and hadronisation are modelled using the \PYTHIA event generator~\cite{Sjostrand:2014zea}, version 8.230. Pileup events are generated with \PYTHIA and are added according to the expected pileup profile in the observed data. All generated events are passed through a \GEANTfour-based~\cite{Agostinelli:2002hh} simulation of the CMS detector and reconstructed using the same version of the CMS event reconstruction software used for the data.

\section{Event selection}
\label{sec:Select}
In the following, the jets are assumed to be sorted in the order of decreasing \pt. Only events with at least three jets are considered. The terms ``leading jet'', ``second leading jet'' and ``third leading jet'' refer to the first, second and third item in this ordering. Similarly, the first two or three items may be collectively referred to as ``the two'' or ``the three leading jets''.

The search for $\phi \to \bbbar$ decays is based on two different signatures. The fully hadronic (FH) selection requires at least the three leading jets to be \PQb-tagged and is applied for the 2017--2018 data sets. Due to trigger rate limitations, this selection requires relatively high \pt thresholds for at least the two leading jets, which restricts the sensitivity of the analysis to Higgs boson masses of 300\GeV and above. The semileptonic (SL) selection equally requires at least the three leading jets to be \PQb-tagged, but further requires one muon within the cone of any of the two leading jets in the event, compatible with the semileptonic decay of a \PQb hadron. This results in further enrichment of the \PQb jet signature and permits lower \pt thresholds. In absence of a suitable trigger for the 2018 data taking period, this selection is available only for the 2017 data set; it is only applied for Higgs boson masses up to 700\GeV, beyond which its sensitivity is negligible compared to the FH channel. Combining both selections allows probing neutral Higgs bosons in the mass range starting as low as 125\GeV and extending up to 1800\GeV.
While the signal process may contain four \PQb jets, the selection requires only at least three since one of these four \PQb jets is frequently outside of the acceptance or does not pass the kinematic thresholds.
The analysis aims to reconstruct $\phi$ candidates by searching for a peak in the invariant mass distribution, \mjj, of the two leading jets, which are expected to originate from the Higgs boson decay in the majority of the signal events. For $\mPhi \geq 160\GeV$, the efficiency for correctly pairing the Higgs boson daughter jets exceeds 50\% and reaches about 96\% at the highest mass point, while it ranges down to 32\% towards the lowest masses in the 2017 SL selection.
The dominant background arises from the production of heavy-flavour multijet events containing either three \PQb jets, or two \PQb jets plus a third jet that is misidentified as a \PQb jet and originates from either a charm quark, a light-flavour quark, or a gluon. Eventually, the results from the 2017--2018 data sets are combined with published results from the 2016 data set~\cite{CMS:2018hir}.

For the 2017 (2018) data set, FH events are selected at the trigger level by requiring at least two jets in the range of $\abs{\eta} < 2.3$ with $\pt > 100\,(112)\GeV$. An event is accepted only if the absolute value of the difference in pseudorapidity, $\Delta \eta$, between any two of these jets is less than or equal to 1.6, since this requirement further reduces the trigger rates while preserving a high trigger efficiency in the probed mass range of the Higgs bosons. At least two jets having $\pt > 80\GeV$ must be identified as \PQb jets. 
This \PQb jet identification uses the \textsc{DeepCSVv2} algorithm~\cite{BTV-16-002} with slightly tighter requirements than for the offline analysis.
For the SL events, the online selection is similar, but the \pt threshold for the leading jets is lowered to 40\GeV, at least two jets with $\pt > 30\GeV$ must be identified as \PQb jets, and, in addition, at least one muon with $\pt > 12\GeV$ is required. The muon requirement considerably reduces the trigger rates and thus allows the reduction of the leading jet \pt thresholds. The resulting SL trigger rate is thus only about 80\% larger than the rate of the FH trigger.

\begin{figure}[!htbp]\centering
    \includegraphics[width=0.49\textwidth]{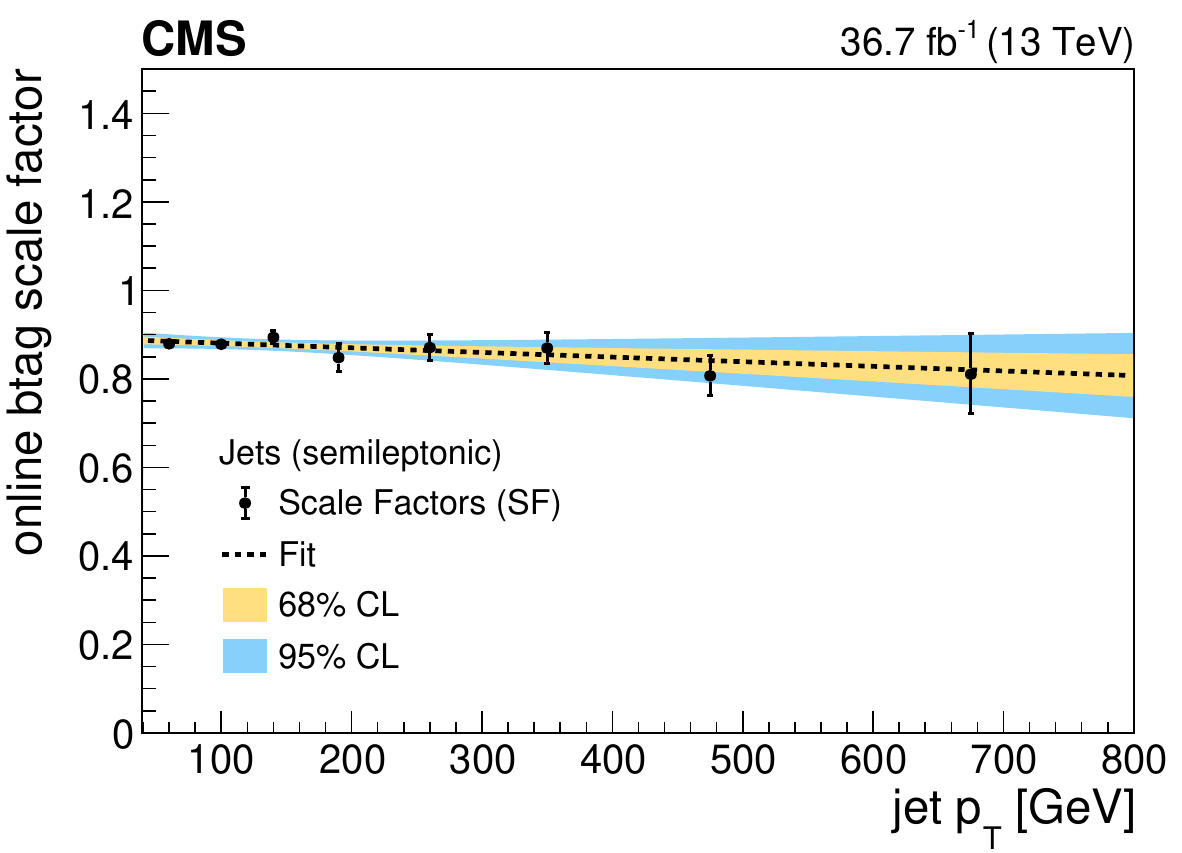}
    \includegraphics[width=0.49\textwidth]{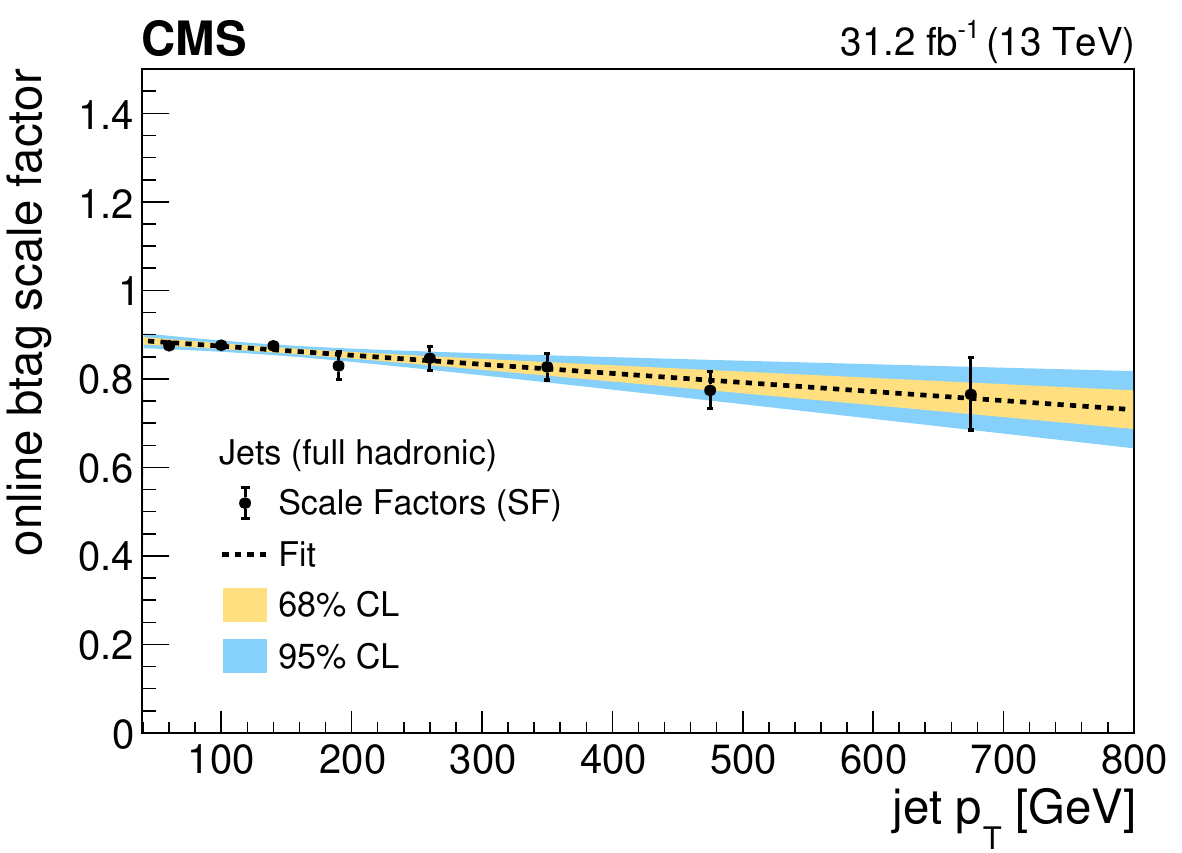}
    \includegraphics[width=0.49\textwidth]{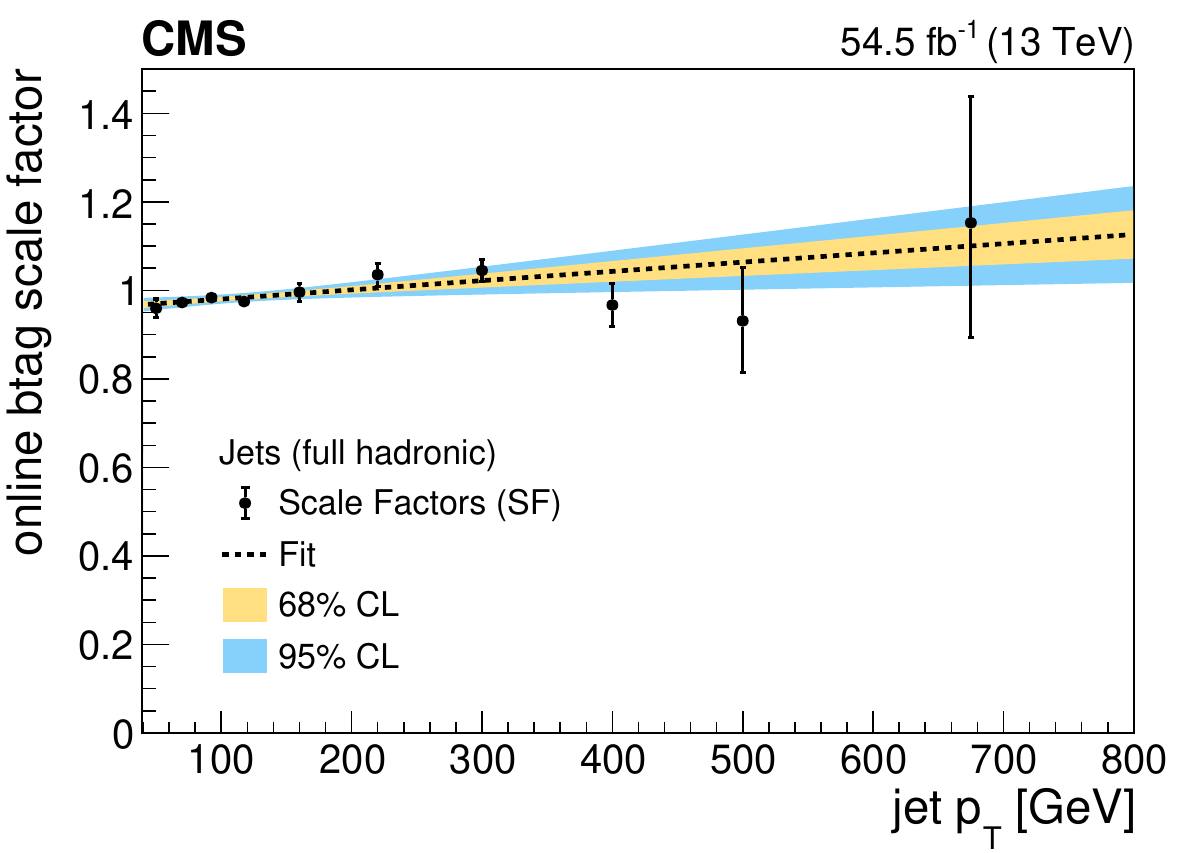}
    \includegraphics[width=0.49\textwidth]{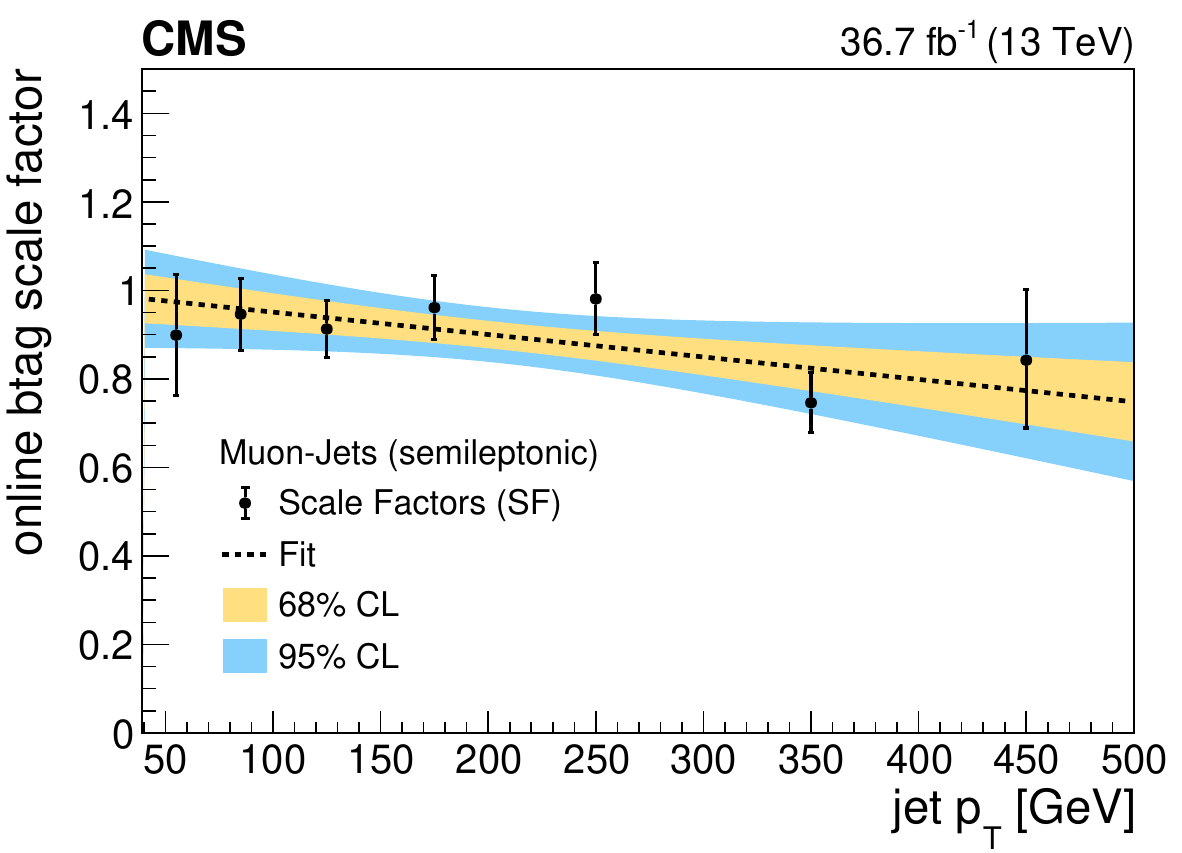}
	\caption{Online \PQb tagging scale factors for \PQb jets in the 2017 SL (upper left), 2017 FH (upper right) and 2018 FH (lower left) analyses, and for \PQb jets with muons in the 2017 SL analysis (lower right). The results of linear fits are also shown, together with  ${\pm}1\sigma$ and ${\pm}2\sigma$ uncertainty bands.}
\label{fig:onlineBtagSF}
\end{figure}
The efficiency of the jet \pt requirements in the trigger is estimated as a function of the jet \pt and $\eta$ from data collected with a prescaled single-jet trigger with lower \pt threshold. This trigger is fully efficient in the kinematic range under consideration, and the fraction of such events fulfilling also the requirement of the nominal jet trigger measures its efficiency. The same procedure is applied to simulated events of multijet production. The ratio of these two efficiencies determines the scale factors that are used to correct all simulated samples in the analysis. The scale factor displays a sharp turn-on versus jet \pt just above the nominal threshold, and saturates quickly at unity. 

The online \PQb tagging efficiencies relative to the offline \PQb tagging selection are obtained from data using prescaled dijet triggers with a single-\PQb-tag requirement. A tag-and-probe method~\cite{CMS:2014pkt} is employed to determine the online \PQb tagging efficiency as a function of jet \pt. The two leading jets are required to pass an offline kinematic selection and \PQb tagging similar to the final analysis selection as described above. The second-leading jet must always pass the online \PQb tagging requirement to ensure that it has passed the trigger requirement. It is then checked whether the first-leading jet also satisfies the online \PQb tagging requirement; the fraction of such cases is a direct measure of the relative online \PQb tagging efficiency. This procedure is applied in both data and simulation. The ratio of these efficiencies between the data and the simulation provides the online \PQb tagging scale factor, displayed in Fig.~\ref{fig:onlineBtagSF}, and it is found to range within 0.8--1.1 across the relevant jet \pt range.

Using a similar approach, the tag-and-probe method is applied to evaluate the muon trigger efficiency as a function of muon \pt using the $\JPsi \to \PGmp\PGmm$ resonance peak~\cite{CMS:2018rym}. The data are collected with a prescaled single-muon trigger with lower \pt threshold in the double-muon data set and compared with simulations of multijet production. Events are selected in the \JPsi mass window to obtain a pure sample of muons with a soft \pt spectrum. As a result, the efficiency is estimated from the overall normalisation factor of the simultaneous invariant mass fit in the bins of muon \pt. The single-muon trigger efficiency exceeds 90\% over the full $\eta$ range, and the efficiency to reconstruct and identify muons is greater than 96\%. Scale factors are obtained to correct all simulated samples, and fall within 0.90--1.05 across the \pt range of the probe muon.

In the FH channel, the offline selection requires at least three jets within $\abs{\eta} < 2.2$, with the two leading jets having $\pt(j_1) > 110\,(130)\GeV$ and $\pt(j_2) > 100\,(130)\GeV$ in the 2017 (2018) data set, and the third-leading jet having $\pt(j_3) > 40\GeV$. The $\eta$ selection is applied to benefit from optimal \PQb tagging performance, and is safely within the $\eta$ acceptance of the trigger. The three leading jets have to pass the \textsc{DeepJet} \PQb tagging requirement of the medium working point~\cite{BTV-16-002}. 
The $\abs{\Delta \eta}$ between the two leading jets must be less than 1.5, and a pairwise separation of $\Delta R > 1$ between each two of the three leading jets is imposed to suppress background from \bbbar pairs arising from gluon splitting, where $\Delta R = \sqrt{\smash[b]{(\Delta\eta)\,^2 + (\Delta\phi)\,^2}}$ with $\Delta\phi$ being the separation in the azimuthal angle.

In the SL channel, because of lower \pt requirements in the trigger path, events can be selected offline with lower \pt thresholds: $\pt(j_1) > 60 \GeV$, $\pt(j_2) > 50 \GeV$, and $\pt(j_3) > 30 \GeV$ for the three leading jets, respectively. As in the FH selection, the three leading jets have to pass the \textsc{DeepJet} \PQb tagging requirement of the medium working point.
One nonisolated muon passing tight identification~\cite{CMS:2018rym} is required with $\pt(\mu) > 13\GeV$ and $\abs{\eta} < 2.2$. This muon must lie within the cone of either of the two most energetic \PQb jets in the event, satisfying $\Delta R(\mu,j) < 0.4$.

To avoid event duplications between the FH and SL selections, events containing at least one reconstructed muon fulfilling the criteria of the SL selection are vetoed from the FH channel in the 2017 data set. 
Apart from the kinematic requirements above, all objects selected offline are required to match the objects triggered online within an angular separation of $\Delta R < 0.3$. At least two among the selected three leading jets need to match online \PQb jets in both FH and SL channels, while muon matching is required for the SL channel.  In case both of the two leading jets contain a muon, the \PQb jet with the largest muon \pt is required to match the muon requirement in the SL trigger.

Further corrections to improve the mass resolution are applied in the selection process. A \PQb jet energy regression based on a deep neural network allows accounting for missing energy, for example because of neutrinos emitted in semileptonic decays of \PQb hadrons~\cite{CMS:2019uxx}. Furthermore, a final state radiation (FSR) recovery technique is used to account for gluons emitted by \PQb jets before hadronising. Jets not among the three leading jets in the event are treated as FSR candidates. If such a candidate lies within a distance $\Delta R < 0.8$ of one of the three leading jets, its four-momentum is added vectorially to that of the respective leading jet.

The selection detailed above defines the signal region (SR) of this analysis, and events passing this selection are referred to as the ``triple \PQb tag'' sample in the following. To study the features of the multijet background, a control region (CR) is constructed in a similar way except for imposing a \PQb tag veto on the third leading jet. This veto rejects jets that would satisfy a loose \PQb tagging requirement. This requirement is defined by a 10\% probability for light-flavour quark and gluon jets to be misidentified as \PQb jets, and has a \PQb jet identification efficiency of about 70\%. The ``\PQb tag veto'' CR has no overlaps with the triple \PQb tag SR, while it preserves similar kinematic distributions for the three leading jets. In addition, the signal contamination in the CR is negligible. A summary of the offline selection parameters is given in Table~\ref{tab:offlinecuts}.

\begin{table}[htbp]\centering
  \topcaption{Summary of the main parameters of the offline selection for the three data sets 2017 SL, 2017 FH, and 2018 FH, where $j_1$, $j_2$, $j_3$ indicate the first three leading jets. Entries denoted by ``\NA'' indicate that the selection is not applied. Signal region (SR) and control region (CR) only differ in the \PQb\ tag selection, shown in the bottom rows, where ``$>$M'' (``$<$L'') indicate that the respective jet should pass (fail) the requirement of the medium (loose) working point, respectively.}
  \begin{tabular}{lcccccc}
\hline
  Variable & \multicolumn{2}{c}{2017 SL}  
             & \multicolumn{2}{c}{2017 FH} 
             & \multicolumn{2}{c}{2018 FH} \\
\hline
  $\pt(j_1)$ [\GeVns{}] & \multicolumn{2}{c}{$>$60} 
                    & \multicolumn{2}{c}{$>$110} 
                    & \multicolumn{2}{c}{$>$130} \\
  $\pt(j_2)$ [\GeVns{}] & \multicolumn{2}{c}{$>$50} 
                    & \multicolumn{2}{c}{$>$100} 
                    & \multicolumn{2}{c}{$>$130} \\
  $\pt(j_3)$ [\GeVns{}] & \multicolumn{2}{c}{$>$30} 
                    & \multicolumn{2}{c}{$>$40} 
                    & \multicolumn{2}{c}{$>$40} \\
  $\abs{\eta} (j_1, j_2, j_3)$ 
                    & \multicolumn{2}{c}{$<$2.2}
                    & \multicolumn{2}{c}{$<$2.2} 
                    & \multicolumn{2}{c}{$<$2.2} \\
  $\abs{\Delta \eta} (j_1, j_2)$ 
                    & \multicolumn{2}{c}{$<$1.5}
                    & \multicolumn{2}{c}{$<$1.5} 
                    & \multicolumn{2}{c}{$<$1.5} \\
  $\Delta R (j_1, j_2)$, $\Delta R (j_1, j_3)$, $\Delta R (j_2, j_3)$ 
                    & \multicolumn{2}{c}{$>$1}
                    & \multicolumn{2}{c}{$>$1} 
                    & \multicolumn{2}{c}{$>$1} \\
\hline
  $\pt(\mu)$ [\GeVns{}] & \multicolumn{2}{c}{$>$13}    
                    & \multicolumn{2}{c}{$>$13}    
                    & \multicolumn{2}{c}{\NA}  \\
  $\abs{\eta}(\mu)$ & \multicolumn{2}{c}{$<$2.2} 
                    &  \multicolumn{2}{c}{$<$2.2}   
                    &  \multicolumn{2}{c}{\NA}   \\
  min($\Delta R(\mu,j_1)$, $\Delta R(\mu,j_2)$) 
                    & \multicolumn{2}{c}{$<$0.4} 
                    & \multicolumn{2}{c}{$<$0.4}   
                    & \multicolumn{2}{c}{\NA}   \\
  $N(\mu)$          & \multicolumn{2}{c}{$\geq$1}    
                    & \multicolumn{2}{c}{$=$0}    
                    & \multicolumn{2}{c}{\NA}  \\
\hline
                    & SR & CR & SR & CR & SR & CR \\
\hline
  btag ($j_1,j_2$)  & $>$M & $>$M & $>$M & $>$M & $>$M & $>$M\\
  btag ($j_3$)      & $>$M & $<$L & $>$M & $<$L 
                    & $>$M & $<$L \\
\hline
\end{tabular}
\label{tab:offlinecuts}
\end{table}

\section{Signal model}
\label{sec:Signal}
The invariant mass of the two leading jets, \mjj, represents the main analysis variable, and its distribution is used for the extraction of the signal. For each considered mass hypothesis of the Higgs boson, \mPhi, the \mjj distribution is obtained from the corresponding simulated signal sample, for nominal masses in the range of 125--1800\GeV.
  
Various scale factors are measured to correct the signal event simulation to match the data. These include the efficiencies of the trigger kinematic and online \PQb tagging requirements, as described in Section~\ref{sec:Data}, as well as jet energy scale and resolution~\cite{CMS:2016lmd}. Correction factors to account for the different offline \PQb tagging efficiencies~\cite{CMS-DP-2018-058, CMS-DP-2021-004}, and muon identification efficiency~\cite{CMS:2018rym} are also applied.
 
\begin{figure}[htbp]\centering
    \includegraphics[width=0.76\textwidth]{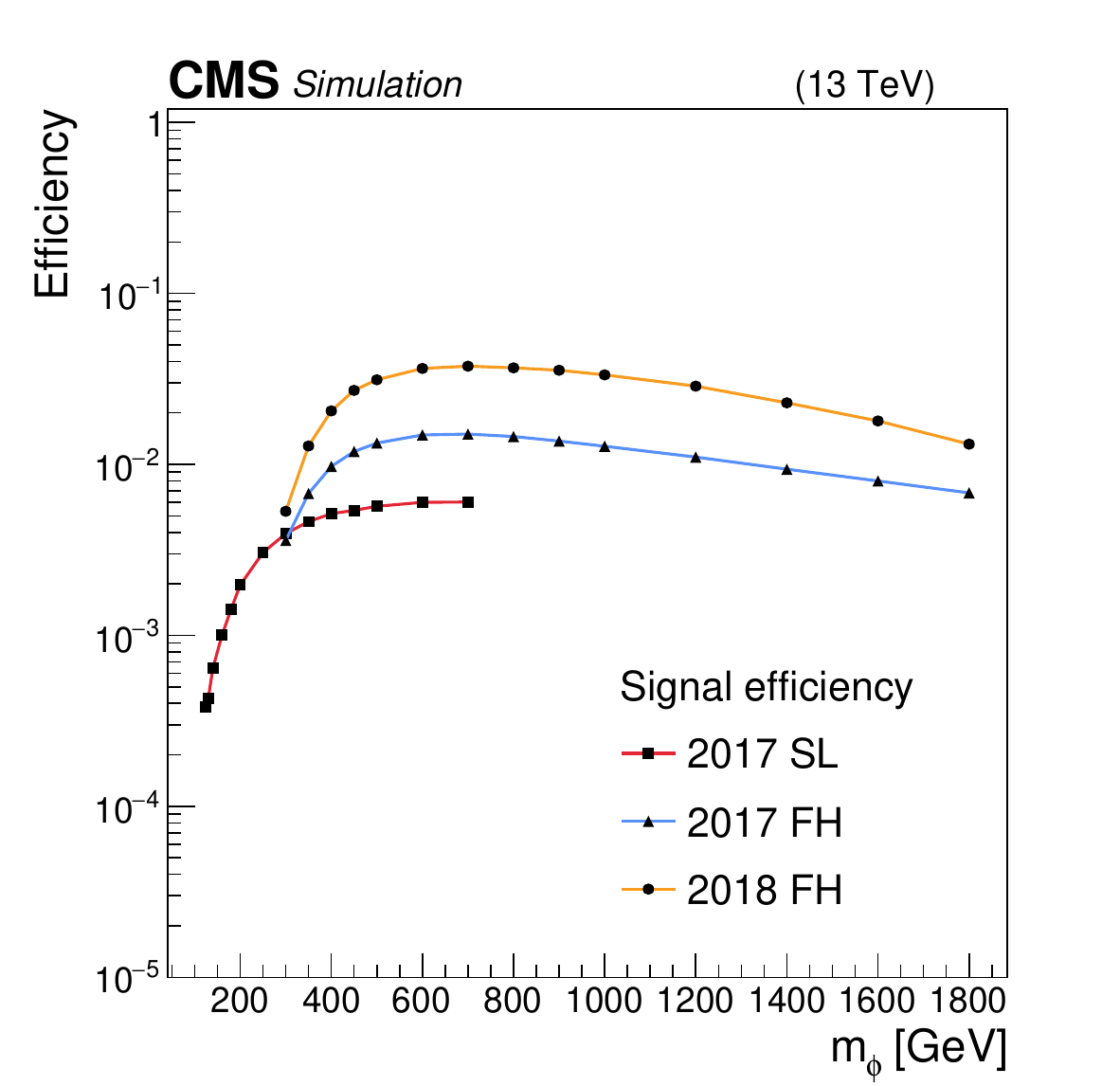}
    \caption{Signal efficiency as a function of the mass \mPhi after triple \PQb tag selection for 2017 SL (squares), 2017 FH (triangles), and 2018 FH (circles) channels.}
\label{fig:signalHEfficiency}
\end{figure}

The signal efficiency after full selection for each value of \mPhi is estimated from simulation and shown in Fig.~\ref{fig:signalHEfficiency}. The total signal efficiency in the SR ranges between 0.3 (0.5) and 1.8 (3.6)\% and peaks around 600\GeV in the 2017 (2018) FH channel. The main limiting factors are the kinematic selection, in particular the \pt thresholds, and the triple \PQb tag requirement, which each reduce the efficiency by about one order of magnitude. A large part of this reduction already occurs in the online selection, whose efficiency ranges from 8 (6) to 16 (23)\% in the 2017 (2018) FH channel, including a sizeable effect from online \PQb tagging. Towards low values of \mPhi, the effect of the \pt thresholds is strongest, while towards large masses, the degradation of the \PQb tagging performance at very large jet \pt is most noticeable. The selection on $\Delta \eta$ of the two leading jets also plays a significant role. The signal efficiency in the 2018 FH data is generally higher compared to 2017 FH because of a less restrictive online \PQb tagging selection.
For the SL channel, the total signal efficiency increases gradually from 0.04 to 0.60\% along the probed mass range of 125--700\GeV, its size compared to the FH channels reflecting the semileptonic branching fraction of \PQb hadrons into final states with muons. In the CR, the signal efficiency is significantly lower than in the SR, typically by a factor of two or more. As the CR is far more populated by background than the SR, by about a factor of five, this results in an order of magnitude relative signal depletion of the CR.

\begin{figure}\centering
    \includegraphics[width=0.48\textwidth]{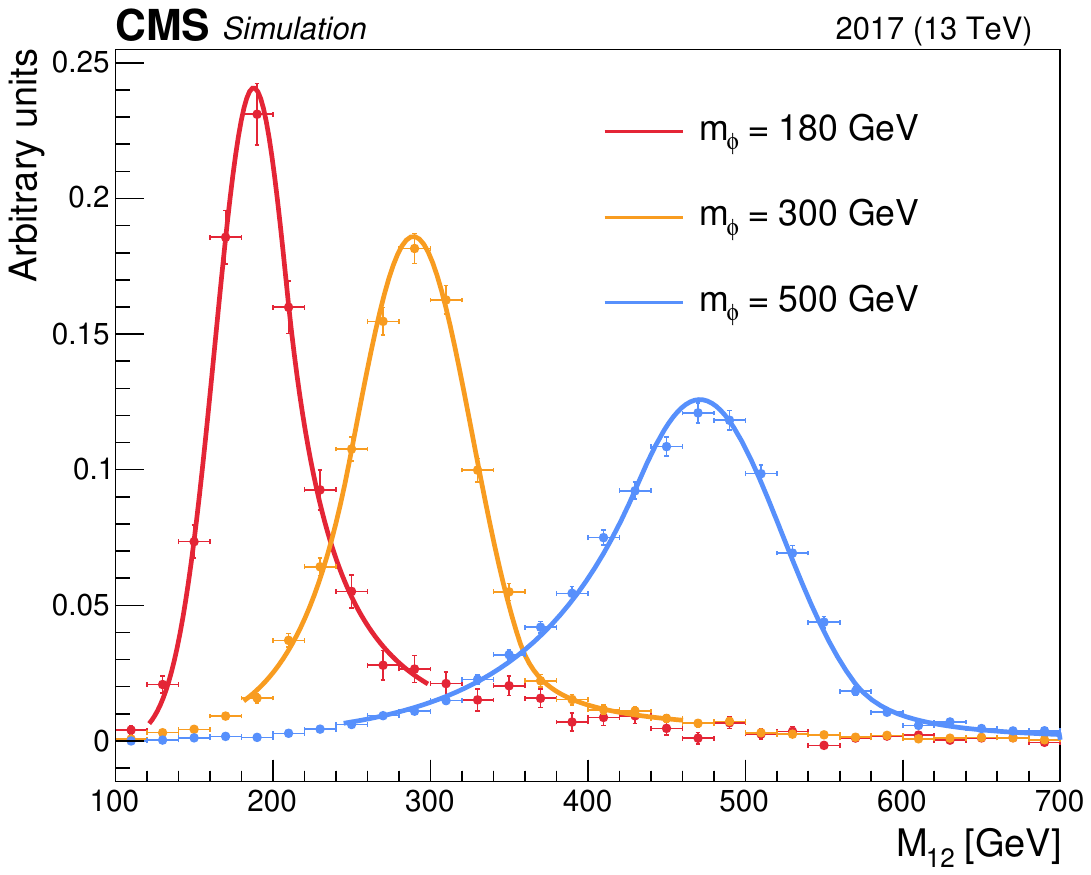}
    \includegraphics[width=0.48\textwidth]{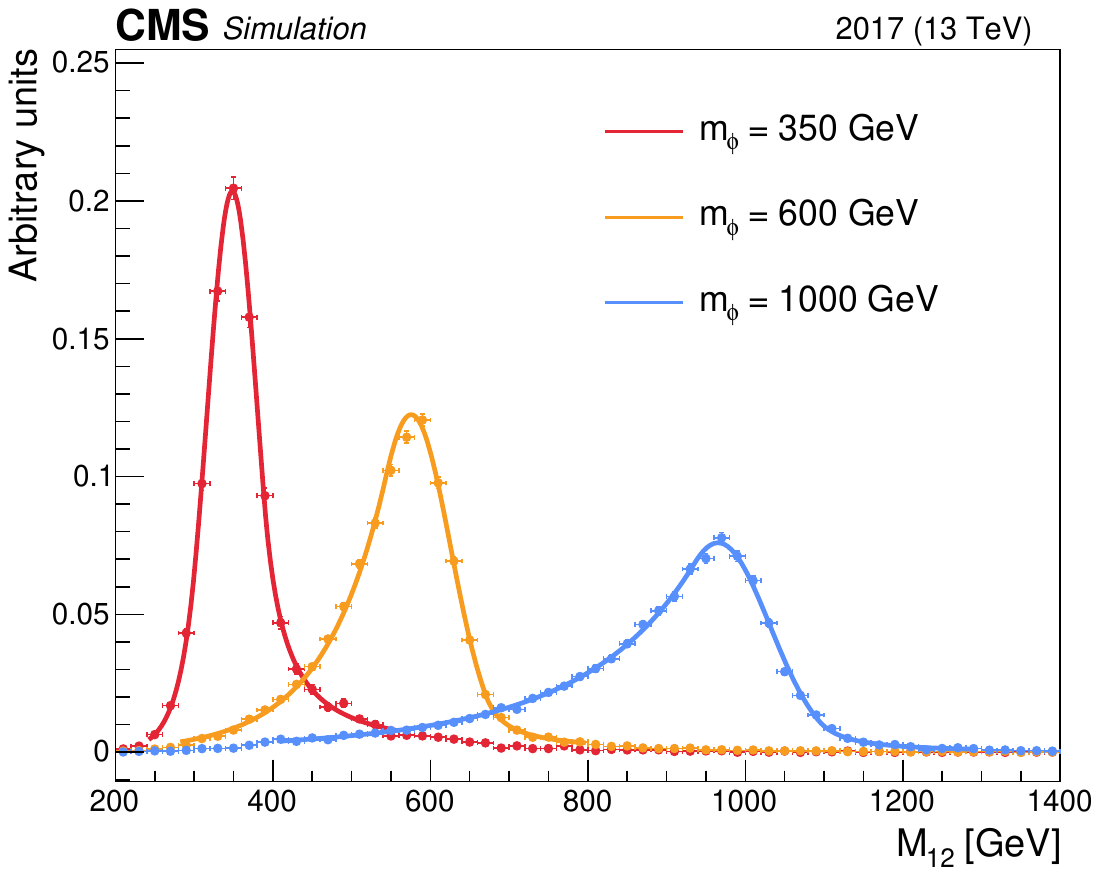}
    \includegraphics[width=0.48\textwidth]{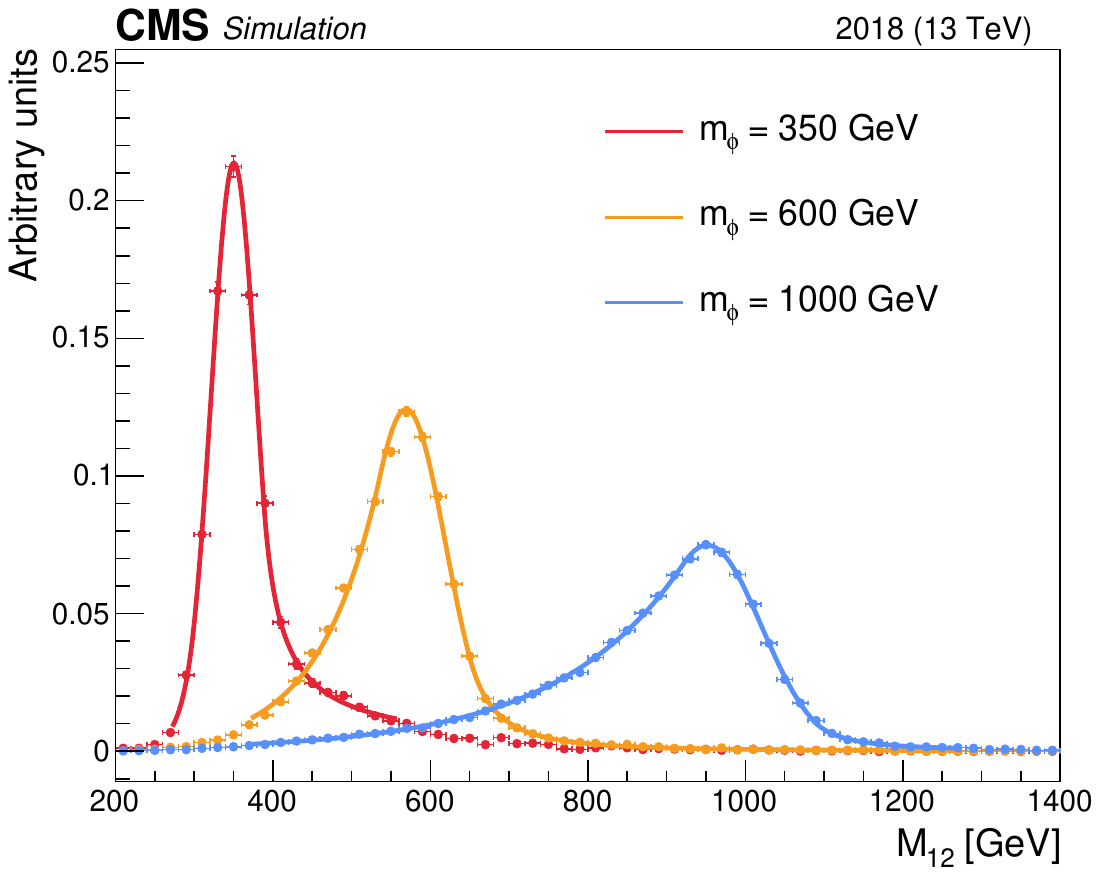}
    \caption{Simulated signal yields normalized to unit area for three representative values of the Higgs boson mass \mPhi in the 2017 SL (upper left), 2017 FH (upper right), and 2018 FH (lower) channels. The solid curves show the signal parameterisations by double-sided Crystal Ball probability density functions.}
\label{fig:signalShapes_2018FH}
\end{figure}

In Fig.~\ref{fig:signalShapes_2018FH}, the distributions of the reconstructed invariant mass of the two leading jets, \mjj, obtained from simulation, are illustrated for the SL and FH channels in the 2017 and 2018 data sets for selected mass points. The natural width expected for an MSSM Higgs boson in the considered mass and \tanb region is negligible compared to the detector resolution~\cite{CMS:2018hir}. Therefore, the observed shape of the mass distribution is largely determined by the experimental resolution and the effects of wrong jet pairing, which occurs when the two leading jets used to compute \mjj are not originating from the Higgs boson decay. Tails towards lower \mjj values are attributed to incomplete reconstruction of the Higgs boson's decay products, particularly due to unaccounted momentum from neutrinos in semileptonic decays of \PQb and \cPqc hadrons. Wrong jet pairing potentially gives rise to tails in both directions, but for the lower mass points the left-side tails are suppressed because of the effect of the jet \pt selection threshold. The resulting full width at half maximum is typically 20\% (25--30\%) of the nominal Higgs boson mass for the FH (SL) channels. 
To describe the \mjj distribution as a smooth shape, each signal is parameterised by a double-sided Crystal Ball probability density function, which has a Gaussian core portion with smoothly-attached low- and high-end power law functions~\cite{Gaiser:1985ix}. The results of the fits are also shown in Fig.~\ref{fig:signalShapes_2018FH}.

\section{Background model}
\label{sec:Background}
The dominant background for this analysis originates from heavy-flavour multijet production yielding at least three energetic jets, of which at least two contain \PQb hadrons. Top quark-antiquark production exhibits an \mjj shape very similar to the multijet process. It is found to be at the level of only a few percent, and is implicitly included in the background model.

In multijet events, the distribution in \mjj falls steeply at large values, while at low values it exhibits a marked turn-on behaviour mainly due to the kinematic selection at the trigger and offline levels. This effect shapes the distributions in the SR and the CR in the same way, since their selections only differ in a \PQb tag requirement versus a \PQb tag veto on the third leading jet. Simulation shows that the \mjj distributions in the SR and CR are indeed very similar in shape~\cite{Vagnerini:2020ajn, Asmuss:2021}. The background model is thus based on the shape of the \mjj distribution in the CR, which is parameterised by a fitted analytic function and multiplied by a function referred to as ``transfer factor'', defined below, that accounts for mild shape differences between the SR and the CR.

For ease of parameterisation, the \mjj range is divided into three (four) overlapping fit ranges for the SL (FH) data sets. These fit ranges are overlapping to ensure that the signal of each mass point, including tails, is sufficiently covered by one mass range. These fit ranges, their boundaries, and the associated Higgs boson mass points are listed in Table~\ref{tab:massrangedefs}.

\begin{table}[htbp]\centering
  \topcaption{Definition of fit ranges for the 2017 SL, 2017 FH, and 2018 FH channels in terms of \mjj and the associated values of the nominal Higgs boson mass, \mPhi, which are probed in this fit range.}
  \begin{tabular}{lccl}
\hline
  Channel & Fit range  &  Mass range (\mjj) [\GeVns{}] & Signal mass (\mPhi) [\GeVns{}] \\
\hline
  2017 SL & 1 & 120--300 & 125, 130, 140, 160, 180, 200 \\
          & 2 & 180--460 & 250, 300, 350 \\
          & 3 & 240--800 & 400, 450, 500, 600, 700 \\[\cmsTabSkip]
  2017 FH & 1 & 240--560   &  300, 350, 400 \\
          & 2 & 280--800   &  450, 500, 600 \\
          & 3 & 400--1300  &  700, 800, 900 \\
          & 4 & 600--2000  &  1000, 1200, 1400, 1600, 1800 \\[\cmsTabSkip]
  2018 FH & 1 & 270--560   & 300, 350 \\
          & 2 & 320--800   & 400, 450, 500 \\
          & 3 & 390--1270  & 600, 700, 800, 900 \\
          & 4 & 500--2000  & 1000, 1200, 1400, 1600, 1800\\
\hline
\end{tabular}
\label{tab:massrangedefs}
\end{table}

\begin{figure}[htbp!]\centering
    \includegraphics[width=0.48\textwidth]{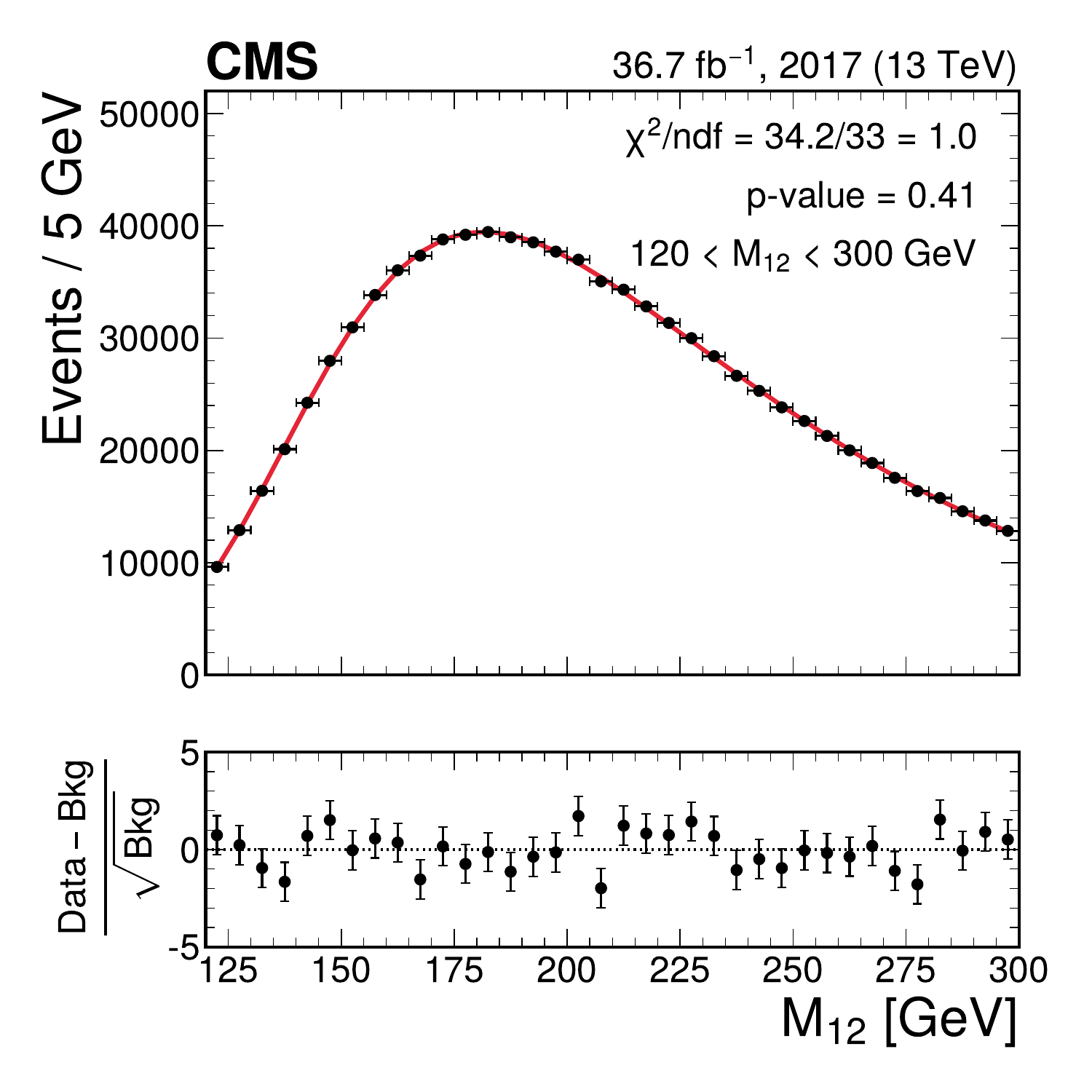}
    \includegraphics[width=0.48\textwidth]{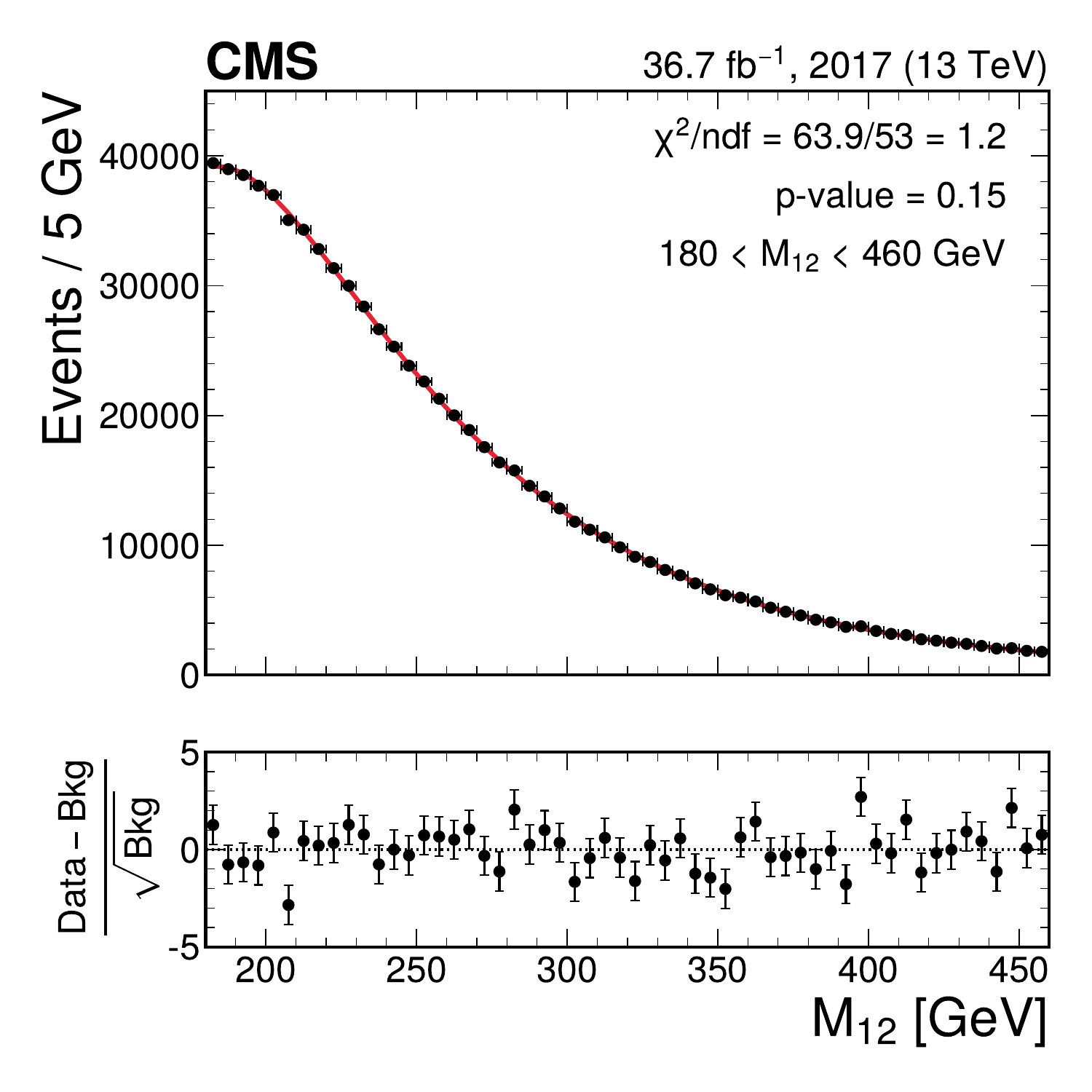}
    \includegraphics[width=0.48\textwidth]{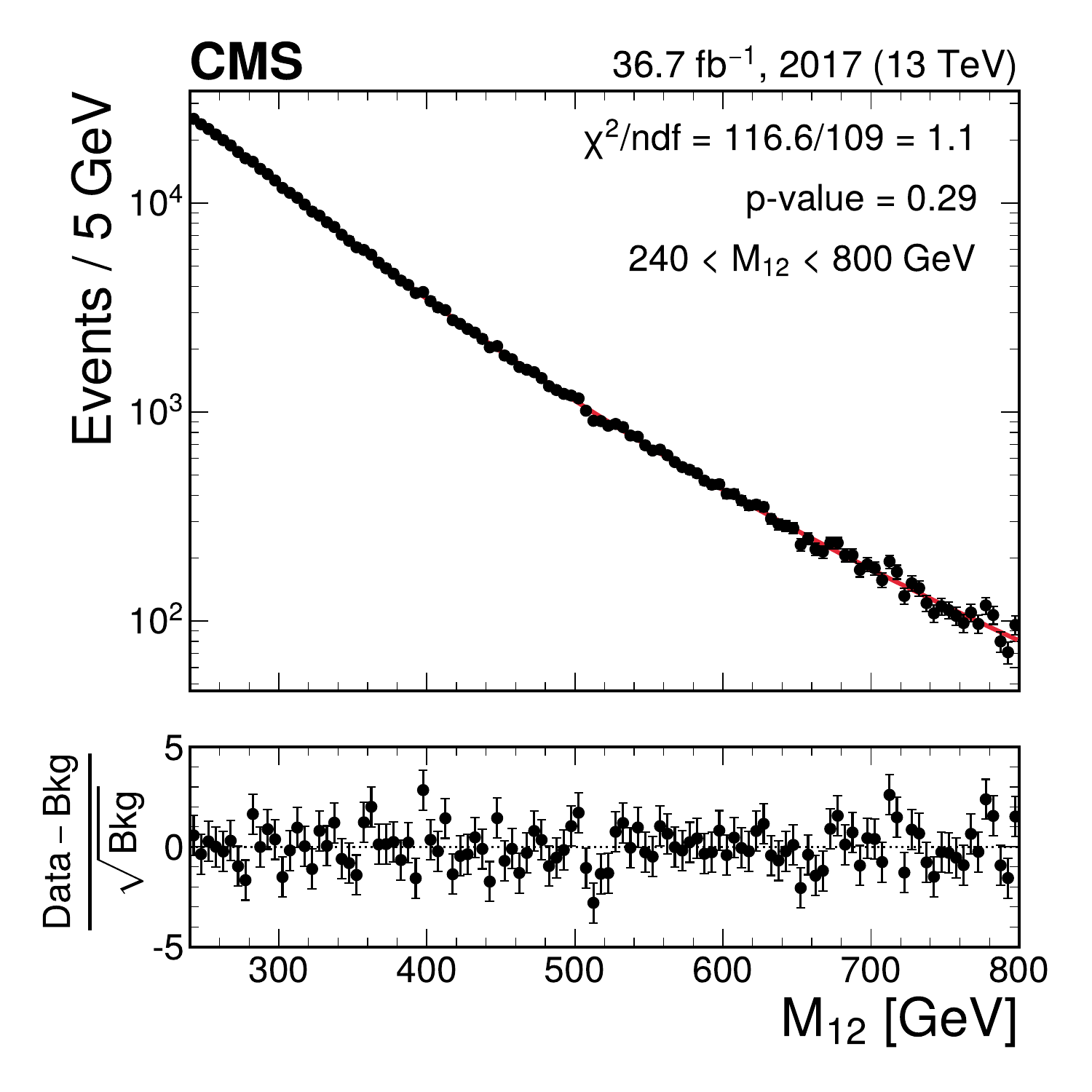}
    \caption{Invariant mass distributions of the three fit ranges in the \PQb tag veto CR for the 2017 SL channel, overlaid with the fitted functions. The distributions are fitted in the \mjj ranges of 120--300\GeV (upper left), 180--460\GeV (upper right), and 240--800\GeV (lower). The $\chi^2$ and the corresponding p-value obtained from the goodness-of-fit test are displayed on each plot. The lower panels show the difference between the data and the fitted function, divided by the estimated statistical uncertainty for each bin. Good agreement between the fitted functions and the data is achieved.}
\label{fig:bbnb_fit_SR1to3_SL2017}
\end{figure}

\begin{figure}[htbp!]\centering
    \includegraphics[width=0.48\textwidth]{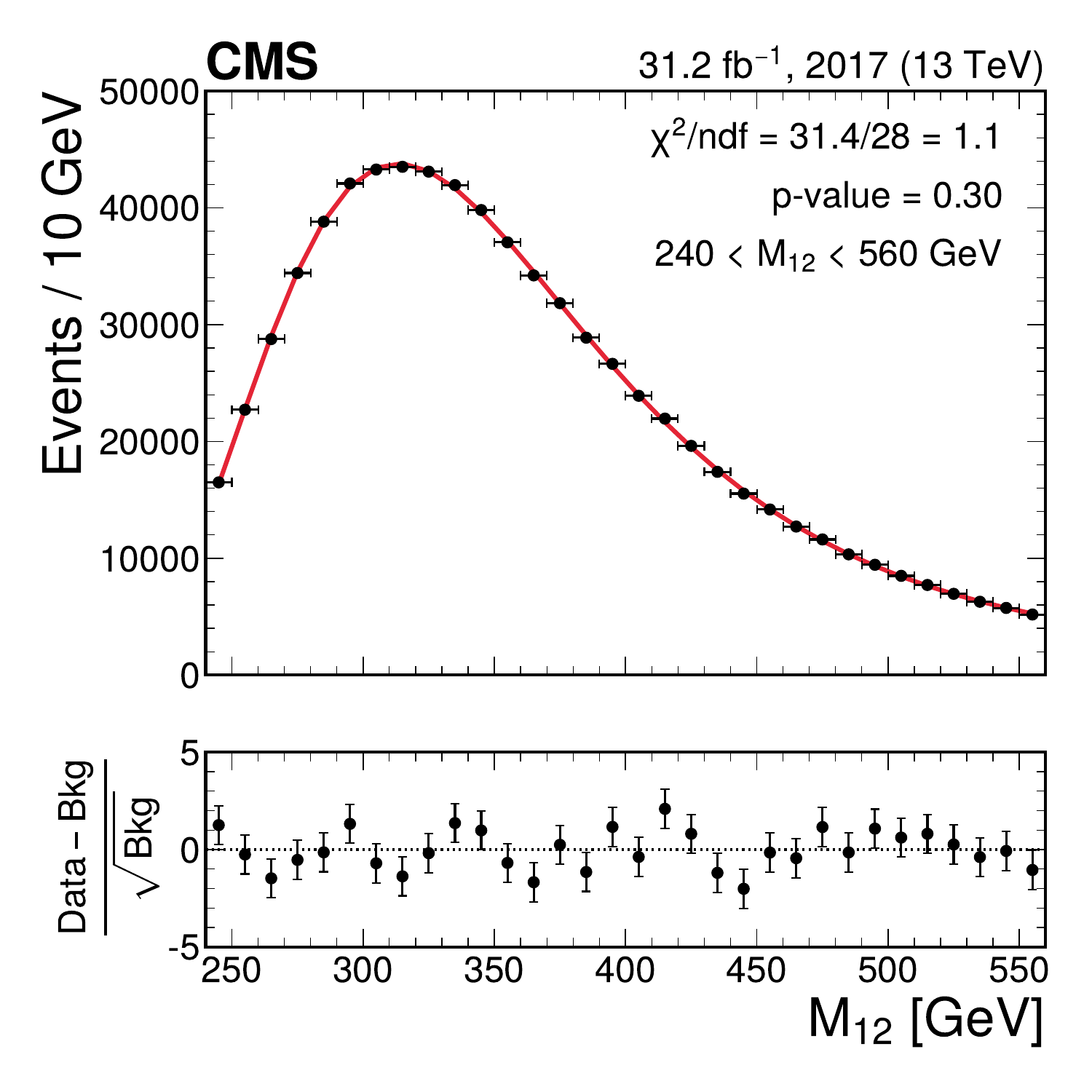}
    \includegraphics[width=0.48\textwidth]{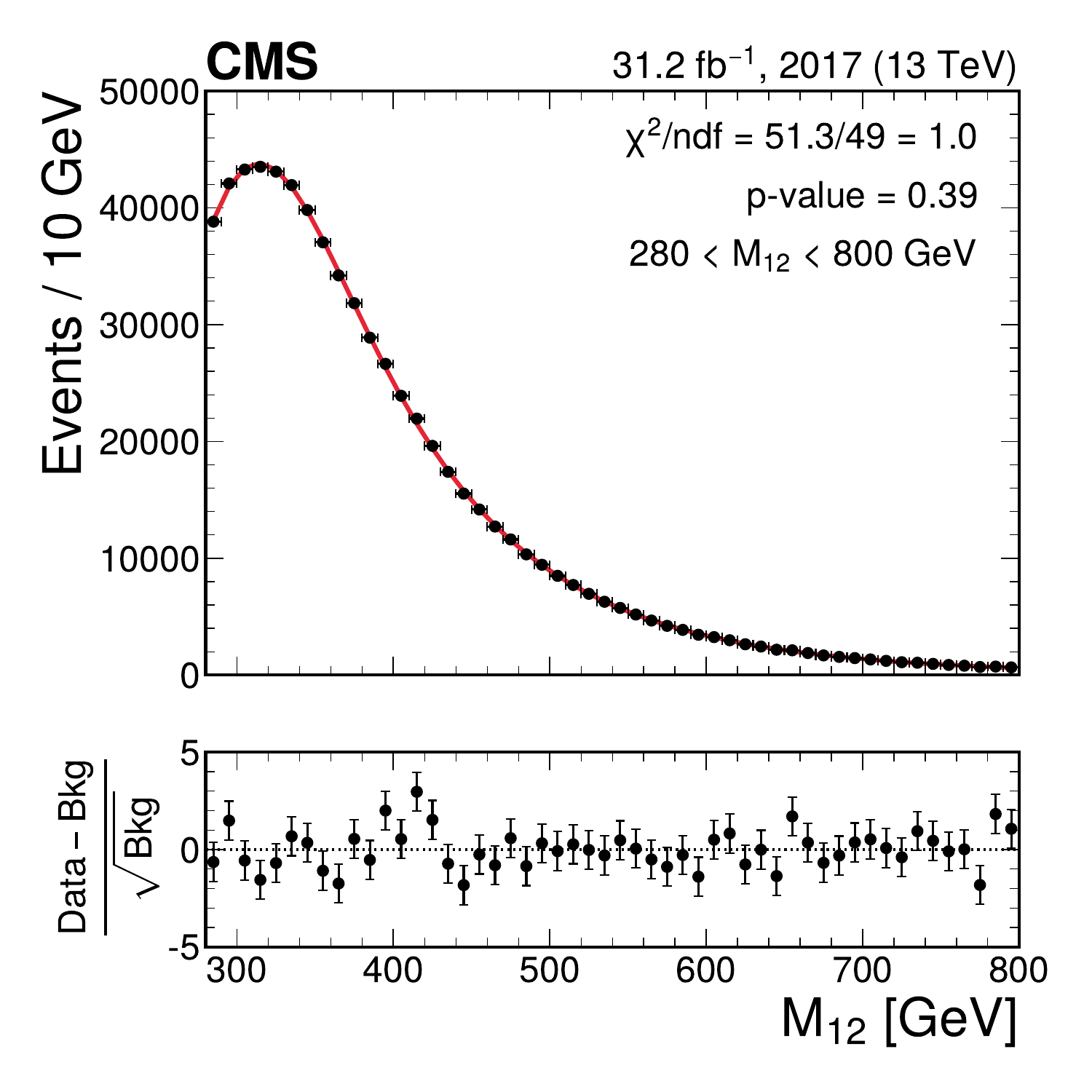}
    \includegraphics[width=0.48\textwidth]{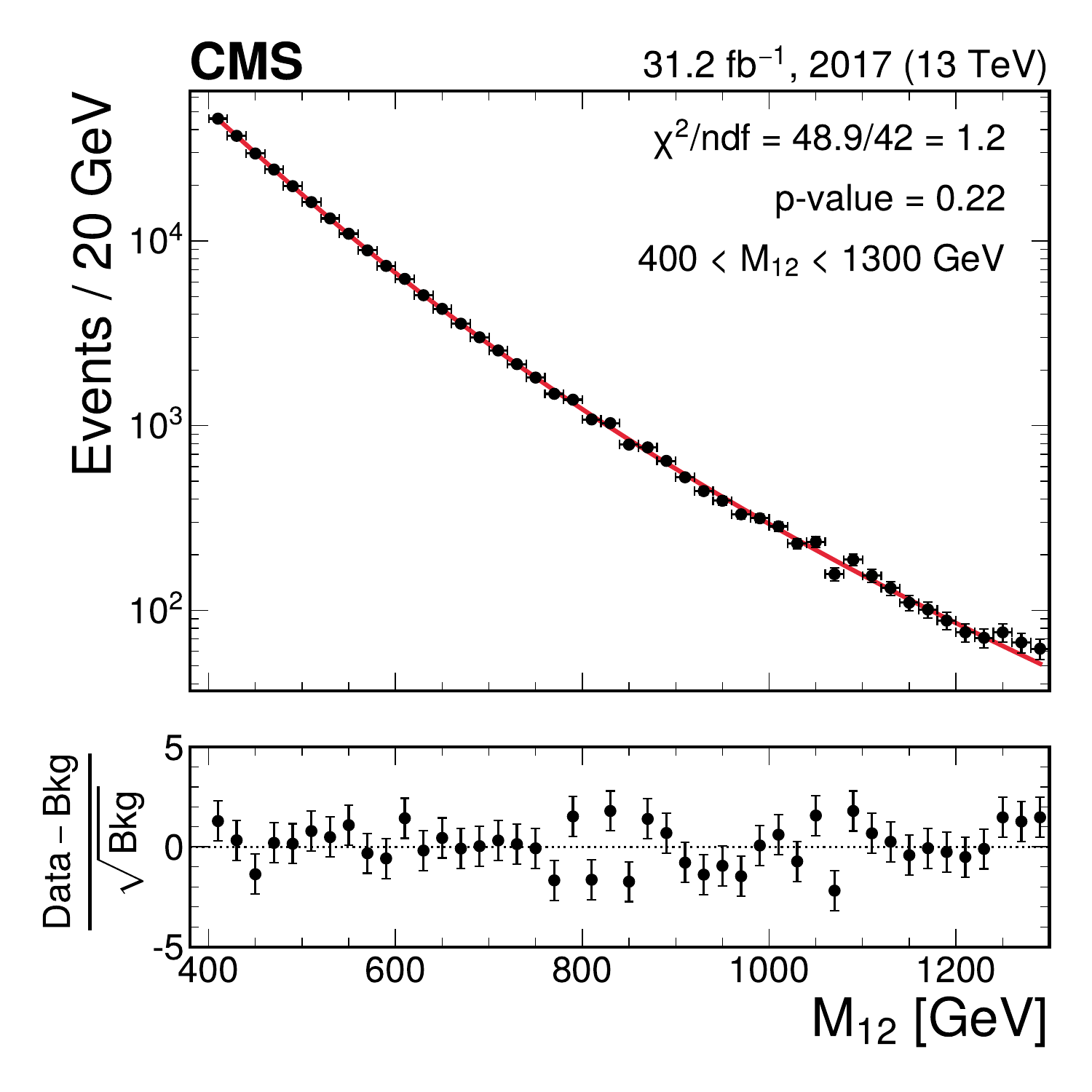}
    \includegraphics[width=0.48\textwidth]{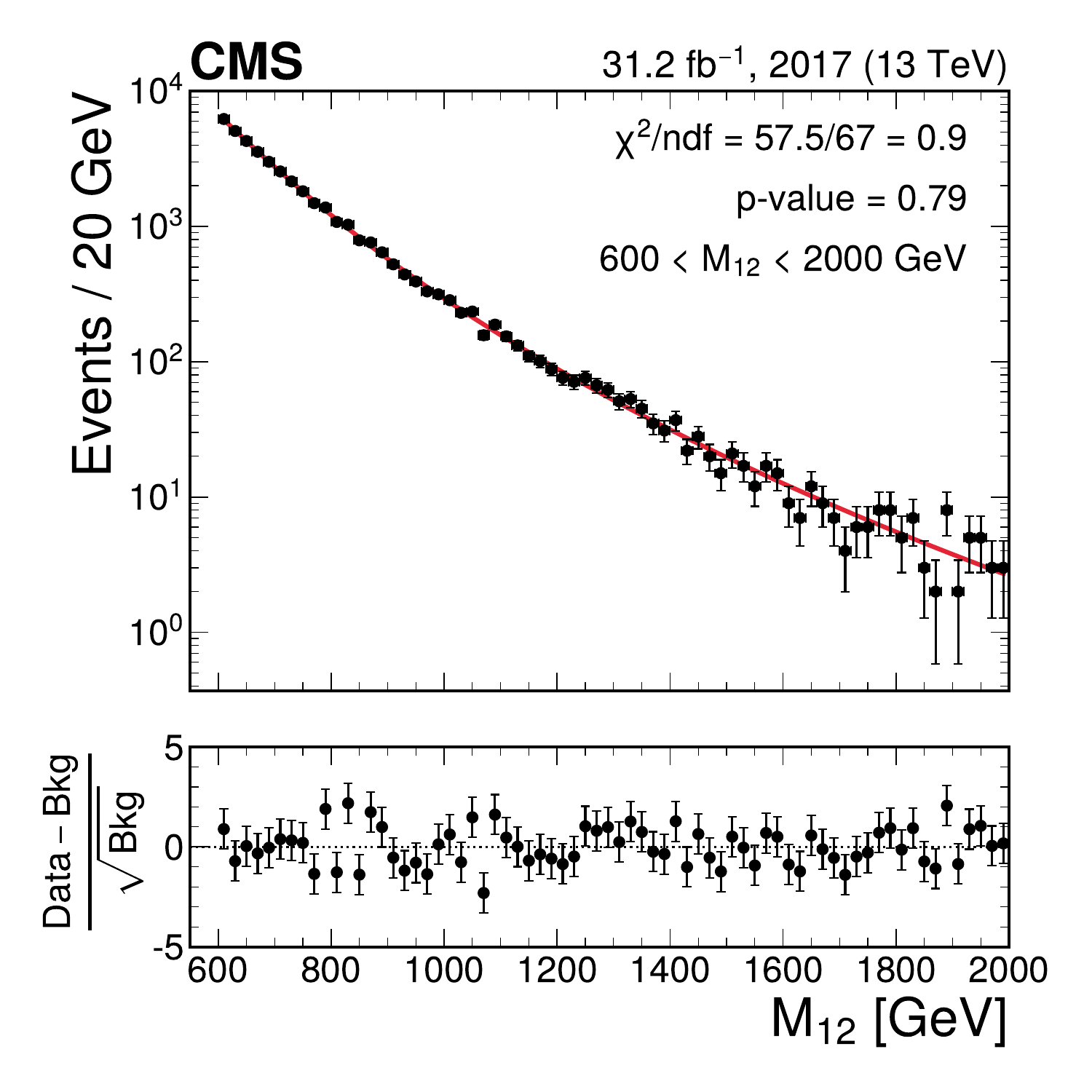}
    \caption{Invariant mass distributions of the four fit ranges in the \PQb tag veto CR for the 2017 FH channel, overlaid with the fitted functions. The distributions are fitted in the \mjj ranges of 240--560\GeV (upper left), 280--800\GeV (upper right), 400--1300\GeV (lower left), and 600--2000\GeV (lower right). The $\chi^2$ goodness-of-fit test and corresponding p-value are indicated in each plot.  The lower panels show the difference between the data and the fitted function, divided by the estimated statistical uncertainty for each bin. Good agreement between the fitted functions and the data is achieved.}
\label{fig:bbnb_fit_SR1to4_FH2017}
\end{figure}

\begin{figure}[htbp!]\centering
    \includegraphics[width=0.48\textwidth]{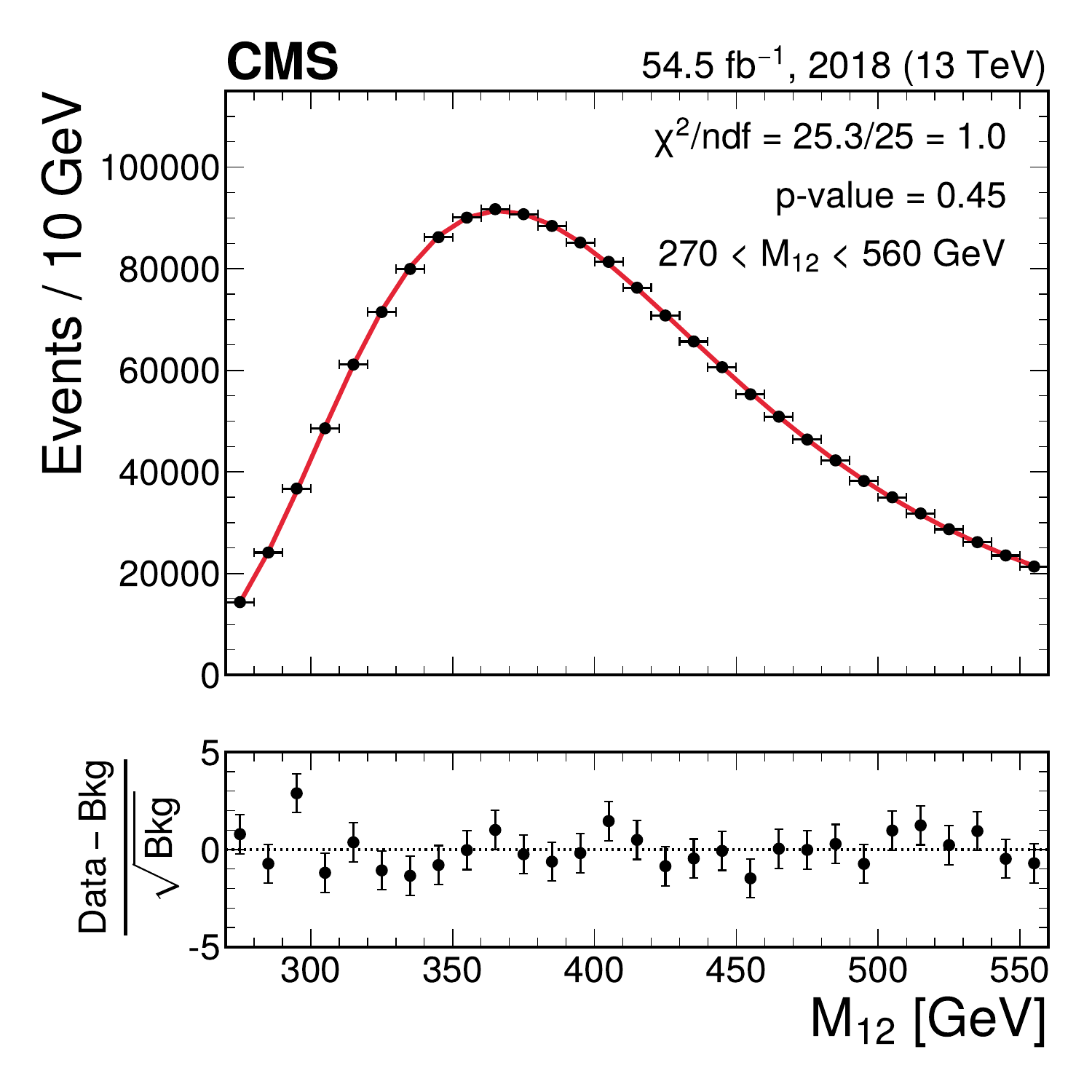}
    \includegraphics[width=0.48\textwidth]{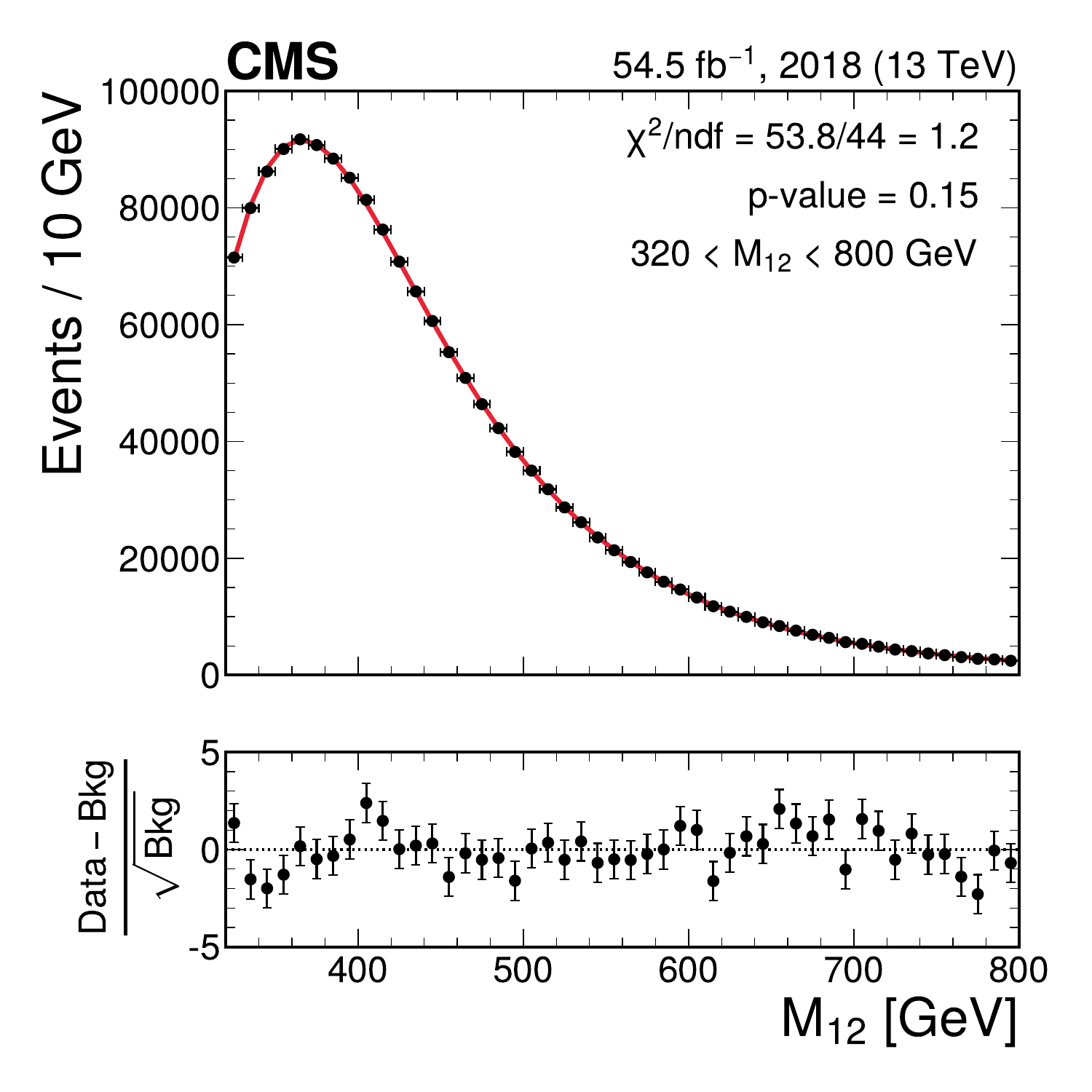}
    \includegraphics[width=0.48\textwidth]{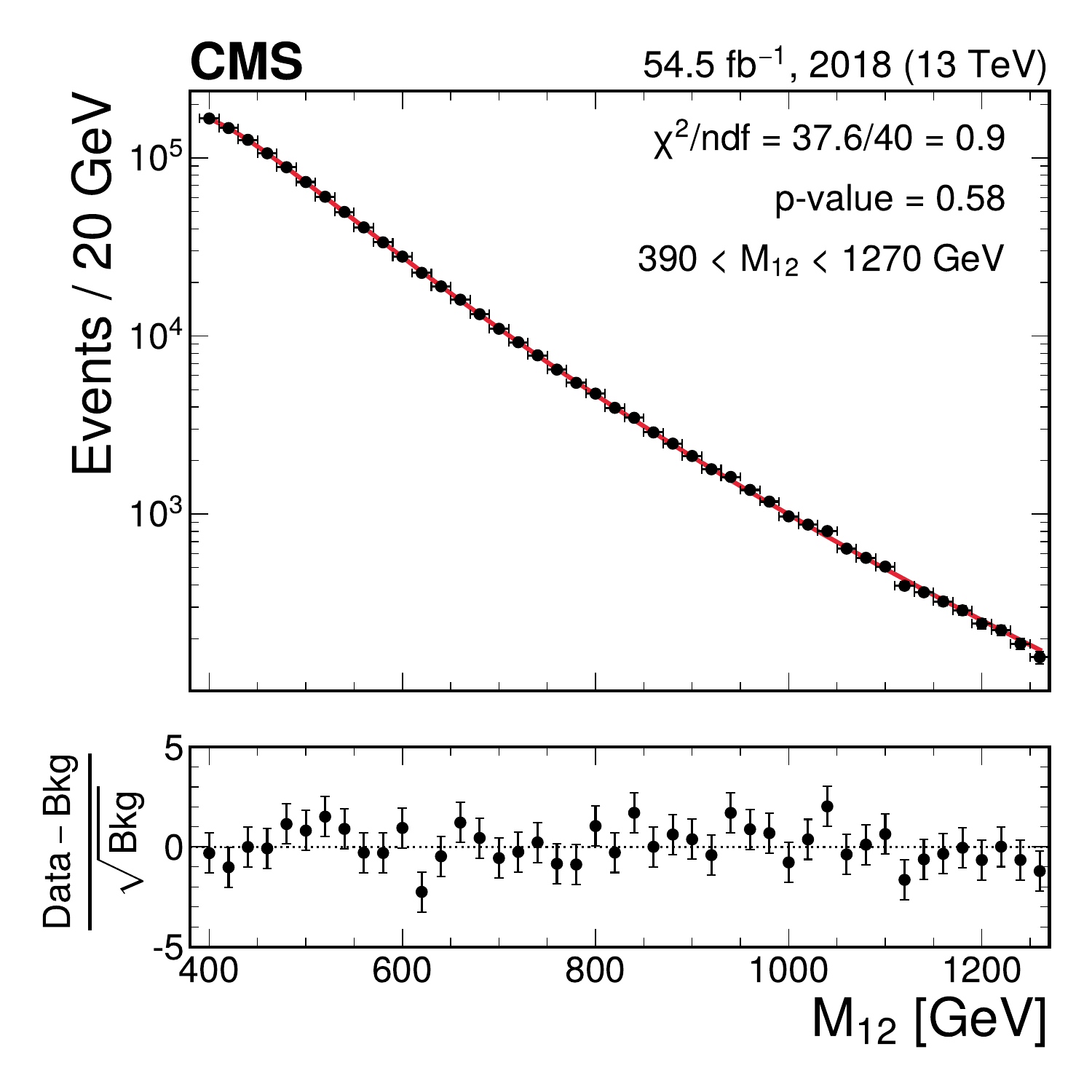}
    \includegraphics[width=0.48\textwidth]{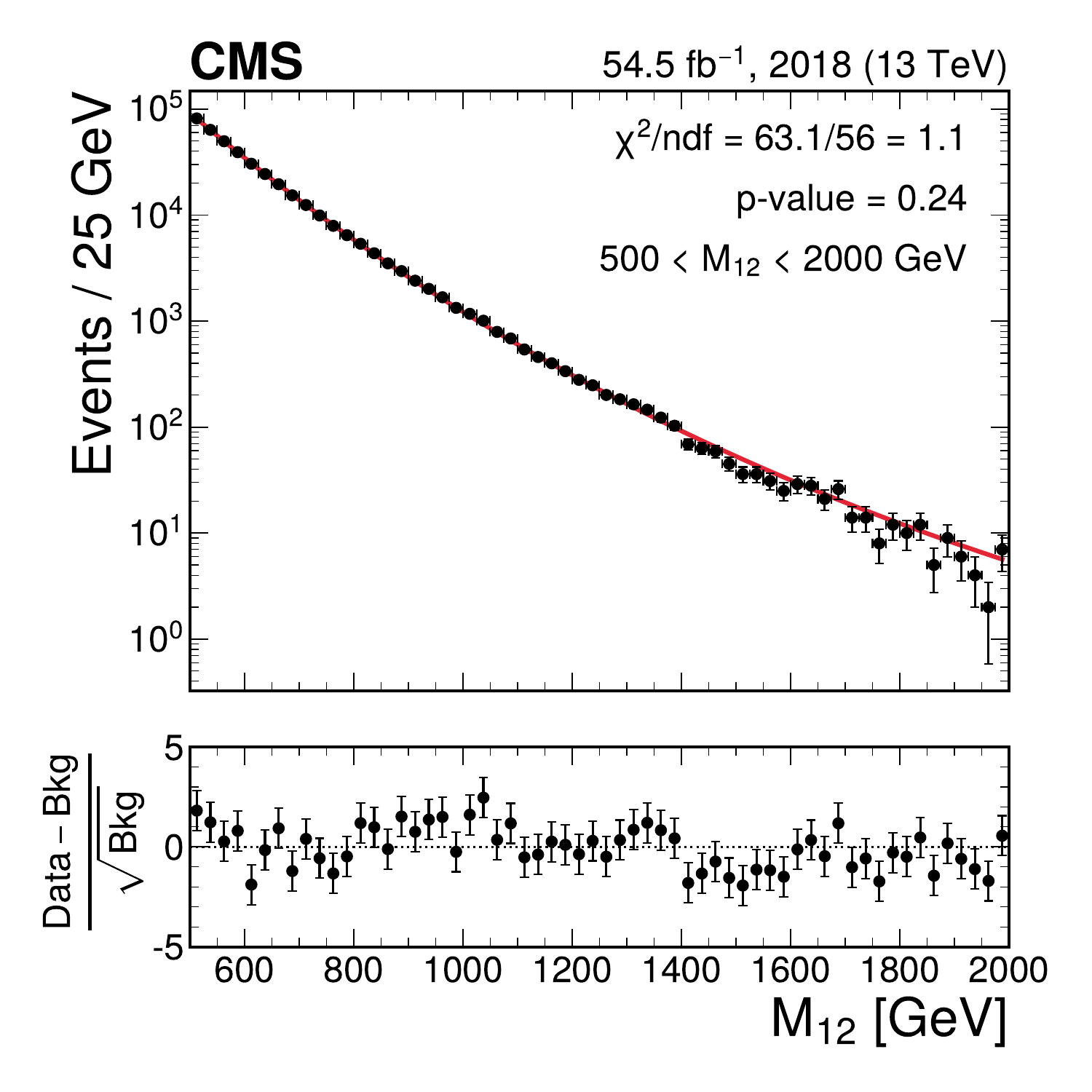}
    \caption{Invariant mass distributions of the four fit ranges in the \PQb tag veto CR for the 2018 FH channel, overlaid with the fitted functions. The distributions are fitted in the \mjj ranges of 270--560\GeV (upper left), 320--800\GeV (upper right), 390--1270\GeV (lower left), and 500--2000\GeV (lower right). The $\chi^2$ goodness-of-fit test and corresponding p-value are indicated in each plot.  The lower panels show the difference between the data and the fitted function, divided by the estimated statistical uncertainty for each bin. Good agreement between the fitted functions and the data is achieved.}
\label{fig:bbnb_fit_SR1to4_FH2018}
\end{figure}
Figures~\ref{fig:bbnb_fit_SR1to3_SL2017}, \ref{fig:bbnb_fit_SR1to4_FH2017}, and \ref{fig:bbnb_fit_SR1to4_FH2018} show the \mjj distributions in the CR in each fit range for the 2017 SL, 2017 FH, and 2018 FH data sets, respectively, along with the parameterisations discussed in the following. The goodness-of-fit estimators indicate adequate descriptions of the distributions.

For the parameterisation of the mass distributions in the CR an extension of the Novosibirsk function is used. The Novosibirsk function was originally devised to describe a Compton spectrum~\cite{Ikeda:1999aq}. Its extended form is defined as:

\begin{equation}
g(\mjj) = p_2 \exp \left( - \frac{1}{2\sigma_{0}^{2}} \ln^{2}\left[ 1 - \frac{\mjj - p_3}{p_4} p_5 -  \frac{(\mjj - p_3)^2}{p_4} p_5 p_6\right] - \frac{\sigma_0^2}{2} \right),
\end{equation}

where $p_2$ is a normalisation parameter, $p_3$ the peak value of the distribution, $p_4$ and $p_5$ are the parameters describing the asymmetry of the spectrum, and $p_6$ is the parameter of the extended term. The variable $\sigma_0$ is defined as:

\begin{equation}
\sigma_0 = \frac{2}{\xi} \sinh^{-1}(p_5 \xi / 2),\ \text{where}\ \xi = 2 \sqrt{\ln 4}.
\end{equation}

In some fit ranges, the effect of the kinematic turn-on at the lower edge exhibits a more complex shape, which cannot be modelled by the extended Novosibirsk function alone. These include the first fit range of the SL channel and the second fit range of the FH channel in the 2017 data set. An additional turn-on factor, represented by a Gaussian error function, is introduced and multiplied by the extended Novosibirsk function. It has the following analytic form:

\begin{equation}
h(\mjj) = 0.5\, \bigl[\erf(p_0[\mjj - p_1])+1\bigr],
\end{equation}
where
\begin{equation}
\erf(x) = \frac{2}{\sqrt{\pi}} \int_0^{x} \re^{-t^2} \rd{}t,
\end{equation}

and the parameters $p_0$ and $p_1$ describe the slope and centre of the turn-on flank, respectively.

The transfer factor is an analytic function that parameterises the ratio of the background mass distributions in SR and CR for each fit range and data set. This function is chosen to be a sum of Chebyshev polynomials up to the third degree following a study with multijet events in simulations. The coefficients of the Chebyshev polynomials are obtained in the final simultaneous fit to the data in the SR and CR, separately for each channel and data set. In addition, the choices of functional form may cause possible bias and give different results for the signal parameters, representing a systematic uncertainty associated with the choice of function. This uncertainty is accounted for with the discrete profiling method~\cite{Dauncey:2014xga}. This method introduces an additional (discrete) nuisance parameter indicating the choice of the transfer factor in this analysis, which is profiled in an analogous way to continuous nuisance parameters.

The background model and the signal extraction method are validated by applying them in a signal-depleted validation region (VR). The selection in the VR proceeds in the same way as in the SR and CR, except that the \PQb tagging score of the third leading jet must lie between the loose and the medium working points. By this definition, there is no overlap between VR, SR, and CR.
The observed limits from these tests are found to be everywhere in agreement with the expected limits within uncertainties, indicating correct performance of the background model. 

\section{Signal extraction and systematic uncertainties}
\label{sec:Syst}
The signal extraction is performed with a simultaneous maximum likelihood fit to the \mjj distributions in the SR and CR, at first separately for the 2017~SL, 2017~FH, and 2018~FH data sets. In this simultaneous fit, the CR data are modelled with the corresponding parameterization as discussed in Section~\ref{sec:Background} with floating parameters. The SR is modelled by a sum of the same parameterization with identical parameters multiplied by the respective transfer factor, and the signal shape multiplied by the signal strength parameter. The parameters of the transfer factor are also floating in the fit. The simultaneous fit ensures that the uncertainties in the CR background function are taken into account with all correlations also in the SR.

Apart from the statistical uncertainties in each data set, systematic uncertainties originating from both theoretical and experimental sources are considered. In the signal extraction fit, all uncertainties are implemented as penalty terms, \ie ``nuisance parameters''~\cite{CMS:2024onh}, to the likelihood function. The following systematic uncertainties are taken into account in the expected signal and background estimation, as well as for the interpretations within the MSSM and generic 2HDMs:

\begin{itemize}
\item[\textit{Integrated luminosity}] The integrated luminosities for the 2017 and 2018 data-taking years have 2.3 and 2.5\% individual uncertainties, respectively~\cite{CMS-PAS-LUM-17-004,CMS-PAS-LUM-18-002}, while the overall uncertainty for the 2016--2018 period is 1.6\%.

\item[\textit{Pileup}] The systematic uncertainty introduced by the total inelastic $\Pp\Pp$ cross section that is used in the simulation of pileup events is quantified by varying the effective total inelastic cross section of 69.2\unit{mb} within its $\pm$4.6\% uncertainty~\cite{CMS:2018mlc}.

\item[\textit{Prefiring correction}] In the 2017 data-taking period, an increase in the offset of the ECAL timing pulse caused the ECAL L1 trigger to accept the event during the previous bunch crossing (referred to as ``prefiring'', producing dead time)~\cite{CMS:2024ppo}. This results in a relative reduction of the signal efficiency by 0.3--2\%. The uncertainty in the efficiency correction is taken into account.

\item[\textit{Online kinematic efficiency}] The efficiency of the jet selection at the trigger level is determined from the data using single-jet control triggers. The uncertainty is very small, at most at the percent level, and only relevant for the lowest mass points that are closest to the kinematic thresholds. The uncertainty in the muon identification efficiency at trigger level is determined using the \JPsi signal and amounts to about 5\%.

\item[\textit{Online \PQb tagging efficiency}] The online \PQb tagging efficiency is determined from the data with dedicated control triggers using a tag-and-probe method as described in Section~\ref{sec:Data}. The systematic uncertainty in the online \PQb tagging efficiency increases with increasing jet \pt from about 1 to 5\% per jet over the considered \pt range. For \PQb jets containing muons, the uncertainty varies between 3 and 10\% over the \pt range.

\item[\textit{Jet energy scale and resolution}] Uncertainties in the jet energy scale and resolution affect solely the signal efficiency and the shape of its mass distribution. They are evaluated by varying the energy and resolution corrections for all jets in the signal model within one standard deviation~\cite{CMS:2016lmd}.

\item[\textit{Offline \PQb tagging efficiency}] The uncertainty in the offline \PQb tagging efficiency depends on the jet \pt and ranges from 2--15\% per jet. The combined uncertainty in the \PQb tagging of three jets per selected signal event, together with the additional uncertainty in the online \PQb tagging of two jets, constitutes the total systematic uncertainty in \PQb jet identification~\cite{CMS-DP-2018-058, CMS-DP-2021-004}.

\item[\textit{Muon identification}] The uncertainty in the offline muon identification efficiency ranges from 1--3\%~\cite{CMS:2018rym}.

\item[\textit{Uncertainty in the choice of background parameterisation}] Different forms of transfer factors are considered, described by Chebyshev polynomials up to the third order. They are integrated in the signal extraction procedure by means of the discrete profiling method, as described in Section~\ref{sec:Background}.

\item[\textit{Theory-related uncertainties}] The theoretical uncertainty in the QCD factorisation and renormalisation scales ($\mu_\text{F}$ and $\mu_\text{R}$), and uncertainties in the choice of the PDFs and \alpS are addressed following the recommendations of the LHC Higgs Working Group~\cite{LHCHiggsCrossSectionWorkingGroup:2016ypw}. The uncertainties arising from the parton shower modelling are found to have negligible impact on the analysis.

\end{itemize}

Only the uncertainties in jet energy scale and resolution are found to affect the shape of the signal mass distributions, while the choice of background parameterisation impacts the shape of the background distributions. Other systematic uncertainties only concern the signal yield. The dominant systematic uncertainty lies in the background estimation over most of the mass range, followed by uncertainties in the \PQb tagging efficiency uncertainties. At very large masses, the \PQb tagging efficiency contributes the dominant uncertainty.

\section{Results}
\label{sec:Result}

As mentioned in Section~\ref{sec:Syst}, a binned maximum likelihood fit of the signal plus background parameterisations to the data \mjj distribution is performed to extract the number of potential signal events. The test statistic chosen to determine the signal yield is based on the maximum likelihood ratio profiled as a function of the signal strength, \ie, the ratio of the signal yield to the one predicted by the signal model as defined by Eq. (25) in~\cite{CMS:2024onh}. The fit is performed simultaneously to the triple \PQb tag SR and the \PQb tag veto CR, first separately for the 2017~SL, 2017~FH, and 2018~FH  data sets, and then a combined fit over all three data sets is performed for the final results. The results have been determined using the CMS statistical analysis tool, \textsc{Combine}~\cite{CMS:2024onh}.

\subsection{Cross section results}
\begin{figure}[!htbp]\centering
    \includegraphics[width=0.49\textwidth]{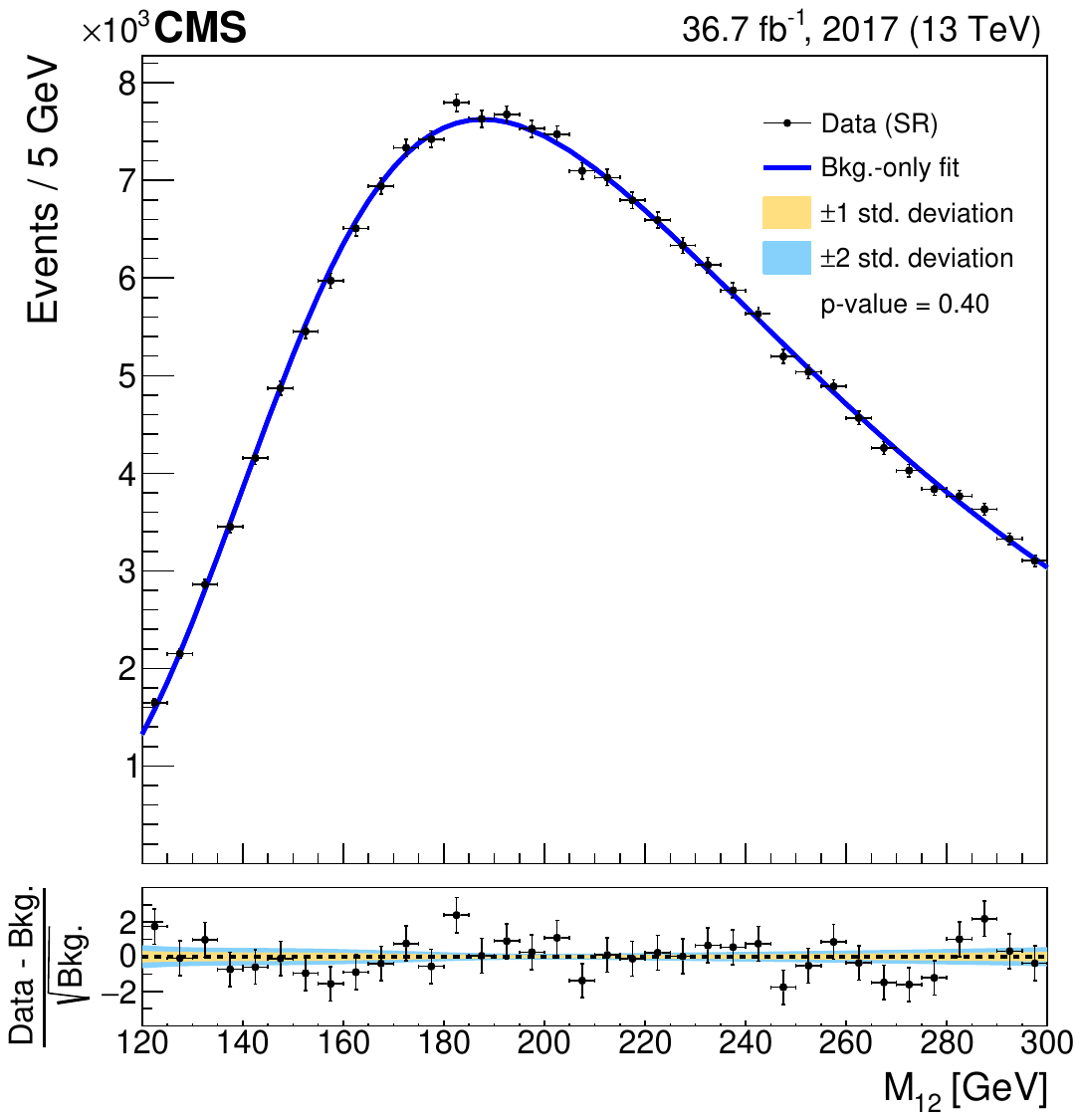}
    \includegraphics[width=0.49\textwidth]{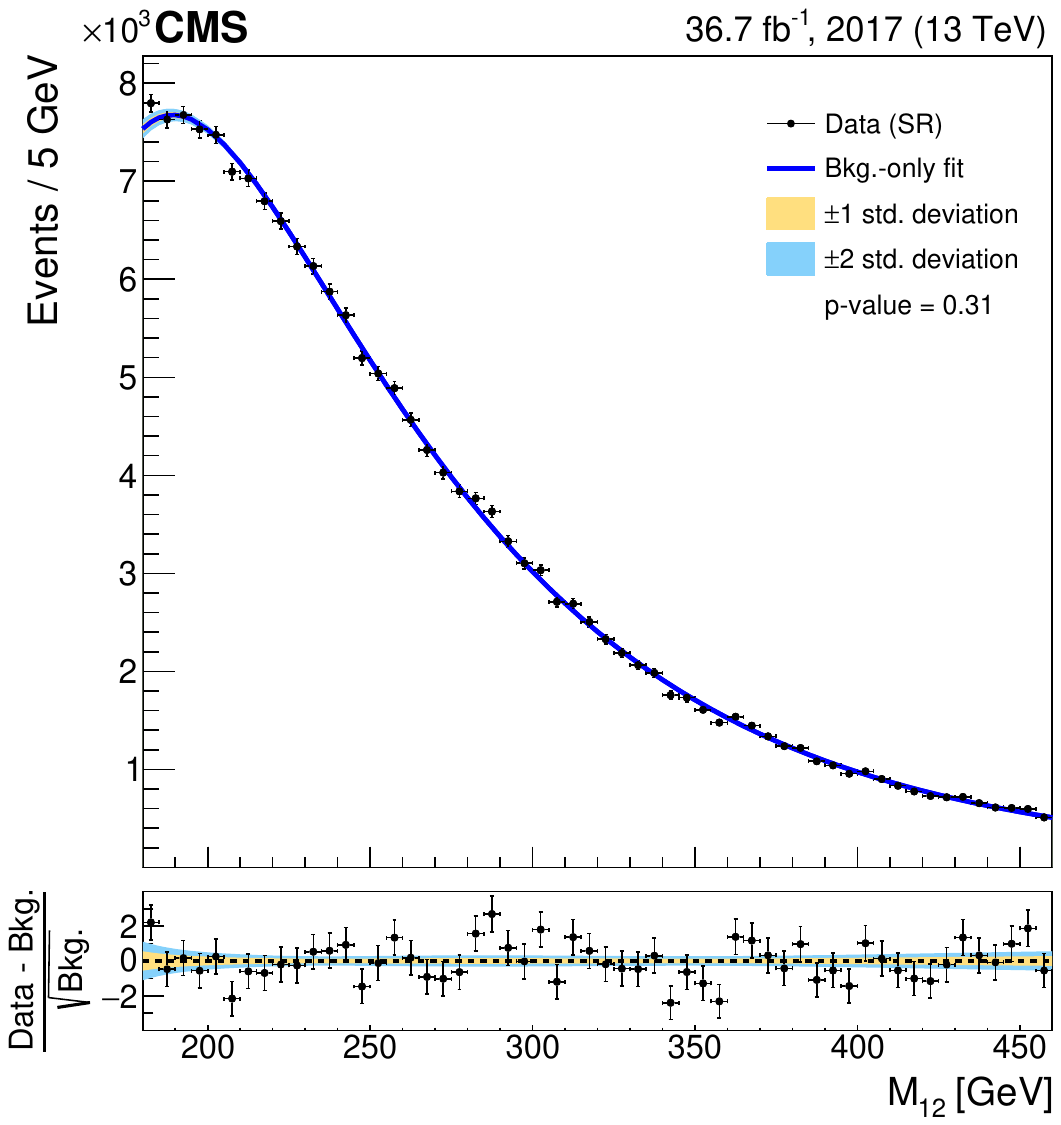}
    \includegraphics[width=0.49\textwidth]{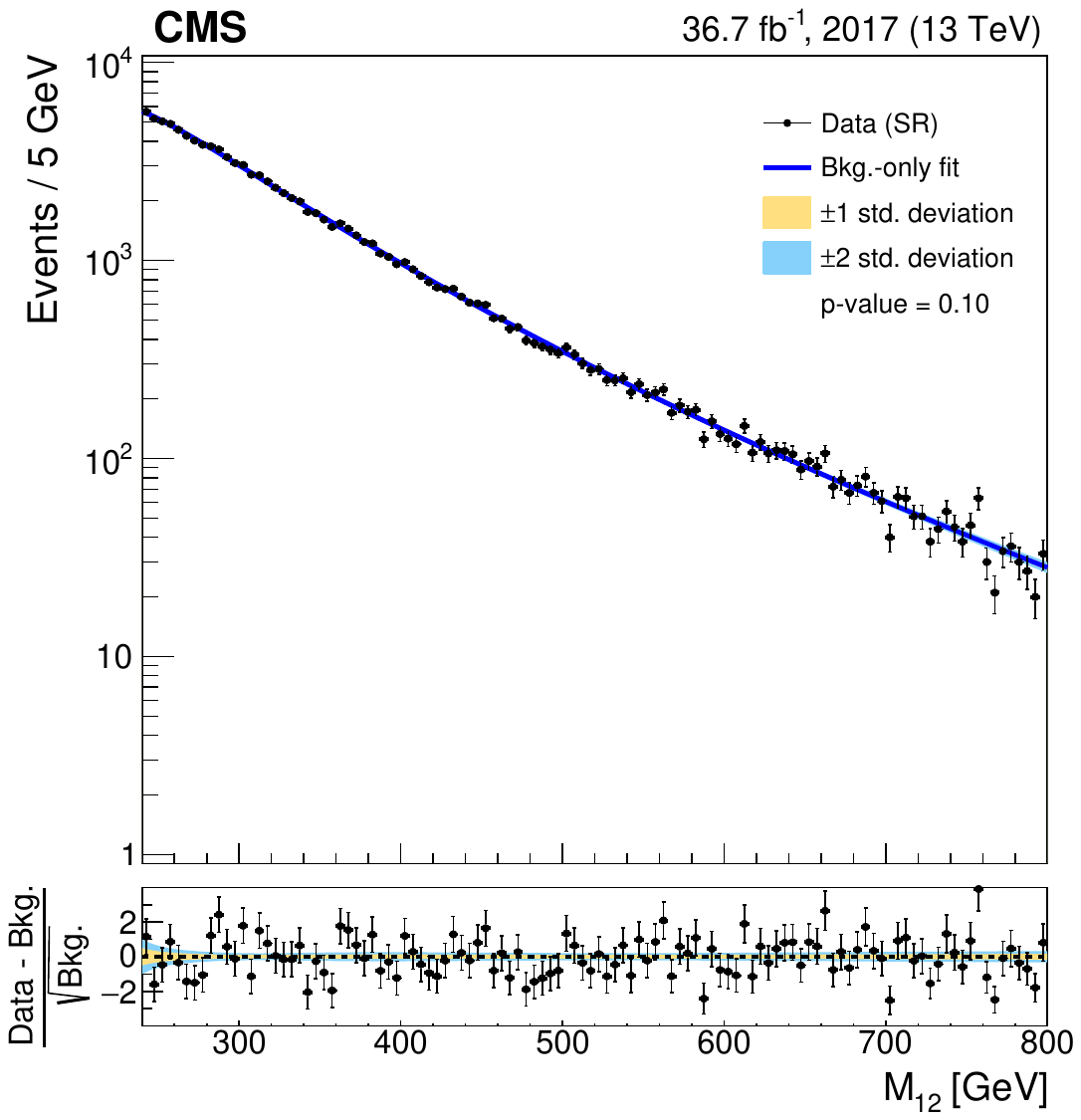}
    \caption{Background-only fits to the \mjj~distribution in each fit range of the 2017 analysis in the SL category, shown together with ${\pm}1\sigma$ and ${\pm}2\sigma$ uncertainty bands extracted from the fit in the upper panels. The lower panels show the difference between data and fitted background, divided by the statistical uncertainty of the latter. The distributions are fitted in the \mjj ranges of 120--300\GeV (upper left), 180--460\GeV (upper right), and 240--800\GeV (lower).}
\label{fig:bonlyFits_2017SL}
\end{figure}

\begin{figure}[!htbp]\centering
    \includegraphics[width=0.49\textwidth]{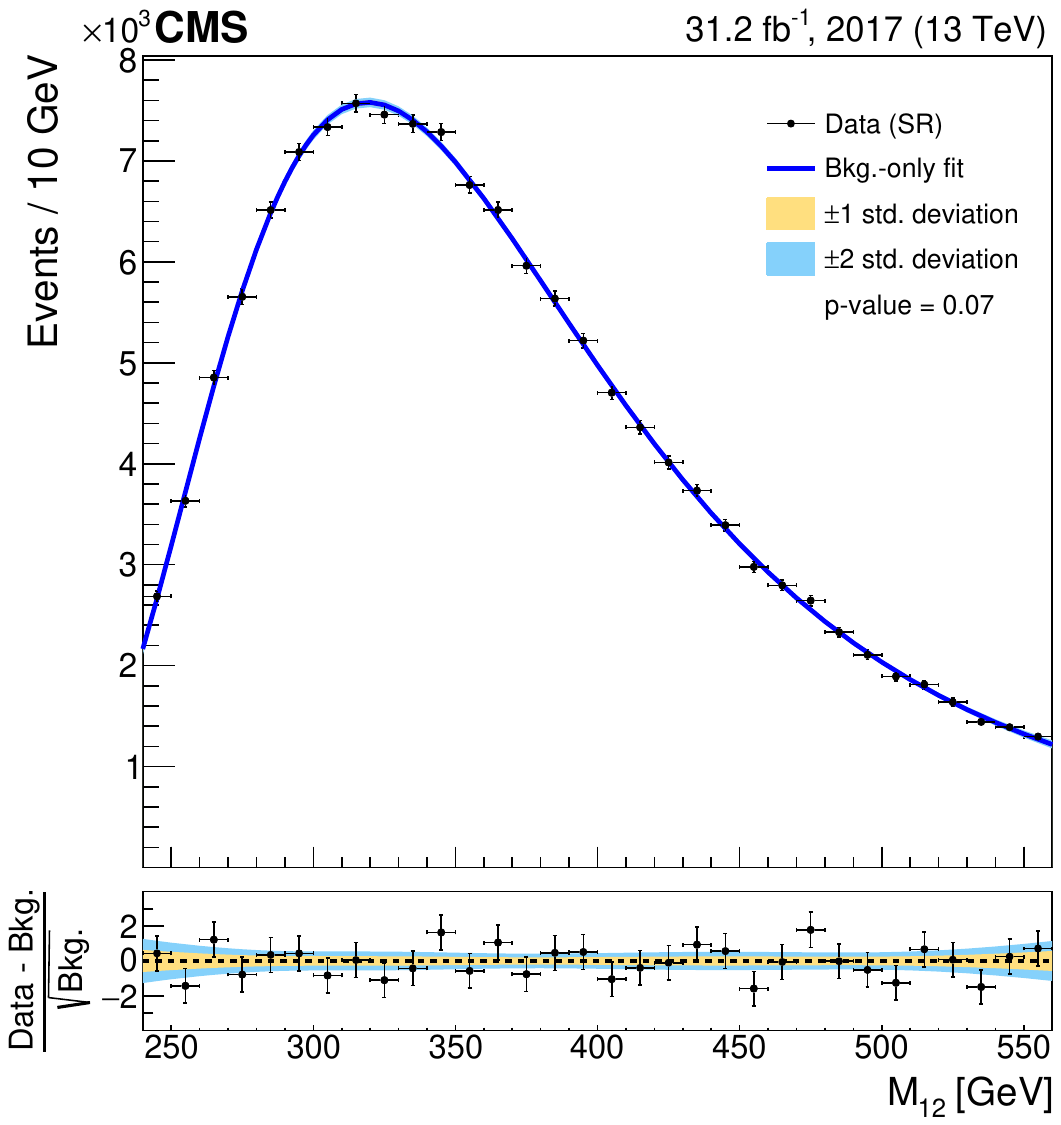}
    \includegraphics[width=0.49\textwidth]{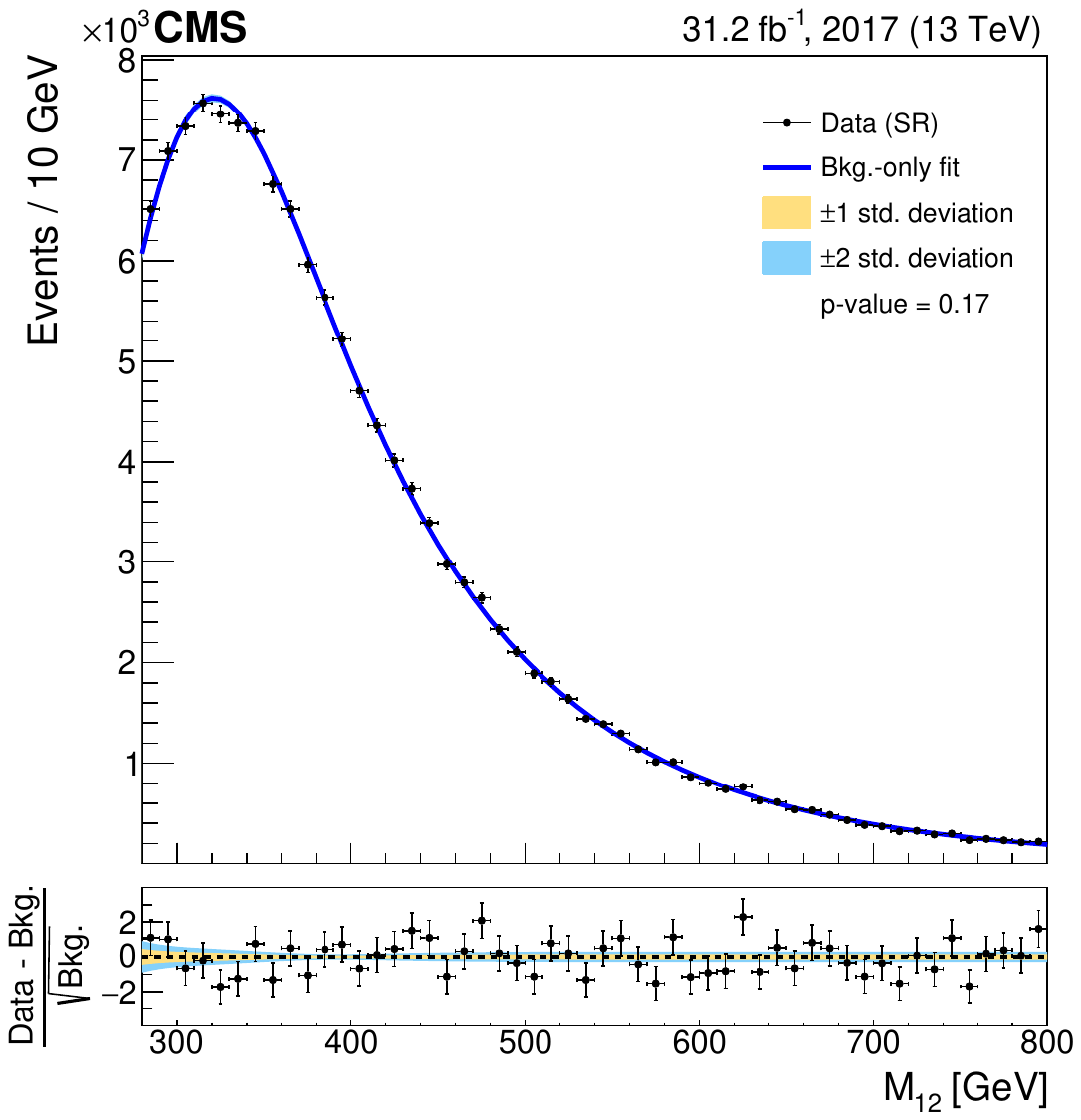}
    \includegraphics[width=0.49\textwidth]{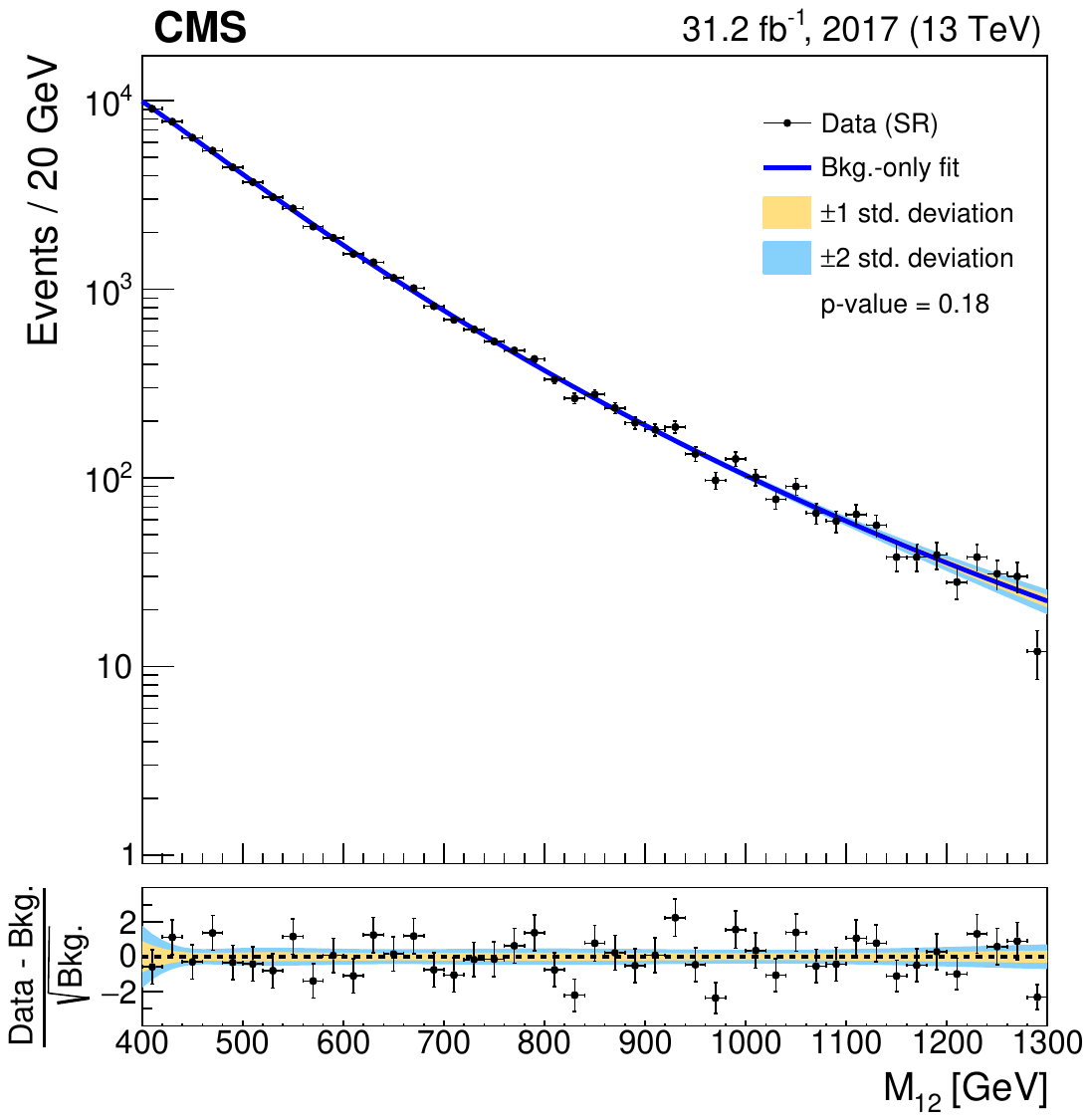}
    \includegraphics[width=0.49\textwidth]{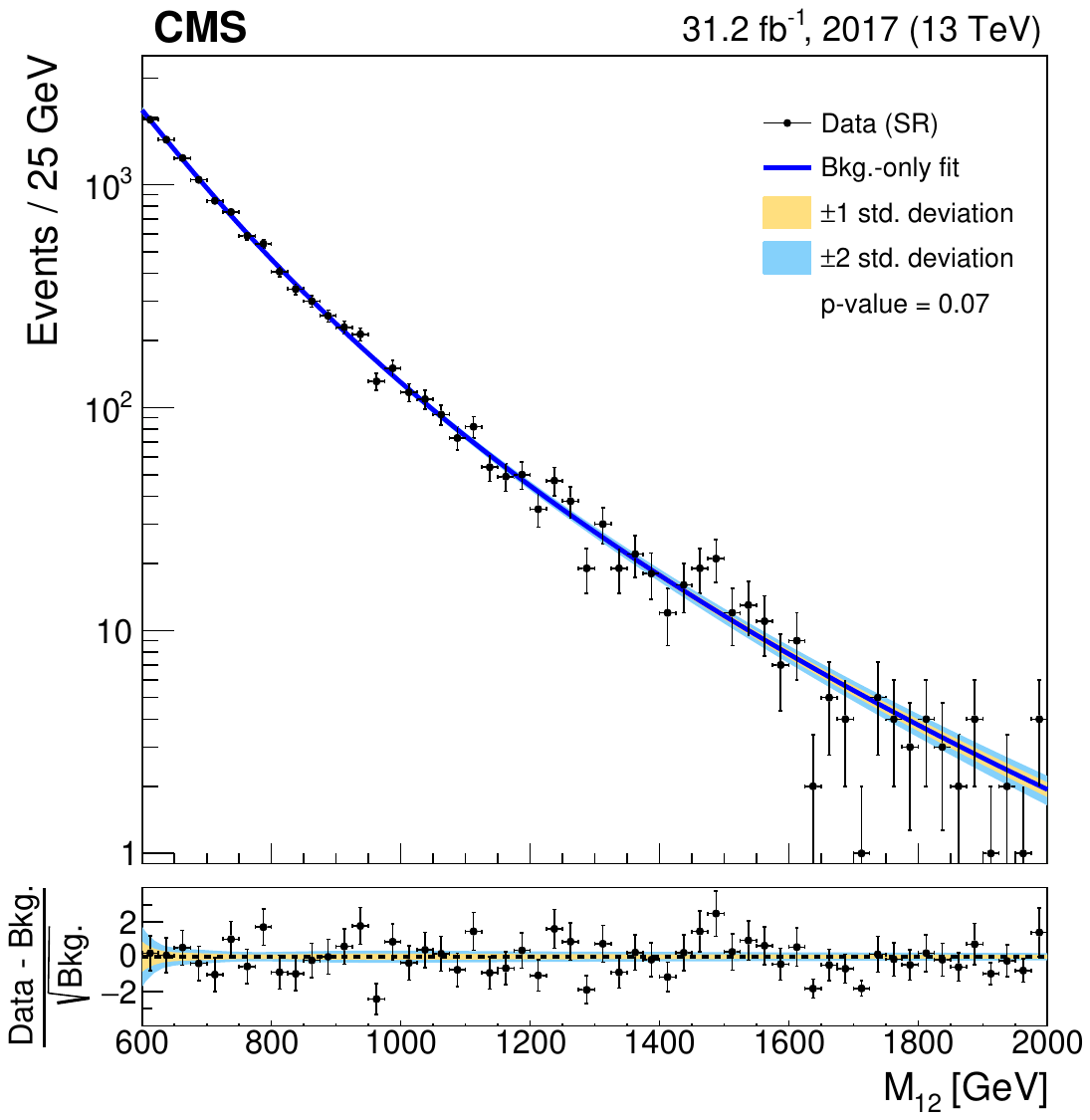}
    \caption{Background-only fits to the \mjj~distribution in each fit range of the 2017 analysis in the FH category, shown together with ${\pm}1\sigma$ and ${\pm}2\sigma$ uncertainty bands extracted from the fit in the upper panels. The lower panels show the difference between data and fitted background, divided by the statistical uncertainty of the latter. The distributions are fitted in the \mjj ranges of 240--560\GeV (upper left), 280--800\GeV (upper right), 400--1300\GeV (lower left), and 600--2000\GeV (lower right).}
\label{fig:bonlyFits_2017FH}
\end{figure}

\begin{figure}[!htbp]\centering
    \includegraphics[width=0.49\textwidth]{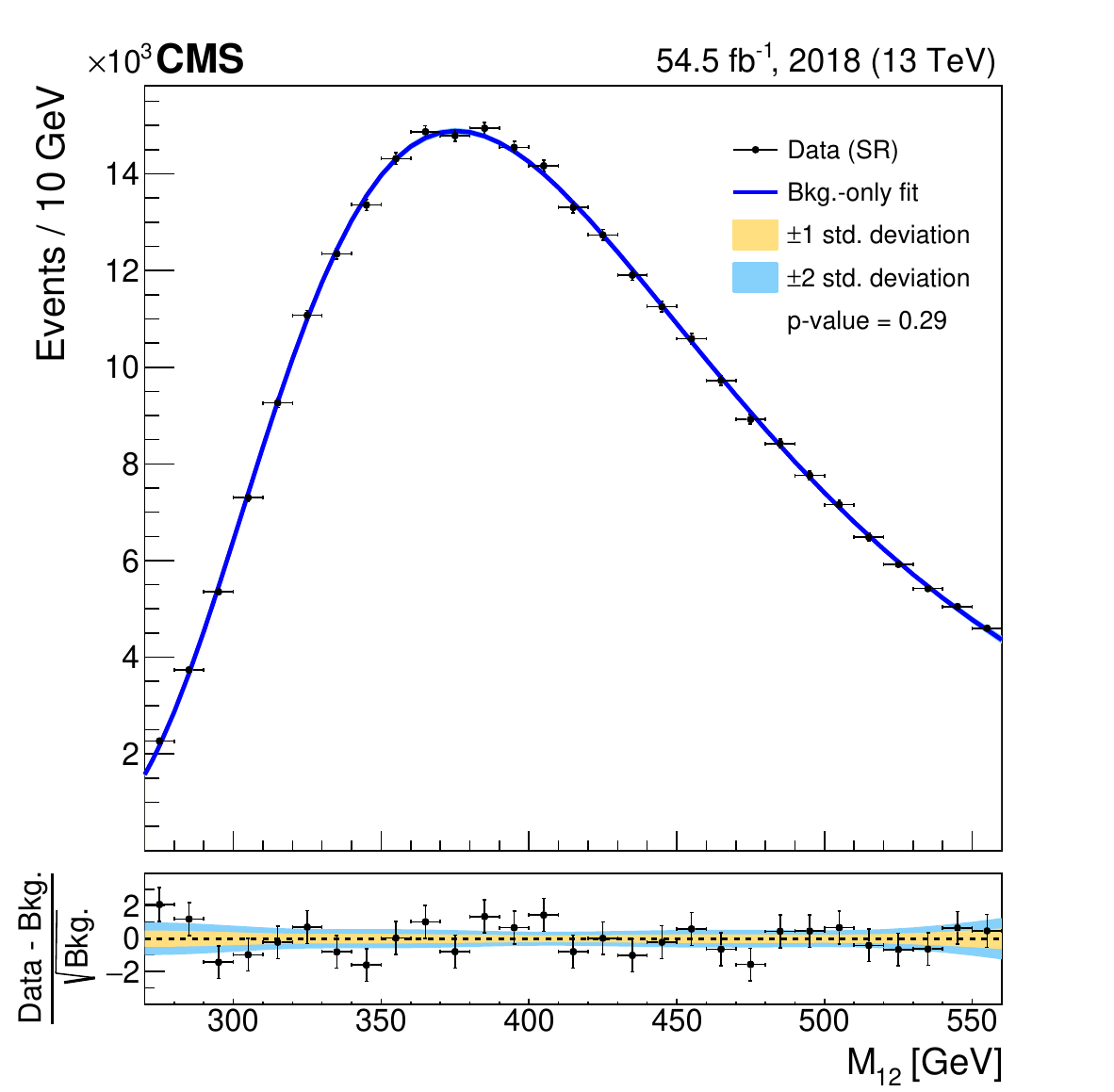}
    \includegraphics[width=0.49\textwidth]{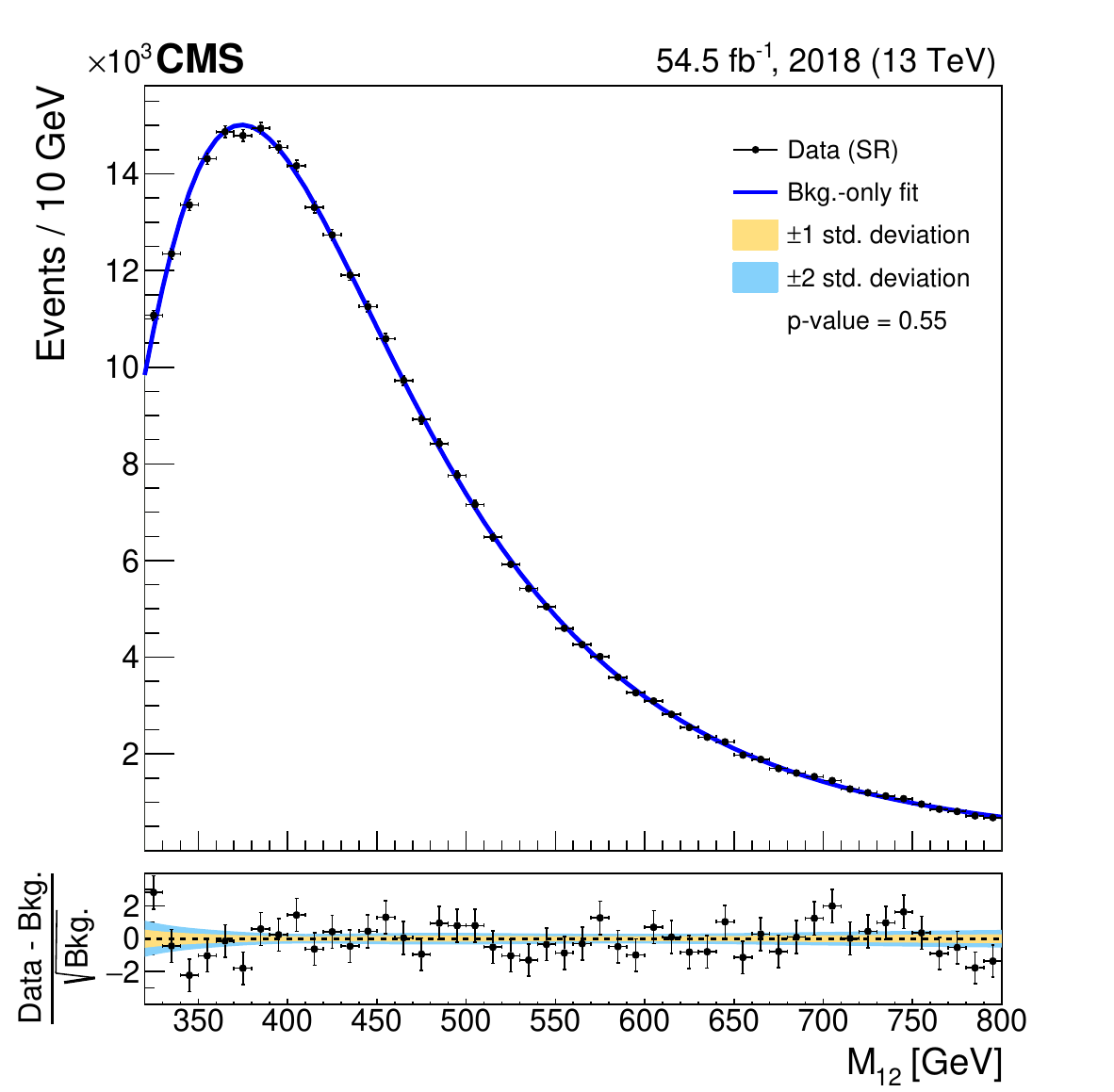}
    \includegraphics[width=0.49\textwidth]{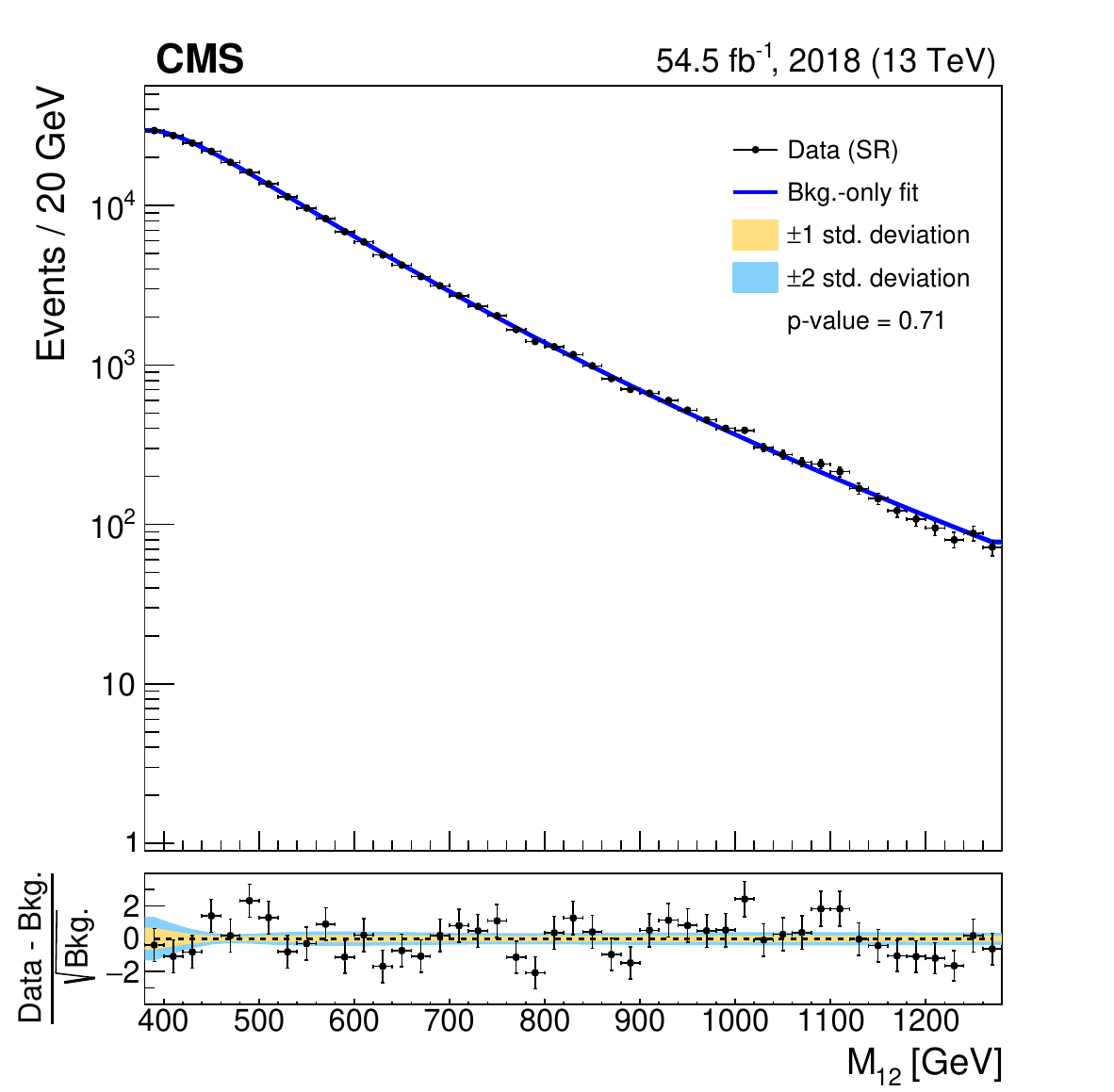}
    \includegraphics[width=0.49\textwidth]{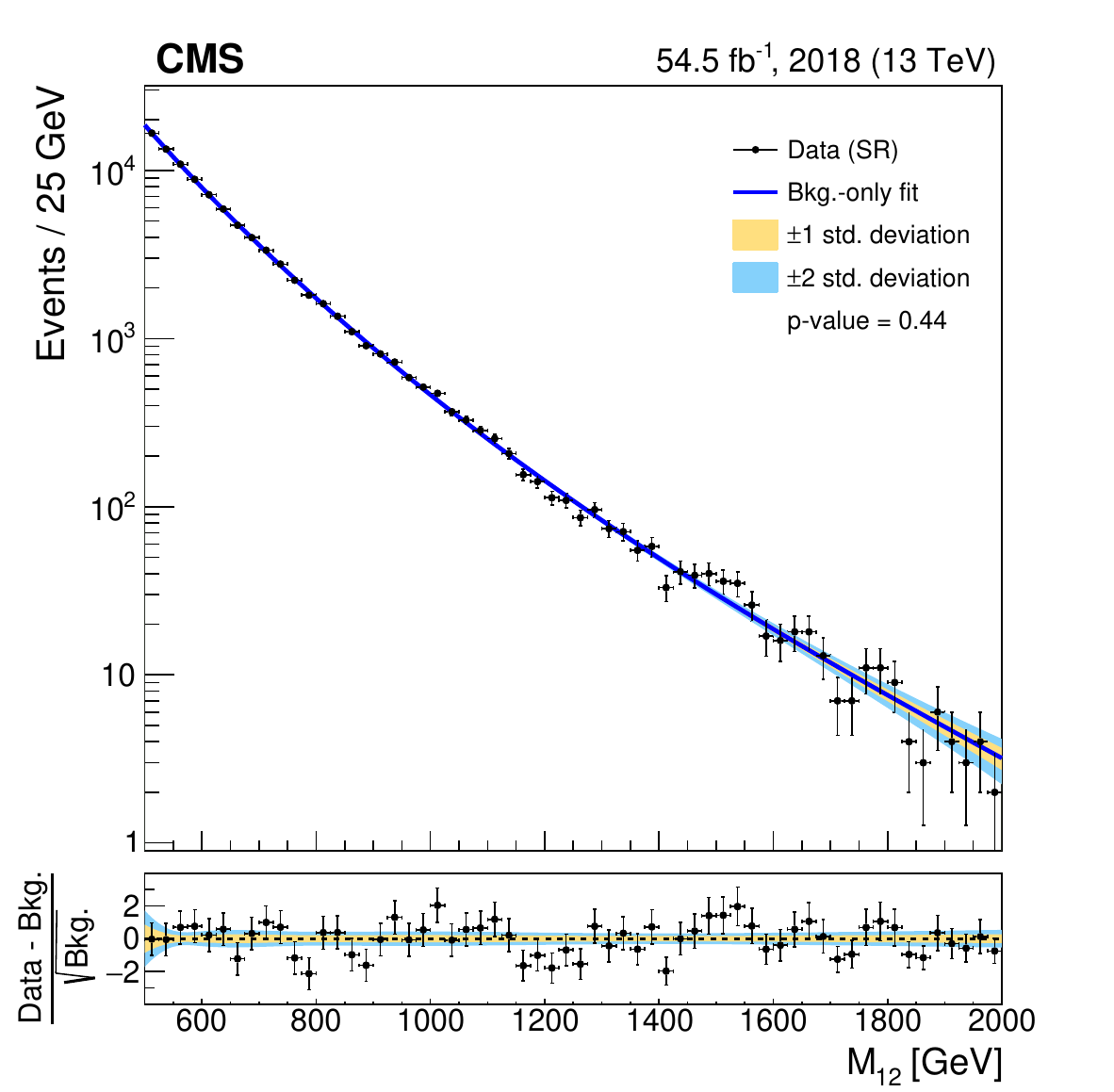}
    \caption{Background-only fits to the \mjj~distribution in each fit range of the 2018 analysis in the FH category, shown together with ${\pm}1\sigma$ and ${\pm}2\sigma$ uncertainty bands extracted from the fit in the upper panels. The lower panels show the difference between data and fitted background, divided by the statistical uncertainty of the latter. The distributions are fitted in the \mjj ranges of 270--560\GeV (upper left), 320--800\GeV (upper right), 390--1270\GeV (lower left), and 500--2000\GeV (lower right).}
\label{fig:bonlyFits_2018FH}
\end{figure}

The invariant mass distributions in the SR from the three different data sets are shown in Figs.~\ref{fig:bonlyFits_2017SL}--\ref{fig:bonlyFits_2018FH}, including the result of the simultaneous CR and SR fit under the background-only hypothesis. In general, the fits show good agreement with the background-only hypothesis.

No significant excess of signal events is observed over the SM background estimation. For each signal mass hypothesis, the results are used to set 95\% confidence level (\CL) upper limits on the production cross section times branching fraction ${\sigma(\Pp\Pp\to\PQb \phi+\mathrm{X}) \mathcal{B}(\phi \to \bbbar)}$ using the modified frequentist \CLs criterion~\cite{Junk:1999kv, Read:2002hq}, and making use of the asymptotic approximation~\cite{Cowan:2010js}. Systematic uncertainties are treated as nuisance parameters and profiled in the statistical inference using log-normal priors for uncertainties in the signal yield, and Gaussian priors for shape uncertainties.

\begin{figure}[!htbp]\centering
    \includegraphics[width=0.80\textwidth]{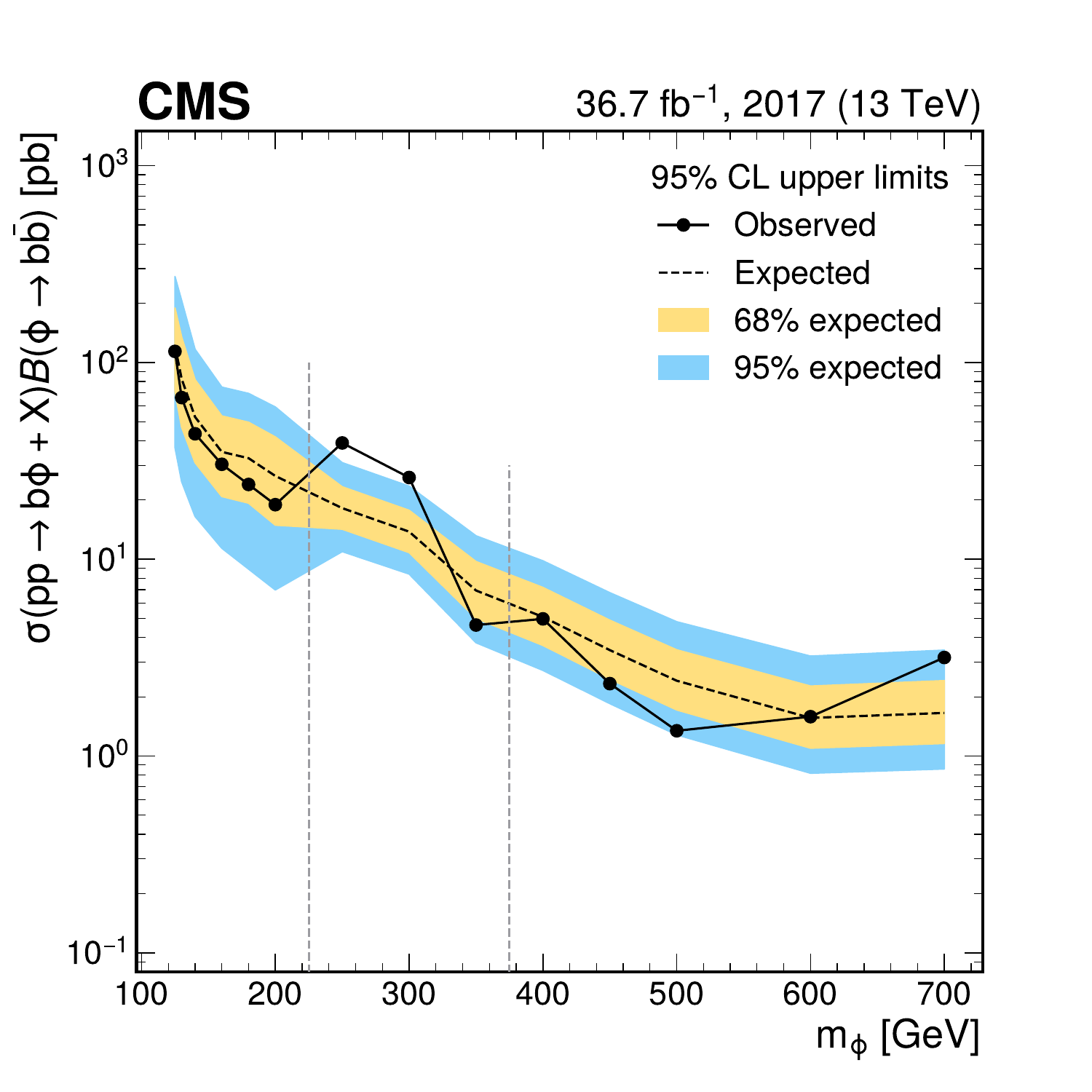}
    \caption{Expected and observed upper limits for the \PQb-quark-associated Higgs boson production cross section times branching fraction of the decay into a \PQb quark pair at 95\% CL as functions of \mPhi for the 2017 SL category. The vertical dashed lines indicate the boundaries of usage of the different fit ranges, as reflected in the rightmost column of Table~\ref{tab:massrangedefs}.}
\label{fig:SR_XSec_2017SL}
\end{figure}

\begin{figure}[!htbp]\centering
    \includegraphics[width=0.80\textwidth]{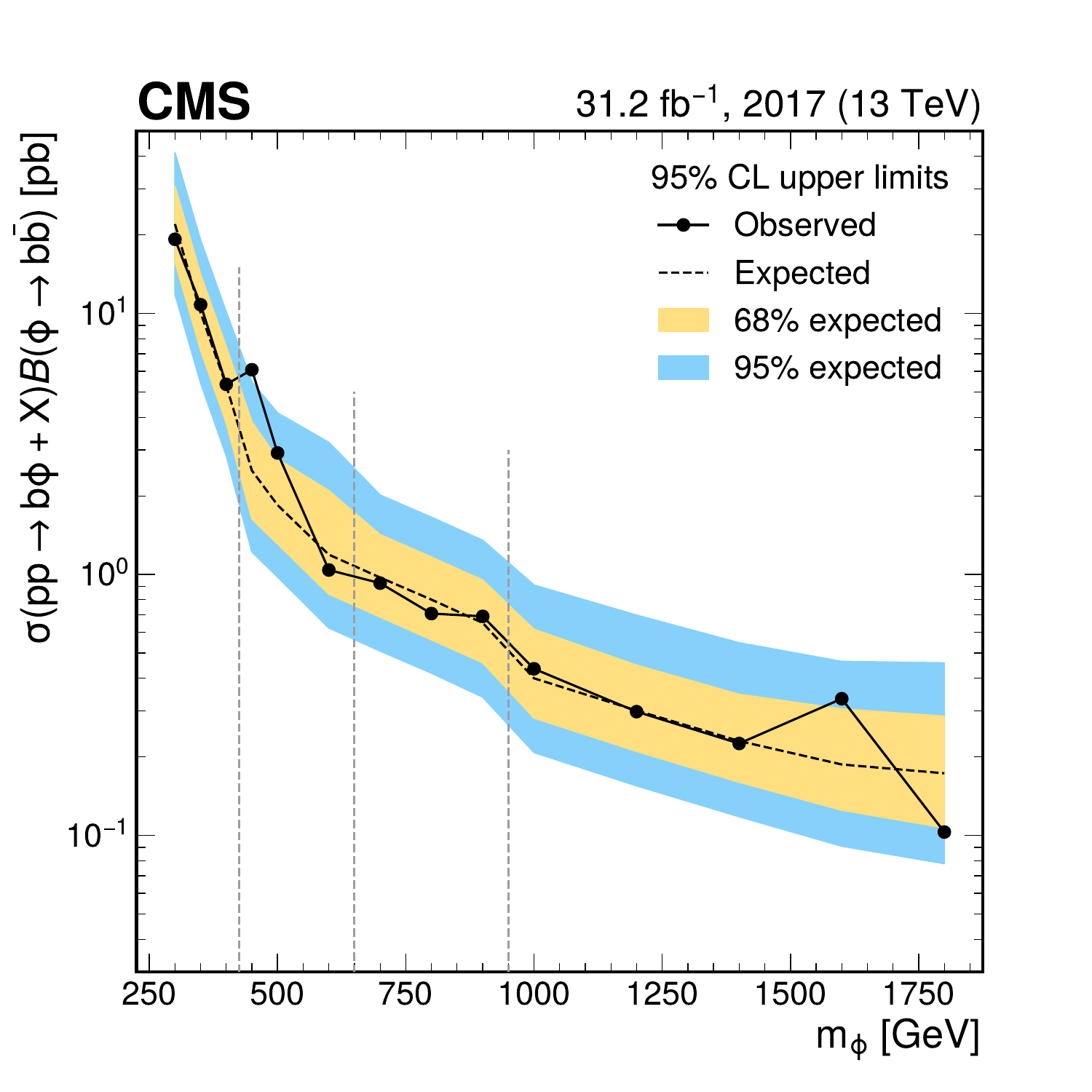}
    \caption{Expected and observed upper limits for the \PQb-quark-associated Higgs boson production cross section times branching fraction of the decay into a \PQb quark pair at 95\% CL as functions of \mPhi for the 2017 FH category. The vertical dashed lines indicate the boundaries of usage of the different fit ranges, as reflected in the rightmost column of Table~\ref{tab:massrangedefs}.}
\label{fig:SR_XSec_2017FH}
\end{figure}

\begin{figure}[!htbp]\centering
    \includegraphics[width=0.80\textwidth]{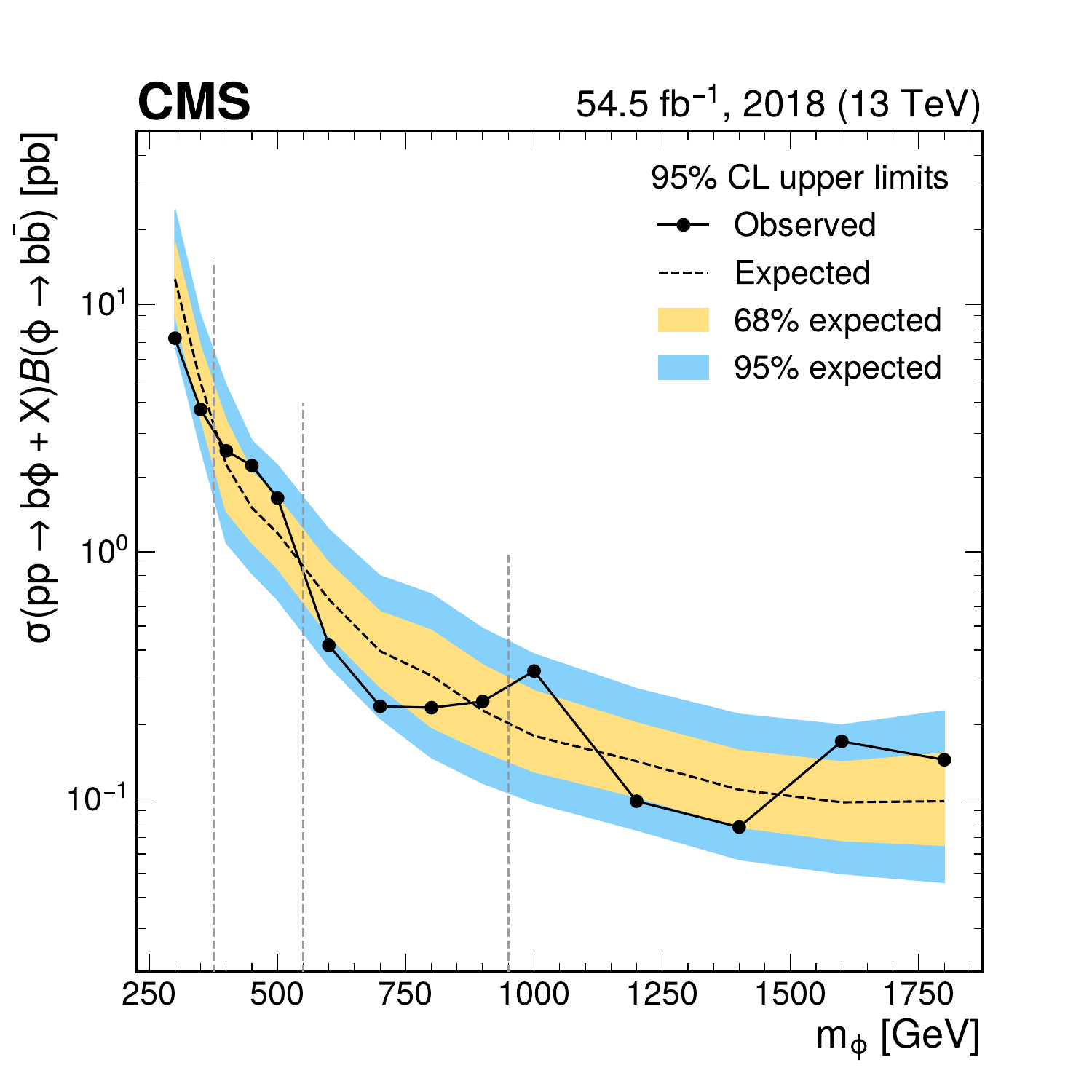}
    \caption{Expected and observed upper limits for the \PQb-quark-associated Higgs boson production cross section times branching fraction of the decay into a \PQb quark pair at 95\% CL as functions of \mPhi for the 2018 FH category. The vertical dashed lines indicate the boundaries of usage of the different fit ranges, as reflected in the rightmost column of Table~\ref{tab:massrangedefs}.}
\label{fig:SR_XSec_2018FH}
\end{figure}

Upper limits on the cross section times branching fraction for the three different data sets are shown in Figs.~\ref{fig:SR_XSec_2017SL}, \ref{fig:SR_XSec_2017FH}, and \ref{fig:SR_XSec_2018FH}. The observed limits are generally within or near ${\pm}2$ standard deviations ($\sigma$) of the expected limit. The largest excesses are observed in the 2017 SL analysis with local (global) significances of 3.2 (2.4)~$\sigma$  at $\mPhi=250\GeV$, and 2.7 (1.9)~$\sigma$ at $\mPhi=300\GeV$, where the global significance accounts for the look-elsewhere effect~\cite{Gross:2010qma} within the 2017 SL, 2017 FH, and 2018 FH analyses.


\begin{figure}[!htbp]\centering
    \includegraphics[width=0.80\textwidth]{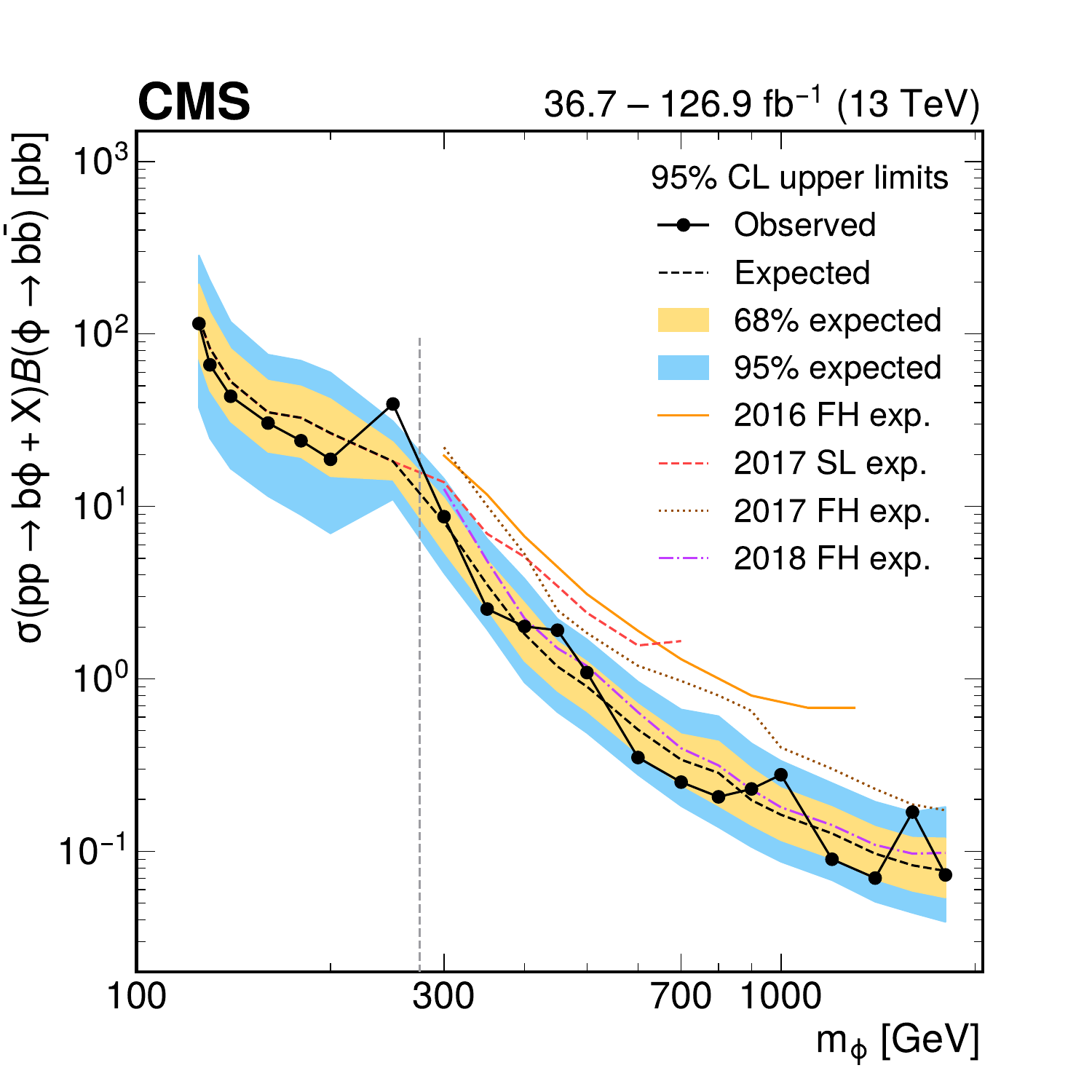}
    \caption{Expected and observed upper limits for the \PQb-quark-associated Higgs boson production cross section times branching fraction of the decay into a \PQb quark pair at 95\% CL as functions of \mPhi, corresponding to the combination with the 2016 data. The vertical dashed line separates the mass range where only the 2017 SL category contributes on its left, from the region where also the 2017 FH and 2018 FH categories contribute on its right. The expected limits from the 2017~SL, 2017~FH, and 2018~FH data sets as well as from the previously published result based on the 2016 data set are also shown as colored lines.}
\label{fig:SR_XSec_Run2Comb}
\end{figure}
The results from the three data sets 2017~SL, 2017~FH, and 2018~FH are combined with each other and with the previously published results from the 2016 data~\cite{CMS:2018hir}. The resulting combined upper limits for cross section times branching fraction are shown in Fig.~\ref{fig:SR_XSec_Run2Comb}, and compared to the results from the individual data sets.

The best limits are obtained with the 2018~FH analysis, resulting from the larger integrated luminosity and the less restrictive trigger conditions.  All 2017--2018 results benefit from the upgrade of the CMS pixel detector~\cite{CMSTrackerGroup:2020edz}, which generally improves the reconstruction of tracks near the interaction point. These analyses further benefit from the improved \PQb jet identification performance due to the \textsc{DeepJet} algorithm~\cite{Bols:2020bkb}, the more flexible matching of online and offline \PQb tag selections and the use of the transfer factor method for background modelling. All these improvements are reflected in the comparison with the result from the 2016 data. For $\mPhi = 1400\GeV$, the combined observed limits range down to about 0.07\unit{pb}. The 2017~SL analysis gives unique results for \mPhi in the range of 125--250\GeV, and, at $\mPhi = 300\GeV$, its sensitivity is on a par with the 2017 FH analysis.

\begin{figure}[!htbp]\centering
    \includegraphics[width=0.49\textwidth]{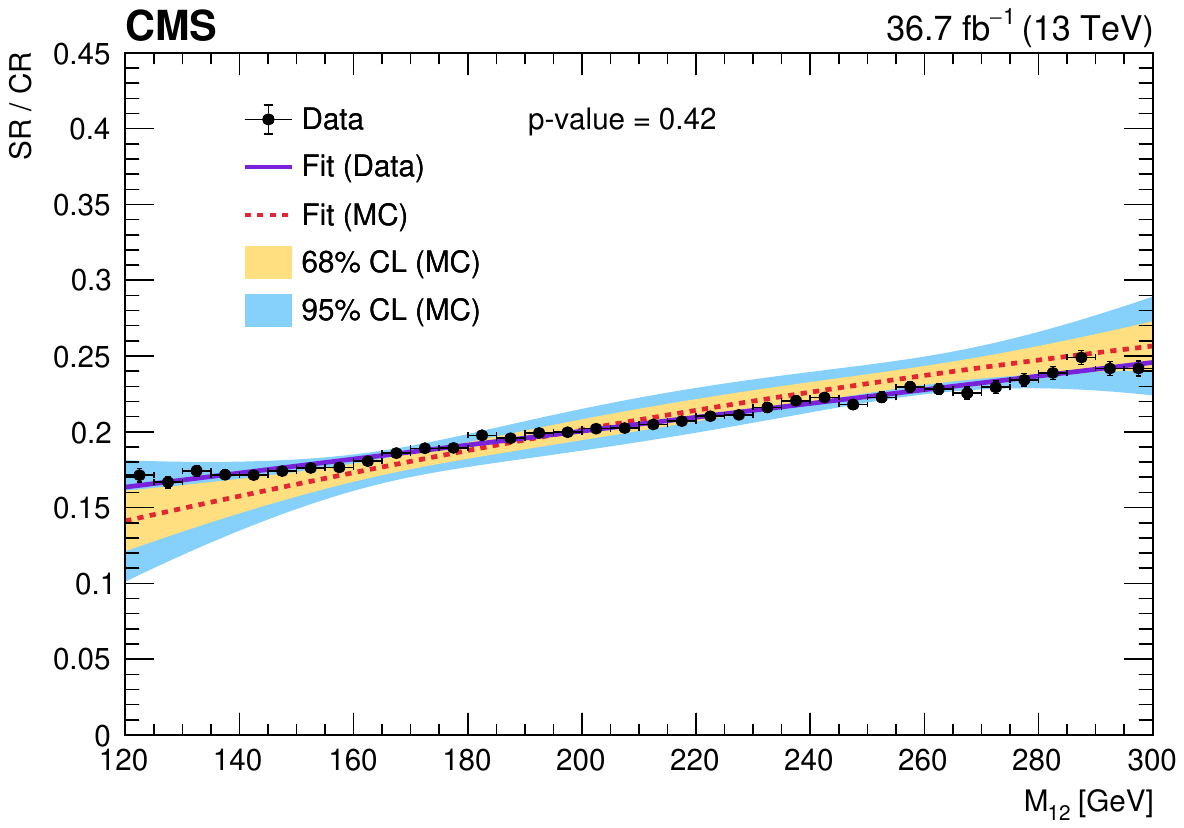}
    \includegraphics[width=0.49\textwidth]{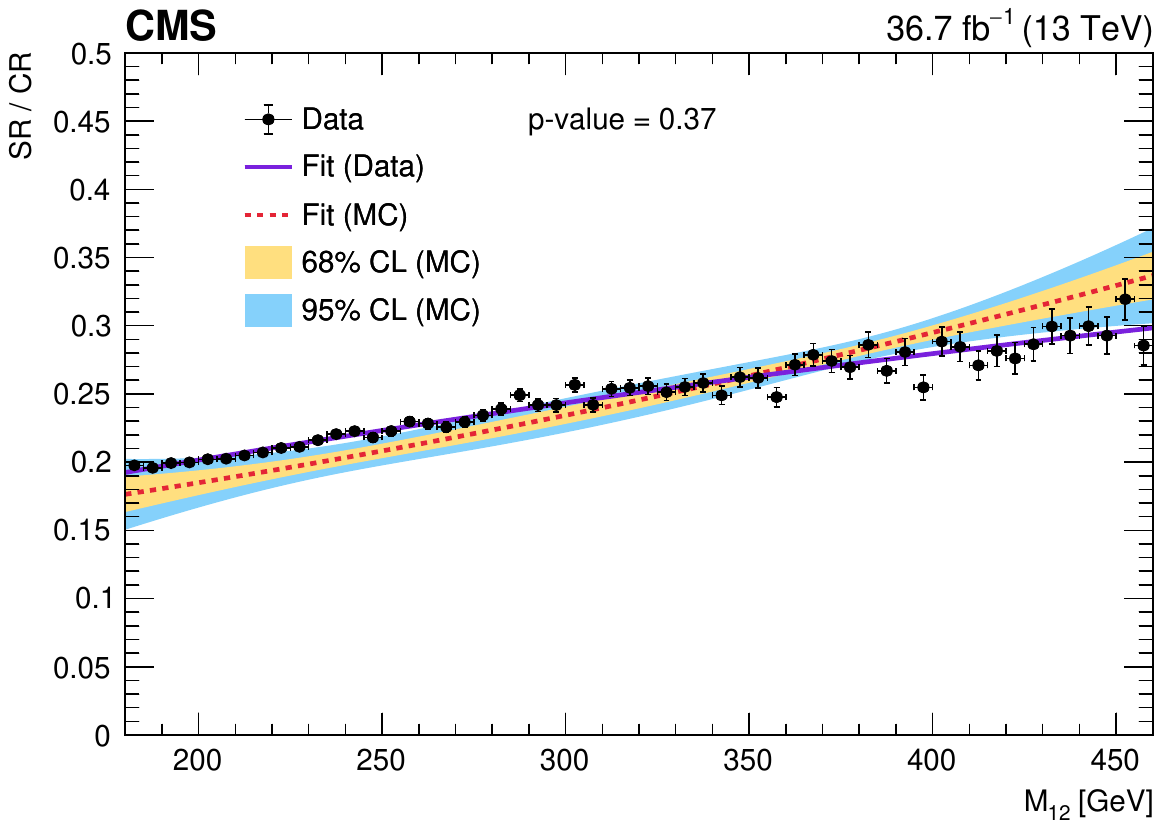}
    \includegraphics[width=0.49\textwidth]{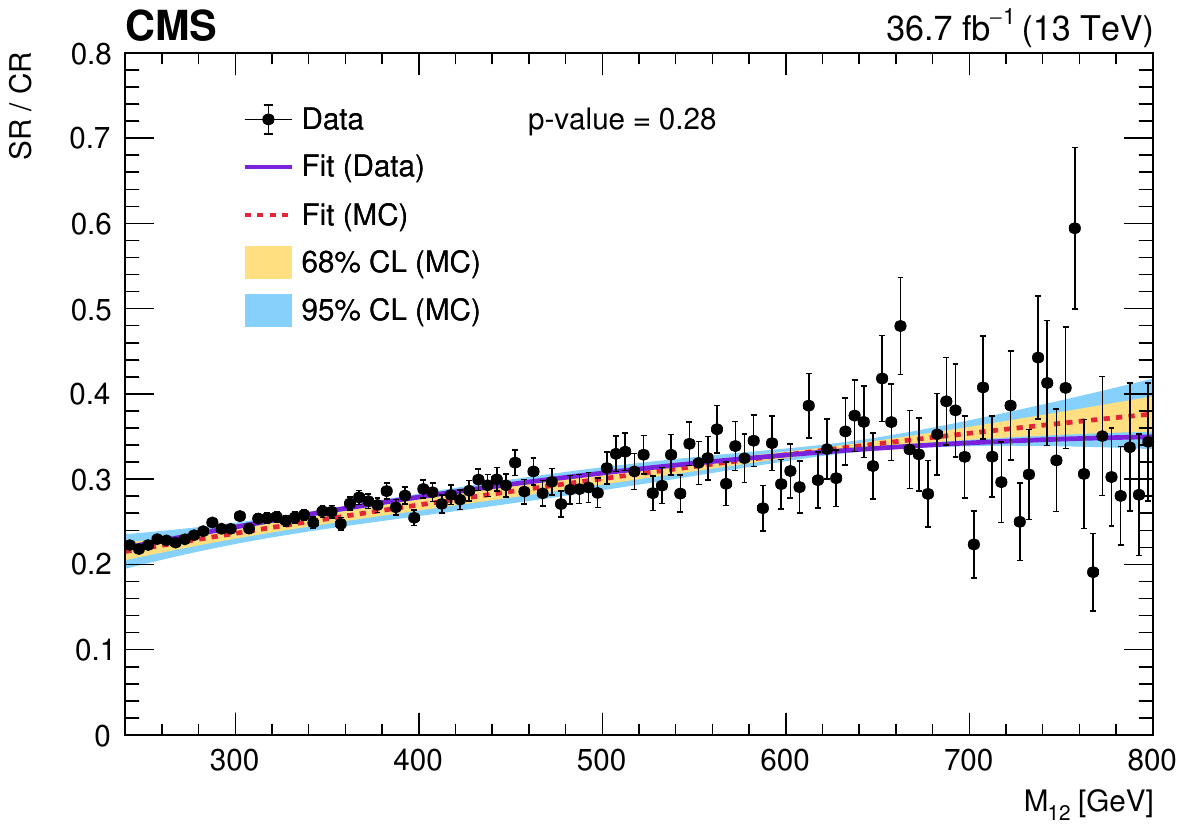}
    \caption{Ratio of \mjj distributions in SR and CR of the 2017 analysis in the SL category, for data (filled circles), fitted with a sum of Chebyshev polynomials up to the second degree (solid line). The ratio is shown in the \mjj ranges of 120--300\GeV (upper left), 180--460\GeV (upper right), and 240--800\GeV (lower). The p value resulting from each fit is also quoted. The result of a similar fit based on QCD multijet simulation is superimposed (dashed line) with $\pm 1\sigma$ and $\pm 2\sigma$ bands of statistical uncertainty extracted from the fit.}
\label{fig:tfDataMC_2017SL}
\end{figure}
The unblinding of the SR allows a direct verification of the assumptions underlying the background model. The 2017 SL data set is shown as a representative example. Figure~\ref{fig:tfDataMC_2017SL} shows the ratio of the \mjj distributions in the SR and the CR, taken from Figs.~\ref{fig:bonlyFits_2017SL} and \ref{fig:bbnb_fit_SR1to3_SL2017}, respectively, for each of the three fit ranges. As expected, the ratio is found to be only slowly varying as a function of \mjj, and it is well modelled by a sum of Chebyshev polynomials as they are used for the transfer factor as described in Section~\ref{sec:Background}. The results of the corresponding fits with simulated QCD multijet events are superimposed. There is qualitative agreement in the sense that data and simulation can be described with similar parameterisations. While the analysis does not rely in any way on a quantitative description of the data by the QCD simulation, the observed level of agreement is worth noting.

\subsection{Interpretation in the MSSM}
\begin{figure}\centering
    \includegraphics[width=0.70\textwidth]{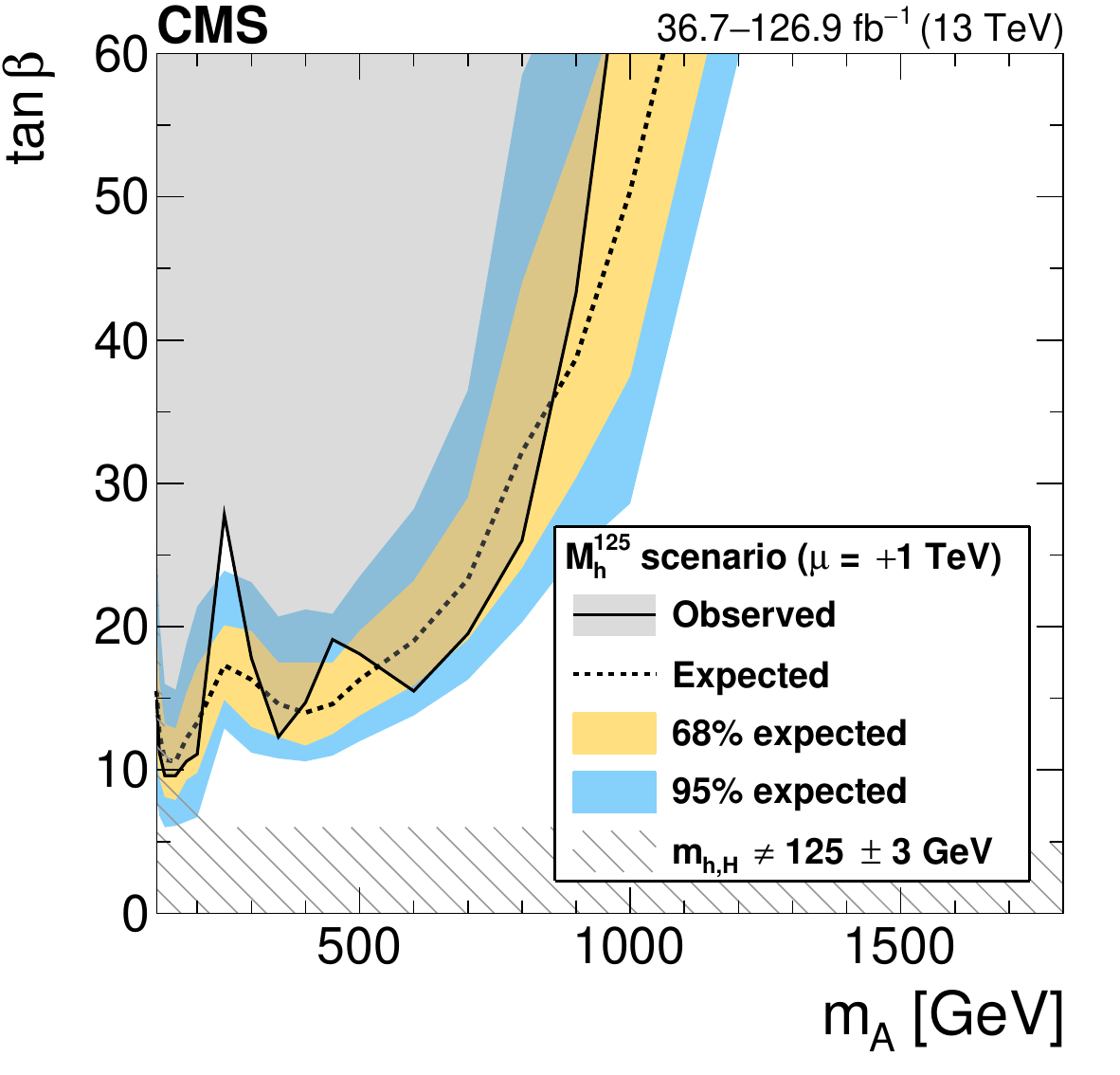}
    \caption{Interpretation in the \mhmass scenario of the MSSM: observed and expected upper limits at 95\% \CL on the parameter \tanb as functions of the mass \mA of the \CP-odd Higgs boson. The higgsino mass parameter has been set to $\mu = +1\TeV$. The hashed area indicates the parameter region in which the mass of the lightest MSSM Higgs boson does not coincide with 125\GeV within a margin of 3\GeV.}
\label{fig:Interp_MSSM_Mh125}
\end{figure}
The results from the combination with the 2016 data are interpreted within various scenarios of the MSSM.
For each benchmark scenario, the cross sections computed centrally by the LHC Higgs Working Group~\cite{LHCHiggsCrossSectionWorkingGroup:2016ypw} for \PQb-associated Higgs boson production are used. They are computed according to the four- and five-flavour schemes at NLO~\cite{Dawson:2003kb,Dittmaier:2003ej} and NNLO~\cite{Harlander:2003ai}, respectively. The predictions of both schemes are harmonized in the total cross section with Santander matching~\cite{Harlander:2011aa} for the hMSSM and \mhmodp scenarios, and with FONLL~\cite{Forte:2015hba,Forte:2016sja} in case of the \mhmass scenario. Branching fractions to the \bbbar final state are based on the \textsc{FeynHiggs}~\cite{Degrassi:2002fi,Heinemeyer:1998yj,Frank:2006yh,Hahn:2013ria,Bahl:2016brp,Bahl:2017aev,Bahl:2018qog} framework for all scenarios except for the hMSSM where \textsc{HDECAY}~\cite{Djouadi:1997yw,Djouadi:2018xqq} was used.

Figure~\ref{fig:Interp_MSSM_Mh125} shows the interpretation within the \mhmass scenario, with the default setting of the higgsino mass parameter of $\mu = +1\TeV$. The most stringent upper limits on the parameter \tanb of about 9 are reached in the mass region of 130--150\GeV. The structure of the expected limit reflects the maximum of the background distribution near a reconstructed invariant mass of 180\GeV and the onset of the fully hadronic channel above 300\GeV. They represent the most stringent limits on \tanb obtained from the \bbbar channel to date.

\begin{figure}\centering
    \includegraphics[width=0.48\textwidth]{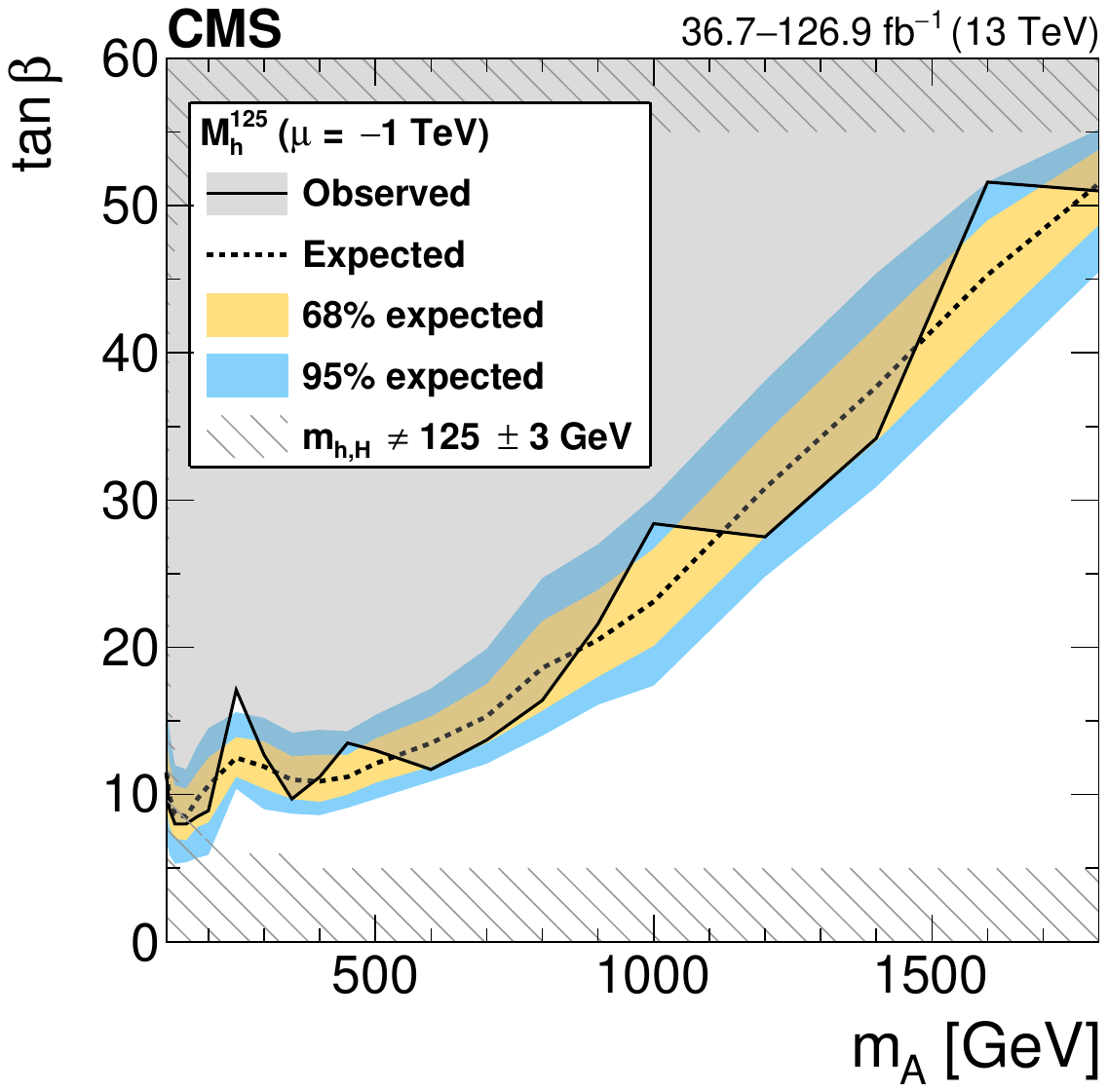}
    \includegraphics[width=0.48\textwidth]{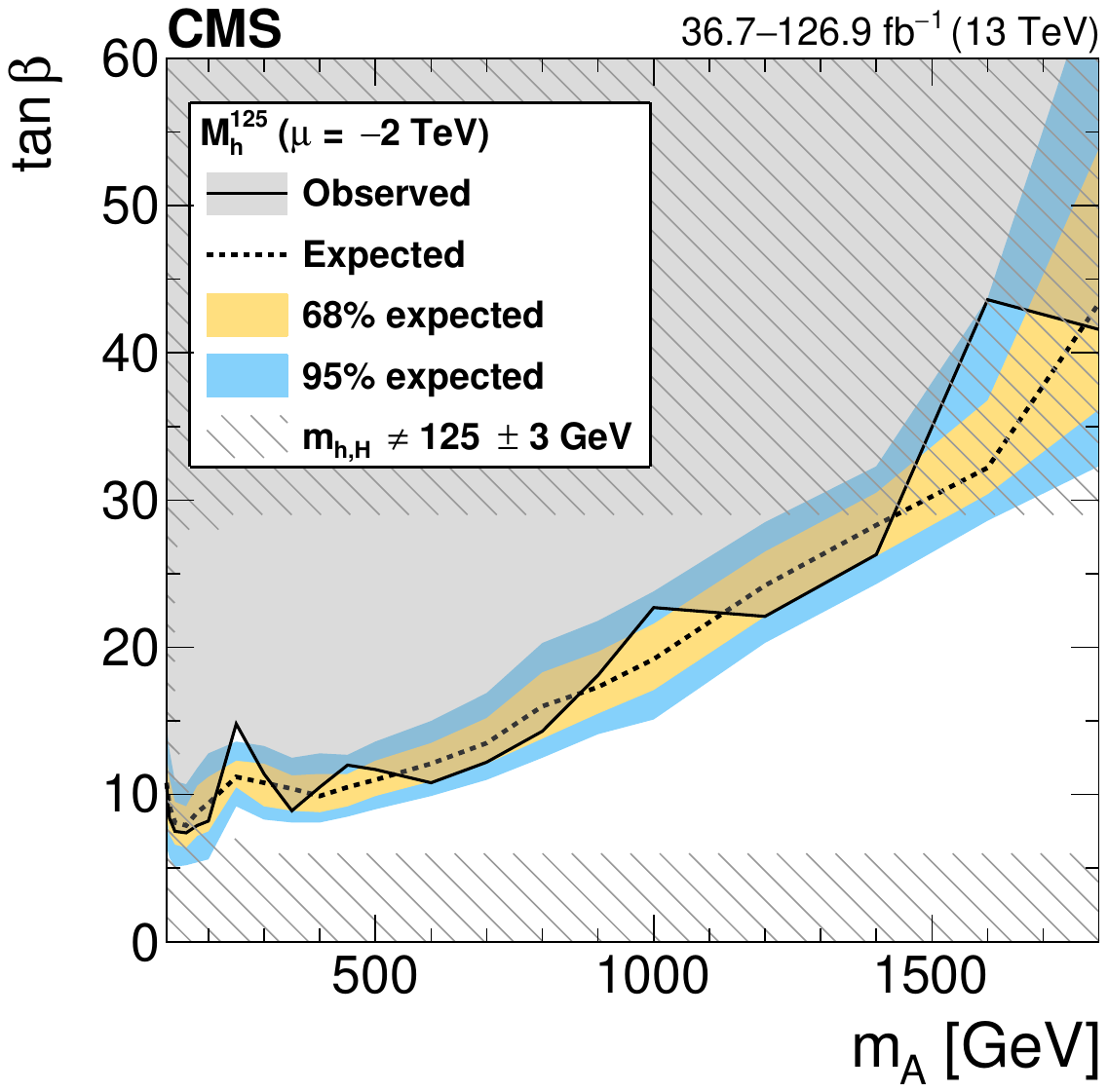}
    \includegraphics[width=0.48\textwidth]{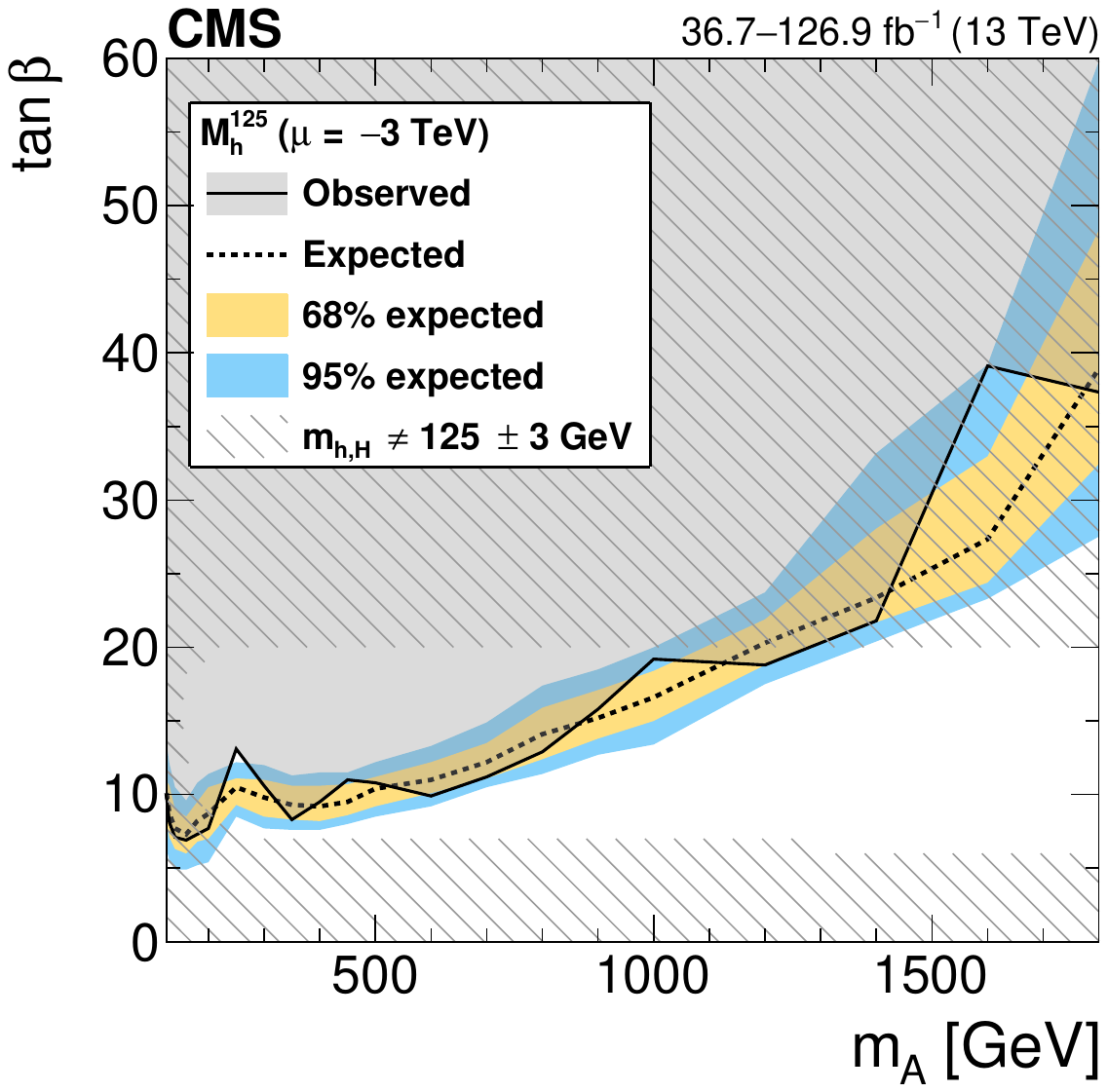}
    \caption{Interpretation in the \mhmass scenario of the MSSM: observed and expected upper limits at 95\% \CL on the parameter \tanb as functions of the mass \mA of the \CP-odd Higgs boson. The higgsino mass parameter has been set to $\mu = -1\TeV$ (upper left), $\mu = -2\TeV$ (upper right), and $\mu = -3\TeV$ (lower). The hashed area indicates the parameter region in which the mass of the lightest MSSM Higgs boson does not coincide with 125\GeV within a margin of 3\GeV.}
\label{fig:Interp_MSSM_Mh125_Higgsino}
\end{figure}
\begin{figure}\centering
    \includegraphics[width=0.48\textwidth]{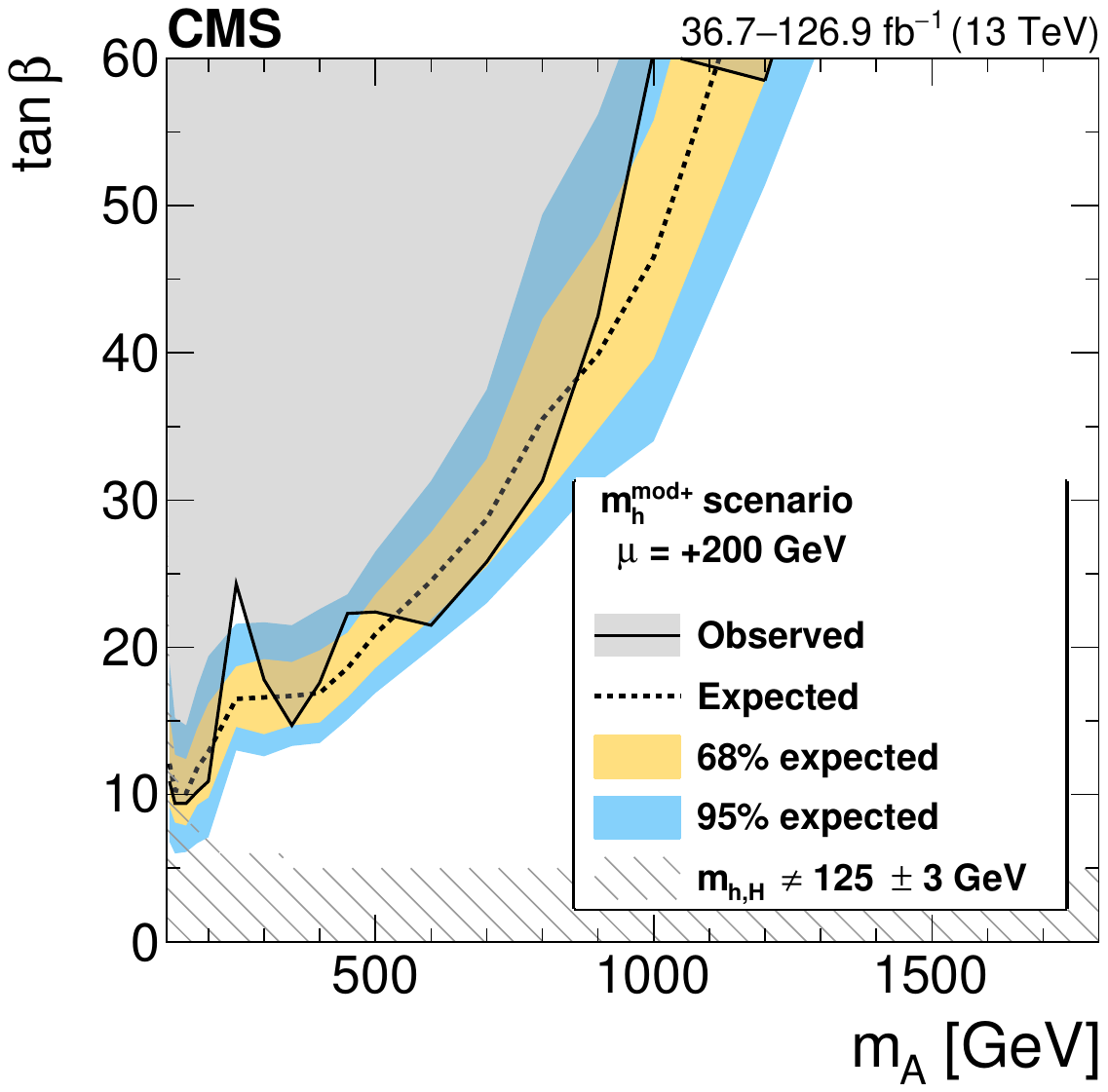}
    \includegraphics[width=0.48\textwidth]{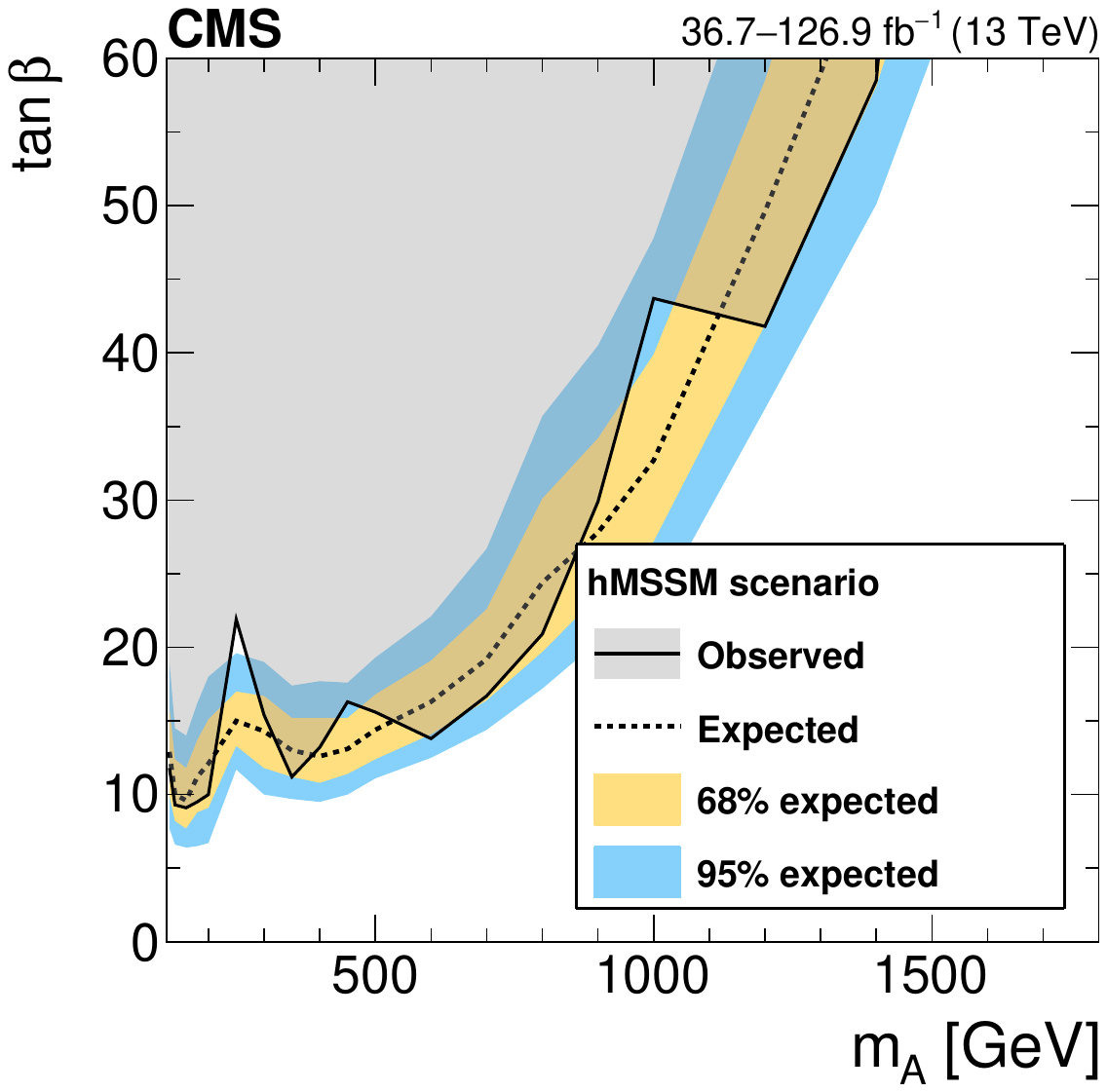}
    \caption{Interpretation in the \mhmodp (left) and hMSSM (right) scenarios of the MSSM: observed and expected upper limits at 95\% \CL on the parameter \tanb as functions of the mass \mA of the \CP-odd Higgs boson. In the left plot, the hashed area indicates the parameter region in which the mass of the lightest MSSM Higgs boson does not coincide with 125\GeV within a margin of 3\GeV.}
\label{fig:Interp_MSSM_mhmodp}
\end{figure}

The $(\PSA, \PH)\to\bbbar$ decay channel is uniquely sensitive to negative values of the higgsino mass parameter $\mu$, as the bottom Yukawa coupling correction $\Delta_\mathrm{b}$~\cite{Carena:2005ek,Asmuss:2021} is directly proportional to $\mu$. Since the product of cross section and branching fraction of this process is proportional to $1 / (1+\Delta_\mathrm{b})^4$, negative values of $\mu$ enhance the total production rate.
Interpretations in \mhmass scenarios considering negative higgsino mass parameters of $\mu = -1$, $-2$, and $-3\TeV$~\cite{Bahl:2020kwe} are shown in Fig.~\ref{fig:Interp_MSSM_Mh125_Higgsino}. The limits on \tanb extend approximately down to 8 at 160\GeV and up to 51 at 1800\GeV in the $\mu = -1\TeV$ case. The hashed region, indicating the parameter space incompatible with the observed Higgs boson mass of $125 \pm 3\GeV$ due to theoretical constraints, grows with the absolute value of $\mu$ in these scenarios. In the  $\mu = -2\,(-3)\TeV$ scenario, \tanb values of above 7.5 (7) are excluded at 160 (180)\GeV, ranging up to nearly 30~(20) in the allowed parameter space at about 1500 (1300)\GeV.
To facilitate comparisons with other studies, Fig.~\ref{fig:Interp_MSSM_mhmodp} also shows interpretations in the \mhmodp and hMSSM scenarios.

\subsection{Interpretation in 2HDMs}
\begin{figure}\centering
    \includegraphics[width=0.48\textwidth]{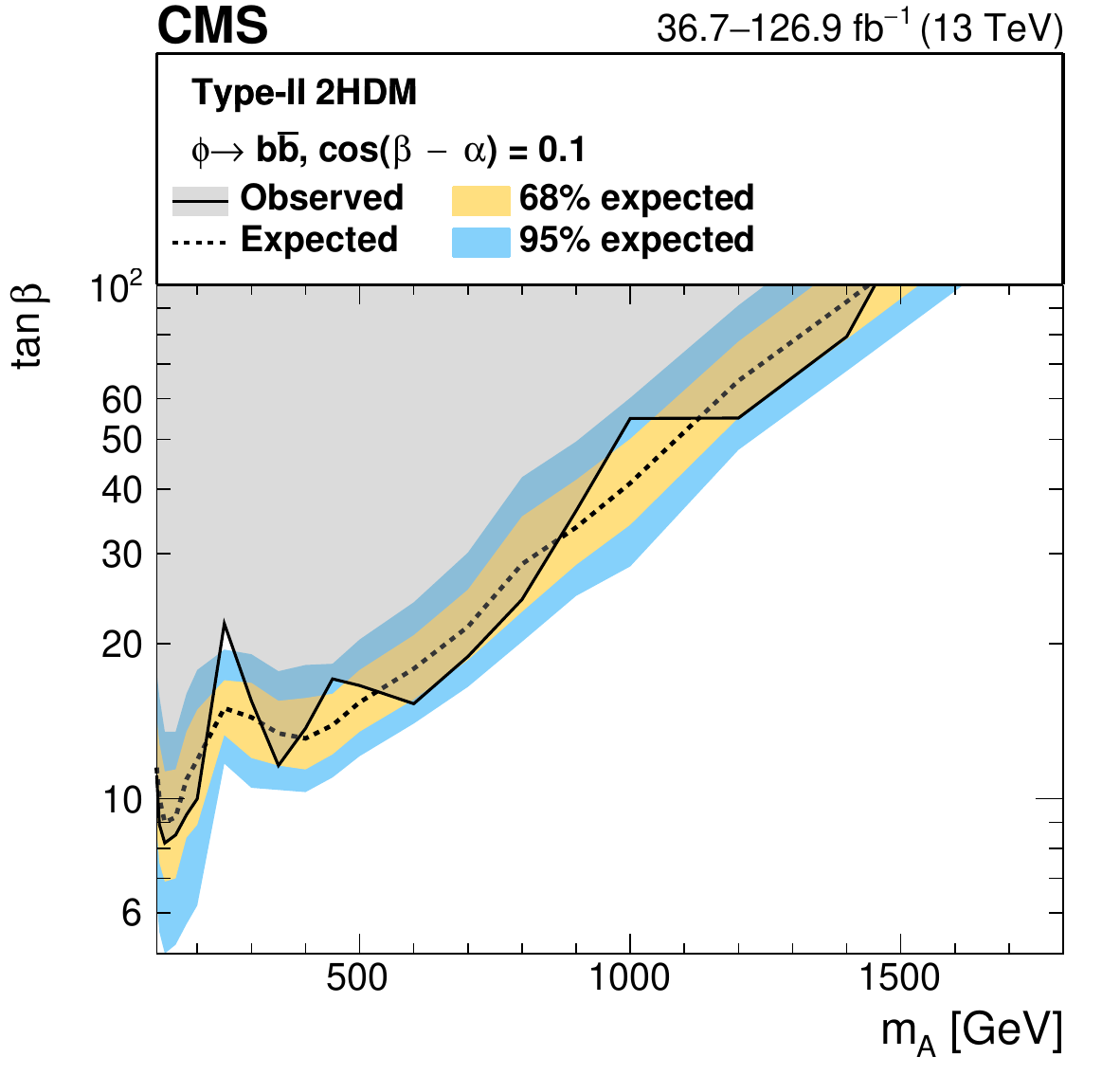}
    \includegraphics[width=0.48\textwidth]{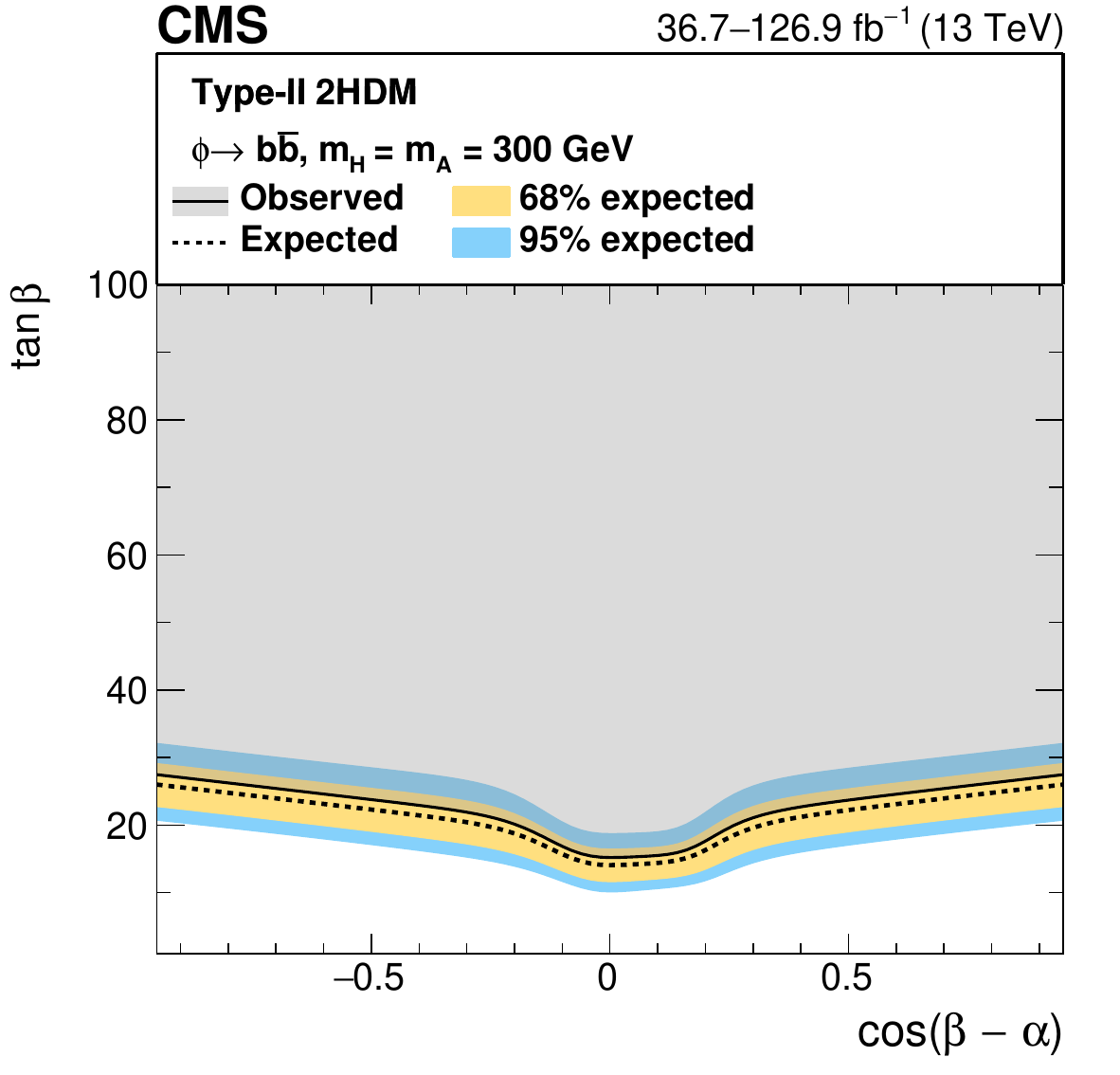}
    \includegraphics[width=0.48\textwidth]{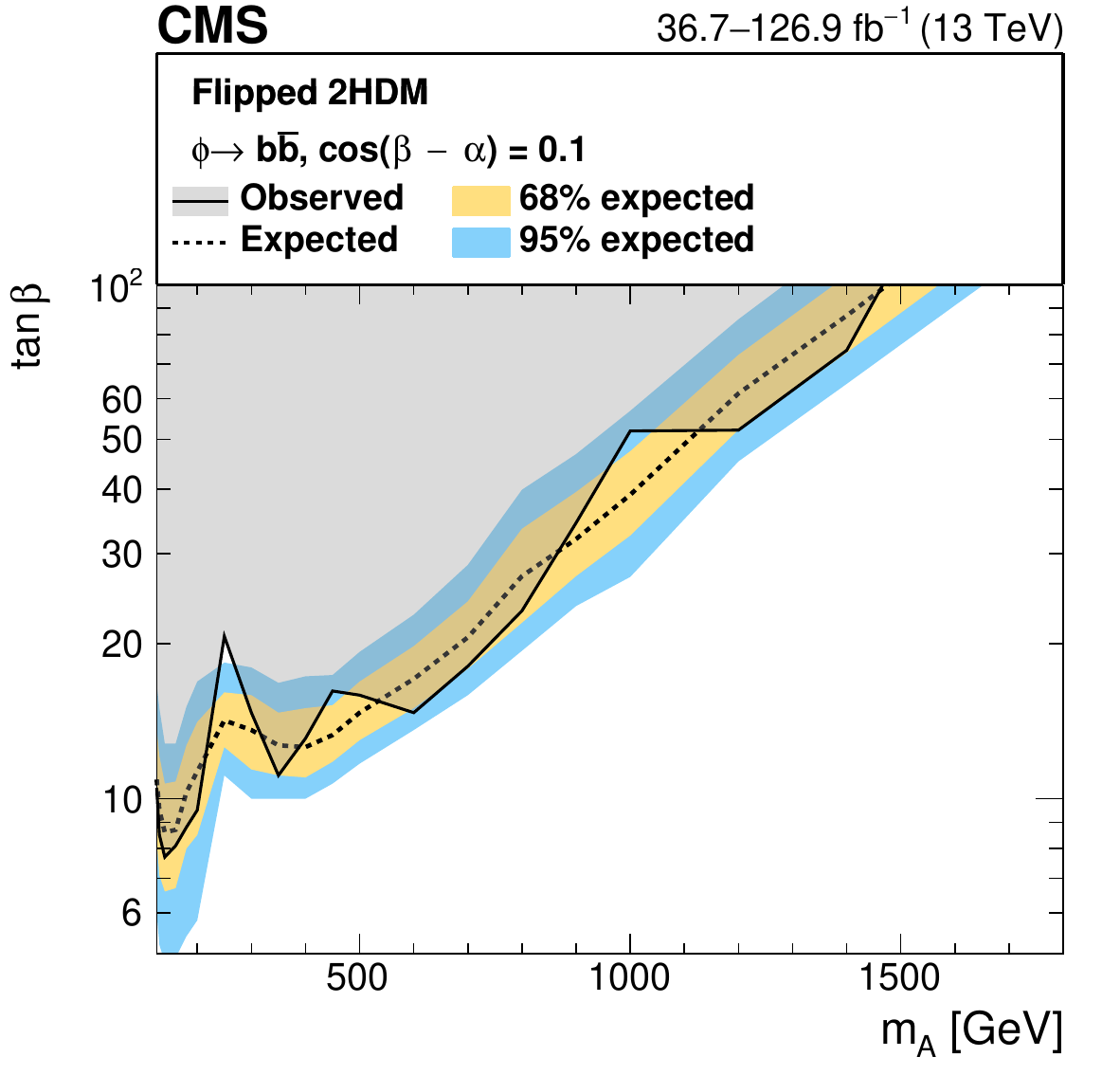}
    \includegraphics[width=0.48\textwidth]{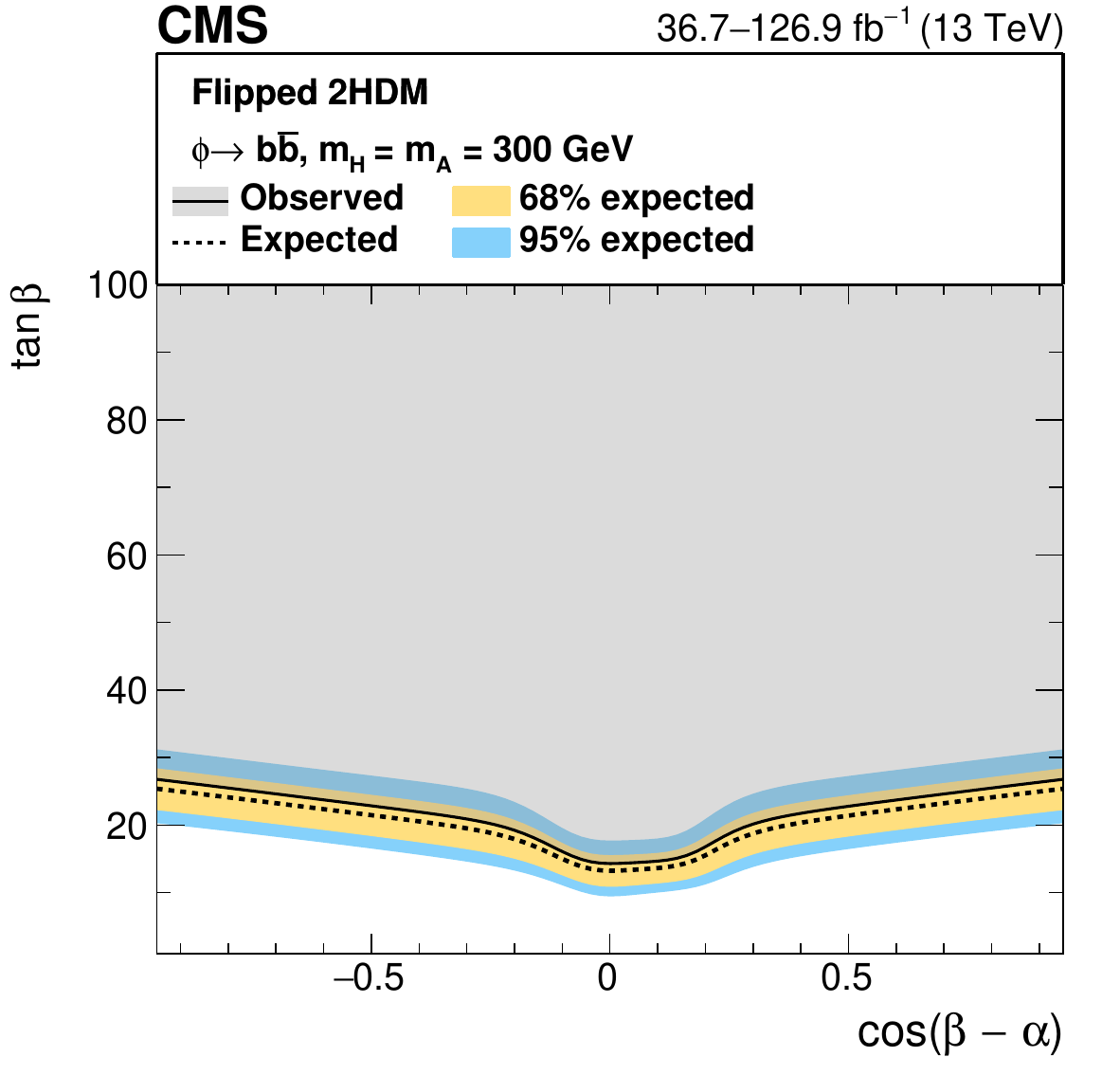}
    \caption{Interpretation in 2HDM scenarios: observed and expected upper limits at 95\% \CL on the parameter \tanb as a function of \mAH for $\cosba=0.1$ (left), and as functions of \cosba for masses of $\mA = \mH = 300\GeV$ (right), for the 2HDM \typetwo scenario (upper), and the 2HDM \flipped scenario (lower).}
\label{fig:Interp_2HDM_M300}
\end{figure}
The results are interpreted in 2HDM scenarios as functions of their parameters. The \PQb-associated production cross sections and the branching fractions at NNLO precision are computed using the \textsc{SusHi}~\cite{Harlander:2012pb,Harlander:2016hcx}, \textsc{2HDMC}~\cite{Eriksson:2009ws} and \textsc{LHAPDF}~\cite{Buckley:2014ana} programs.
Generally, 2HDMs feature a large number of free parameters, which can be reduced down to three parameters according to scenario G defined in Ref.~\cite{Haber:2015pua}, namely \mA, \tanb, and \cosba. The interpretations are performed for the benchmark scenarios where the coupling of the Higgs boson to \PQb quarks is enhanced for $\tanb > 1$, \ie the \typetwo and \flipped benchmark scenarios. In each interpretation, one parameter is fixed and limits are provided in the space of the remaining two parameters. 

The left panels in Fig.~\ref{fig:Interp_2HDM_M300} show the limits in the \typetwo and \flipped models in the \tanb and \mA parameter space for $\cosba=0.1$. 
The limits in both benchmark scenarios are rather similar, and values of \tanb can be excluded down to about 8 at 140\GeV, 12 at 400\GeV, 17 at 600\GeV, 30 at 900\GeV, and 50 at 1200\GeV.  In the \flipped model, the sensitivity of the \bbbar channel is unique since no other difermion mode is sensitive in this area of the parameter space.

\begin{figure}\centering
    \includegraphics[width=0.48\textwidth]{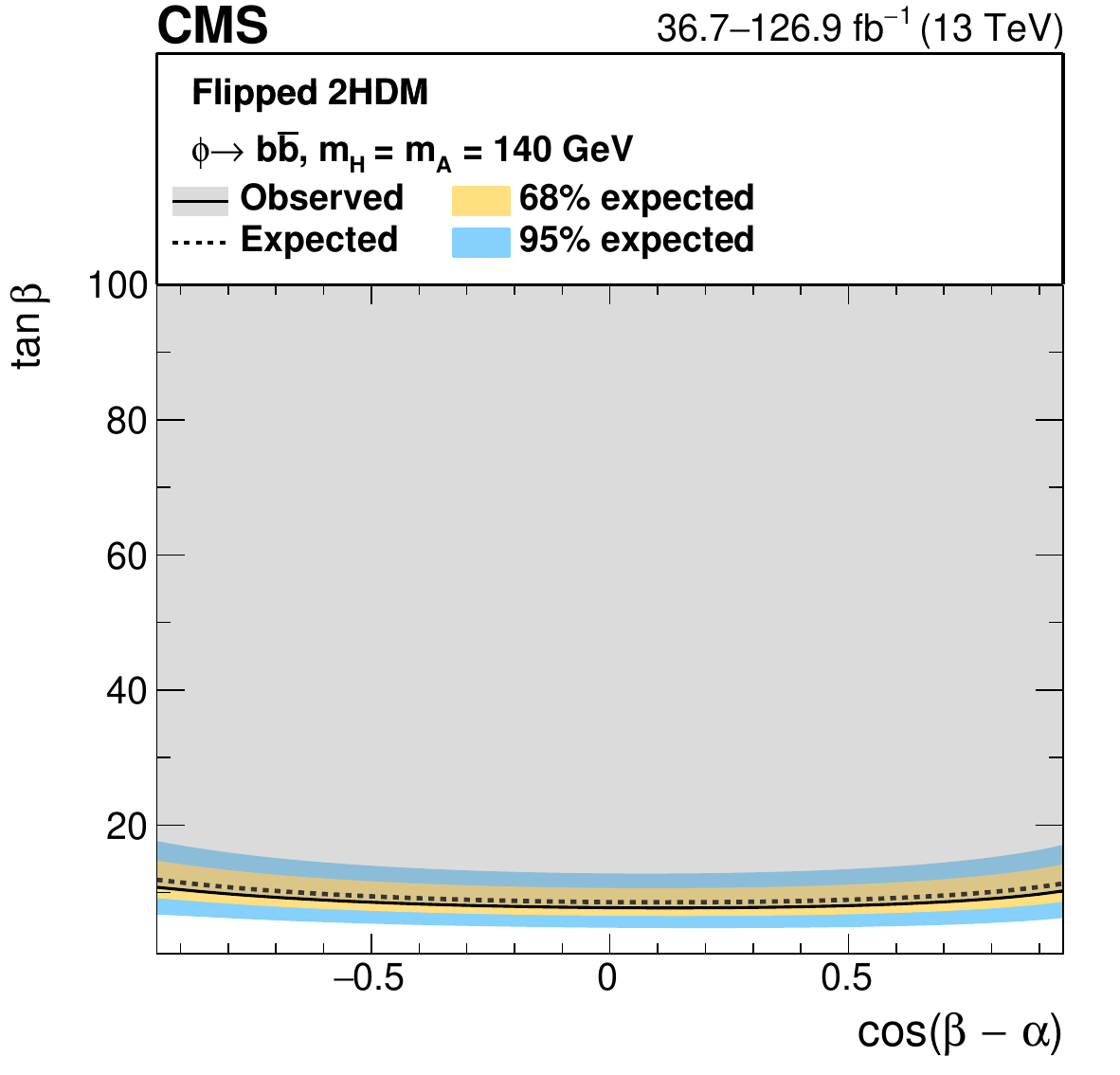}
    \includegraphics[width=0.48\textwidth]{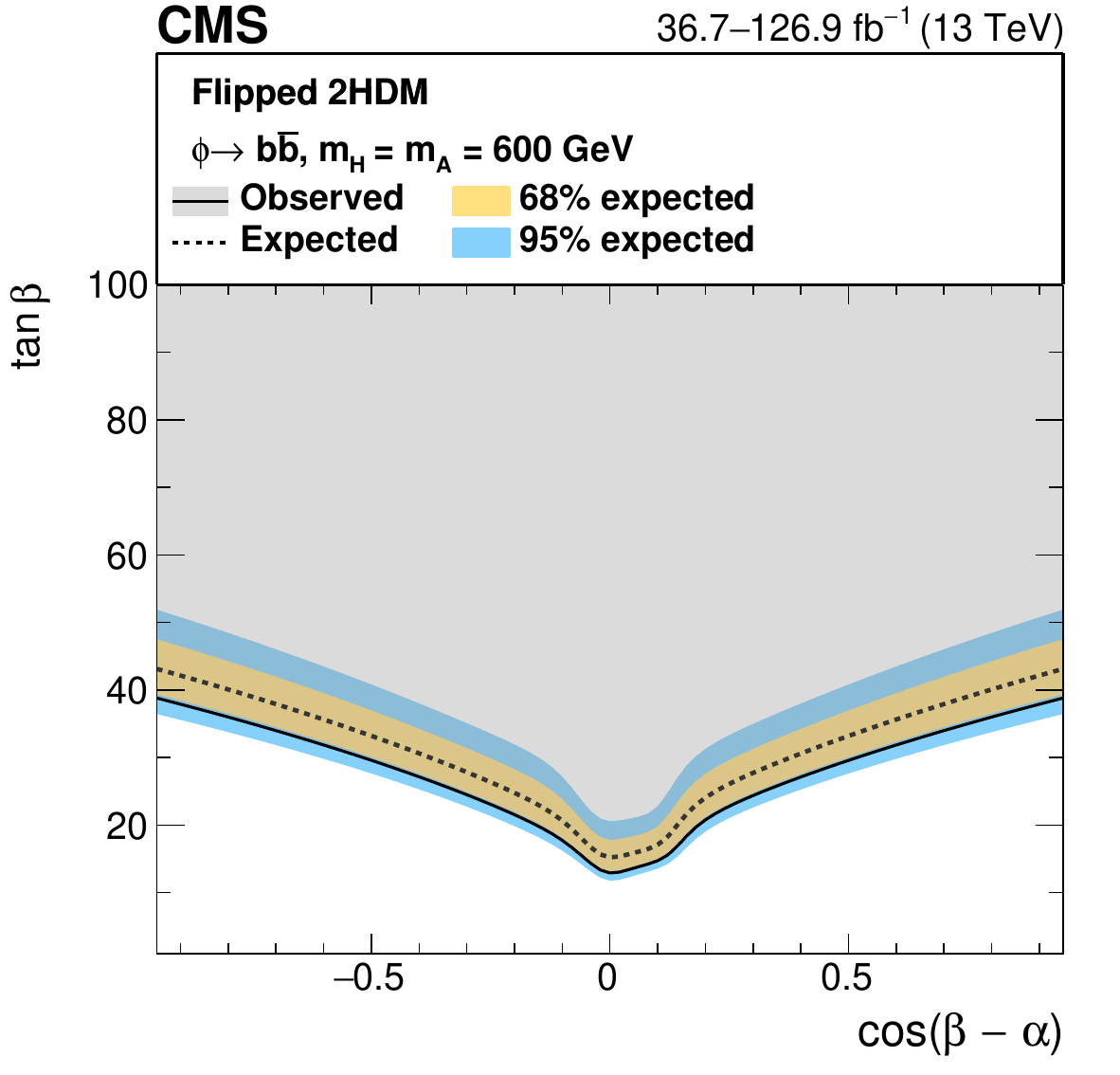}
    \includegraphics[width=0.48\textwidth]{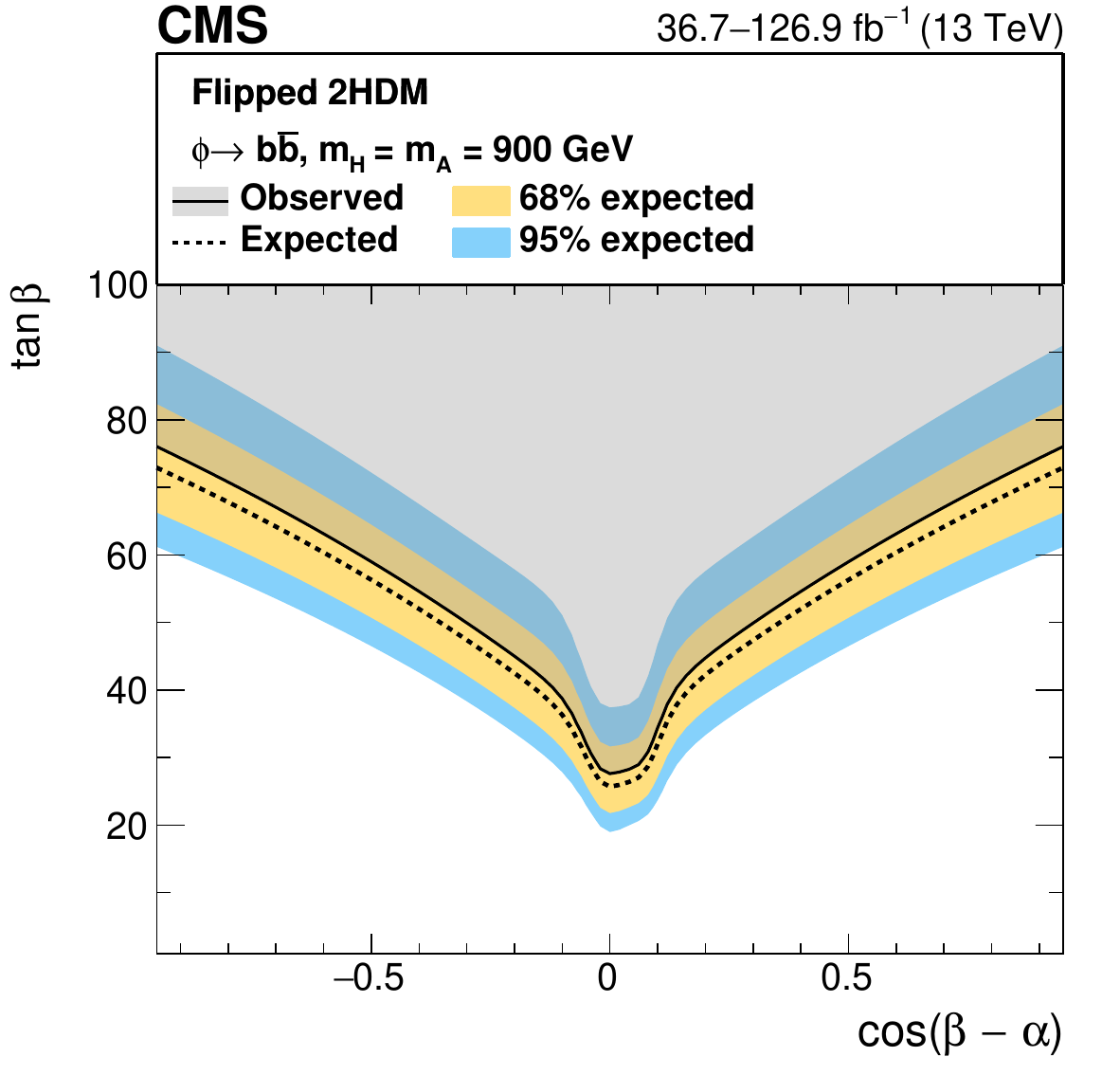}
    \includegraphics[width=0.48\textwidth]{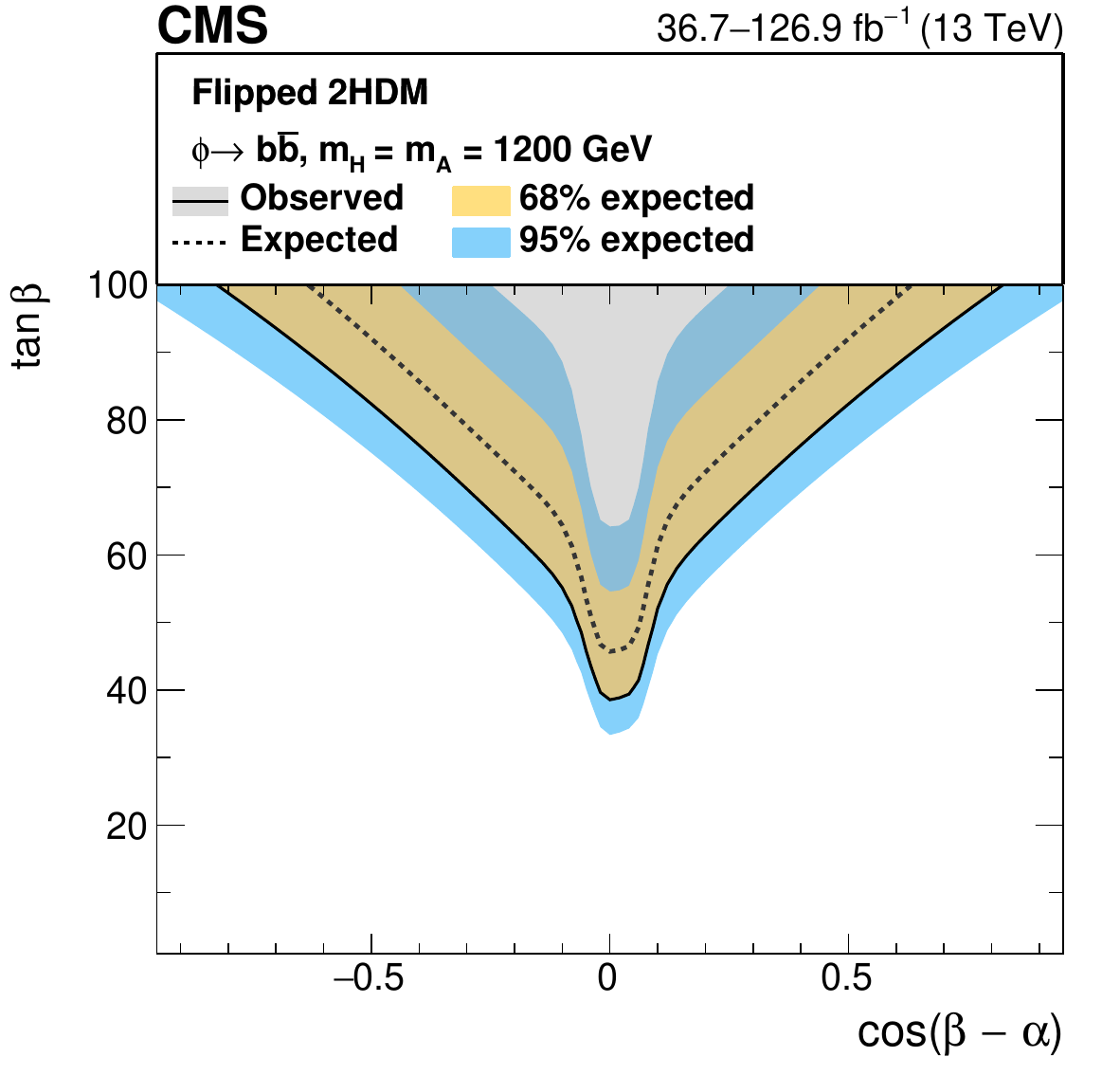}
    \caption{Interpretation in the 2HDM flipped scenario: observed and expected upper limits at 95\% \CL on the parameter \tanb as functions of \cosba for masses of $\mA = \mH = 140$, 600, 900, and 1200\GeV.}
\label{fig:Interp_2HDM_flipped_M140to1200}
\end{figure}

Values of \cosba further away from the alignment limit display a visible deterioration in the \tanb limits that grows with increasing mass, as can be seen in the right panels of Fig.~\ref{fig:Interp_2HDM_M300} for $\mA = \mH = 300\GeV$ in the \typetwo and \flipped models, and in Fig.~\ref{fig:Interp_2HDM_flipped_M140to1200} for four other mass points in the \flipped model, which are very similar to those from the \typetwo model. This effect can be largely attributed to the branching fractions of these Higgs bosons. For the \PH boson, the most important decay mode competing with $\PH \to \bbbar$ is $\PH\to \PSh\PSh$ if the \PH mass is above its kinematic threshold.
Outside of the alignment limit, this channel tends to saturate the total decay width quickly. The dips in the \tanb exclusions near $\cosba = 0$ indicate the region where the \PH boson contribution to the combined sensitivity is still sizable. This dip does not exist for $\mA = \mH = 140\GeV$, which is below the $\PH \to \PSh\PSh$ threshold. For the \PSA boson, on the other hand, the most important competing decay mode is $\PSA \to \PZ\PSh$. The increase of the $\PSA\PZ\PSh$ coupling, at the expense of the $\PSA \to \bbbar$ branching fraction, is reflected in the steady increase of the \tanb limits for larger $\abs{\cosba}$.

\section{Summary}
\label{sec:Summary}
A search for beyond-the-standard-model neutral Higgs bosons, $\phi$, produced in association with \PQb quarks and decaying into a pair of \PQb quarks is presented using a CMS data set of 13\TeV proton-proton collisions, based on an integrated luminosity of 36.7--126.9\fbinv. The multi \PQb quark final state is selected with requirements targeting both fully hadronic and semileptonic \PQb quark decays, allowing for a sensitivity in the mass range extending from 125 to 1800\GeV. No significant excess of events above the expected SM background is observed. Exclusion limits at 95\% confidence level on the production cross section times branching fraction are obtained. The results are also interpreted as constraints in the parameter space of MSSM and 2HDM scenarios to which this search is sensitive. These results represent the most stringent limits to date in the high-mass regime with this final state.

\begin{acknowledgments}
    We congratulate our colleagues in the CERN accelerator departments for the excellent performance of the LHC and thank the technical and administrative staffs at CERN and at other CMS institutes for their contributions to the success of the CMS effort. In addition, we gratefully acknowledge the computing centres and personnel of the Worldwide LHC Computing Grid and other centres for delivering so effectively the computing infrastructure essential to our analyses. Finally, we acknowledge the enduring support for the construction and operation of the LHC, the CMS detector, and the supporting computing infrastructure provided by the following funding agencies: SC (Armenia), BMBWF and FWF (Austria); FNRS and FWO (Belgium); CNPq, CAPES, FAPERJ, FAPERGS, and FAPESP (Brazil); MES and BNSF (Bulgaria); CERN; CAS, MoST, and NSFC (China); MINCIENCIAS (Colombia); MSES and CSF (Croatia); RIF (Cyprus); SENESCYT (Ecuador); ERC PRG, RVTT3 and MoER TK202 (Estonia); Academy of Finland, MEC, and HIP (Finland); CEA and CNRS/IN2P3 (France); SRNSF (Georgia); BMBF, DFG, and HGF (Germany); GSRI (Greece); NKFIH (Hungary); DAE and DST (India); IPM (Iran); SFI (Ireland); INFN (Italy); MSIP and NRF (Republic of Korea); MES (Latvia); LMTLT (Lithuania); MOE and UM (Malaysia); BUAP, CINVESTAV, CONACYT, LNS, SEP, and UASLP-FAI (Mexico); MOS (Montenegro); MBIE (New Zealand); PAEC (Pakistan); MES and NSC (Poland); FCT (Portugal);  MESTD (Serbia); MICIU/AEI and PCTI (Spain); MOSTR (Sri Lanka); Swiss Funding Agencies (Switzerland); MST (Taipei); MHESI and NSTDA (Thailand); TUBITAK and TENMAK (Turkey); NASU (Ukraine); STFC (United Kingdom); DOE and NSF (USA).
      
    \hyphenation{Rachada-pisek} Individuals have received support from the Marie-Curie programme and the European Research Council and Horizon 2020 Grant, contract Nos.\ 675440, 724704, 752730, 758316, 765710, 824093, 101115353, 101002207, and COST Action CA16108 (European Union); the Leventis Foundation; the Alfred P.\ Sloan Foundation; the Alexander von Humboldt Foundation; the Science Committee, project no. 22rl-037 (Armenia); the Fonds pour la Formation \`a la Recherche dans l'Industrie et dans l'Agriculture (FRIA-Belgium); the Beijing Municipal Science \& Technology Commission, No. Z191100007219010 and Fundamental Research Funds for the Central Universities (China); the Ministry of Education, Youth and Sports (MEYS) of the Czech Republic; the Shota Rustaveli National Science Foundation, grant FR-22-985 (Georgia); the Deutsche Forschungsgemeinschaft (DFG), among others, under Germany's Excellence Strategy -- EXC 2121 ``Quantum Universe" -- 390833306, and under project number 400140256 - GRK2497; the Hellenic Foundation for Research and Innovation (HFRI), Project Number 2288 (Greece); the Hungarian Academy of Sciences, the New National Excellence Program - \'UNKP, the NKFIH research grants K 131991, K 133046, K 138136, K 143460, K 143477, K 146913, K 146914, K 147048, 2020-2.2.1-ED-2021-00181, TKP2021-NKTA-64, and 2021-4.1.2-NEMZ\_KI-2024-00036 (Hungary); the Council of Science and Industrial Research, India; ICSC -- National Research Centre for High Performance Computing, Big Data and Quantum Computing and FAIR -- Future Artificial Intelligence Research, funded by the NextGenerationEU program (Italy); the Latvian Council of Science; the Ministry of Education and Science, project no. 2022/WK/14, and the National Science Center, contracts Opus 2021/41/B/ST2/01369 and 2021/43/B/ST2/01552 (Poland); the Funda\c{c}\~ao para a Ci\^encia e a Tecnologia, grant CEECIND/01334/2018 (Portugal); the National Priorities Research Program by Qatar National Research Fund; MICIU/AEI/10.13039/501100011033, ERDF/EU, "European Union NextGenerationEU/PRTR", and Programa Severo Ochoa del Principado de Asturias (Spain); the Chulalongkorn Academic into Its 2nd Century Project Advancement Project, and the National Science, Research and Innovation Fund via the Program Management Unit for Human Resources \& Institutional Development, Research and Innovation, grant B39G680009 (Thailand); the Kavli Foundation; the Nvidia Corporation; the SuperMicro Corporation; the Welch Foundation, contract C-1845; and the Weston Havens Foundation (USA).
\end{acknowledgments}

\bibliography{auto_generated}

\cleardoublepage \appendix\section{The CMS Collaboration \label{app:collab}}\begin{sloppypar}\hyphenpenalty=5000\widowpenalty=500\clubpenalty=5000\input{SUS-24-001-public-authorlist.tex}\end{sloppypar}
\end{document}

%% file: SUS-24-001-public-authorlist.tex
\cmsinstitute{Yerevan Physics Institute, Yerevan, Armenia}
{\tolerance=6000
V.~Chekhovsky, A.~Hayrapetyan, V.~Makarenko\cmsorcid{0000-0002-8406-8605}, A.~Tumasyan\cmsAuthorMark{1}\cmsorcid{0009-0000-0684-6742}
\par}
\cmsinstitute{Institut f\"{u}r Hochenergiephysik, Vienna, Austria}
{\tolerance=6000
W.~Adam\cmsorcid{0000-0001-9099-4341}, J.W.~Andrejkovic, L.~Benato\cmsorcid{0000-0001-5135-7489}, T.~Bergauer\cmsorcid{0000-0002-5786-0293}, S.~Chatterjee\cmsorcid{0000-0003-2660-0349}, K.~Damanakis\cmsorcid{0000-0001-5389-2872}, M.~Dragicevic\cmsorcid{0000-0003-1967-6783}, P.S.~Hussain\cmsorcid{0000-0002-4825-5278}, M.~Jeitler\cmsAuthorMark{2}\cmsorcid{0000-0002-5141-9560}, N.~Krammer\cmsorcid{0000-0002-0548-0985}, A.~Li\cmsorcid{0000-0002-4547-116X}, D.~Liko\cmsorcid{0000-0002-3380-473X}, I.~Mikulec\cmsorcid{0000-0003-0385-2746}, J.~Schieck\cmsAuthorMark{2}\cmsorcid{0000-0002-1058-8093}, R.~Sch\"{o}fbeck\cmsAuthorMark{2}\cmsorcid{0000-0002-2332-8784}, D.~Schwarz\cmsorcid{0000-0002-3821-7331}, M.~Sonawane\cmsorcid{0000-0003-0510-7010}, W.~Waltenberger\cmsorcid{0000-0002-6215-7228}, C.-E.~Wulz\cmsAuthorMark{2}\cmsorcid{0000-0001-9226-5812}
% ...
\par}
\vskip\cmsinstskip
\dag:~Deceased\\
$^{1}$Also at Yerevan State University, Yerevan, Armenia\\
$^{2}$Also at TU Wien, Vienna, Austria\\
% ...
